\documentclass[final,authoryear]{elsarticle}
\usepackage[margin=1in]{geometry}

\usepackage{amssymb}
\usepackage{amsmath}

\usepackage{amsfonts}
\usepackage{booktabs}
\usepackage{subcaption}
\usepackage{multirow}
\usepackage[table]{xcolor}
\usepackage{todonotes}
\usepackage{dirtytalk}
\usepackage{geometry}
\usepackage{pdflscape}

\newcommand{\rand}{{\textup{rand}}}
\newcommand{\kinit}{{k_\textup{init}}}
\newcommand{\kmax}{{k_\textup{max}}}
\newcommand{\ulow}{{u_\textup{l}}}
\newcommand{\uhigh}{{u_\textup{h}}}
\newcommand{\prob}{{\mathbb{P}}}

\newcommand{\ptot}{{p_\textup{tot}}}
\newcommand{\Ttest}{{T_\textup{test}}}
\newcommand{\Tdecision}{{T_\textup{dec}}}
\newcommand{\Tresults}{{T_\textup{res}}}
\newcommand{\Tinactive}{{T_\textup{inactive}}}
\newcommand{\payfix}{{\phi_\textup{fix}}}
\newcommand{\paybonus}{{\phi_\textup{bon}}}
\newcommand{\payavg}{{\phi_\textup{targ}}}
\newcommand{\payrand}{{\phi_\rand}}
\newcommand{\ravg}{{s_\textup{targ}}}
\newcommand{\rrand}{{s_\rand}}
\newcommand{\Rrand}{S^\rand}

\newcommand{\floor}[1]{\left\lfloor #1 \right\rfloor}

\usepackage[acronym]{glossaries}
\newacronym{DPG}{DPG}{Dynamic Population Game}
\newacronym{DeSciL}{DeSciL}{ETH Decision Science Laboratory}
\newacronym{MTurk}{MTurk}{Amazon Mechanical Turk}
\newacronym{MWW}{MWW}{Mann-Whitney-Wilcoxon}
\newacronym[\glsshortpluralkey={MPE}, \glslongpluralkey={Markov Perfect Equilibria}]{MPE}{MPE}{Markov Perfect Equilibrium}
\newacronym{SNE}{SNE}{Stationary Nash Equilibrium}

\journal{Journal of Economic Behavior \& Organization}

\makeatletter
\def\ps@pprintTitle{%
 \let\@oddhead\@empty
 \let\@evenhead\@empty
 \let\@oddfoot\@empty
 \let\@evenfoot\@empty
}
\makeatother

\begin{document}

\begin{frontmatter}



\title{Dynamic Resource Allocation with Karma: \\ An Experimental Study} 


\author[dcs]{Ezzat Elokda} 
\ead{elokdae@kth.se}

\author[ifa]{Saverio Bolognani}
\ead{bsaverio@ethz.ch}

\author[ifa]{Florian D\"orfler}
\ead{dorfler@ethz.ch}

\author[uzh]{Heinrich H. Nax}
\ead{heinrich.nax@uzh.ch}

\affiliation[dcs]{
    organization={Department of Decision and Control Systems, KTH Royal Institute of Technology},
    addressline={Malvinas väg 10}, 
    city={Stockholm},
    postcode={114 28}, 
    country={Sweden}}

\affiliation[ifa]{
    organization={Automatic Control Laboratory, ETH Zurich},
    addressline={Physikstrasse 3}, 
    city={Zurich},
    postcode={8092}, 
    country={Switzerland}}

\affiliation[uzh]{
    organization={Zurich Center for Market Design and SUZ, University of Zurich},
    addressline={Andreastrasse 15}, 
    city={Zurich},
    postcode={8050}, 
    country={Switzerland}}

\begin{abstract}
We perform a behavioral experiment of \emph{karma}, a class of mechanisms for repeated resource allocation with attractive fairness and efficiency properties, in theory.
Individuals in these mechanisms bid non-tradable credits that flow from resource consumers to yielders.
Human subjects recruited on Amazon MTurk are repeatedly and randomly paired to bid karma according to time-varying individual \emph{urgency} to acquire resources.
Treatments varied in the urgency process (frequent moderate versus sporadic high urgency) and the richness of the bidding scheme (binary versus full range).
Benchmarked against random allocation, karma achieves an (almost) Pareto improvement, despite subjects deviating from the theoretically optimal Nash bidding policy:
maximum improvement is attained by subjects deviating by up to one karma bid unit on average, and positive improvement with deviations of up to 3--4 units.
These findings hold across all treatments, among which no significant differences are found, with the exception of the sporadic high urgency process
with binary bidding being (weakly) favorable.
They offer behaviorally robust lower bounds for karma's expected performance in human populations and guide its future implementation.
\end{abstract}




\end{frontmatter}


\section{Introduction}

\begin{figure}[!t]
       \centering
    \includegraphics[width=\textwidth]{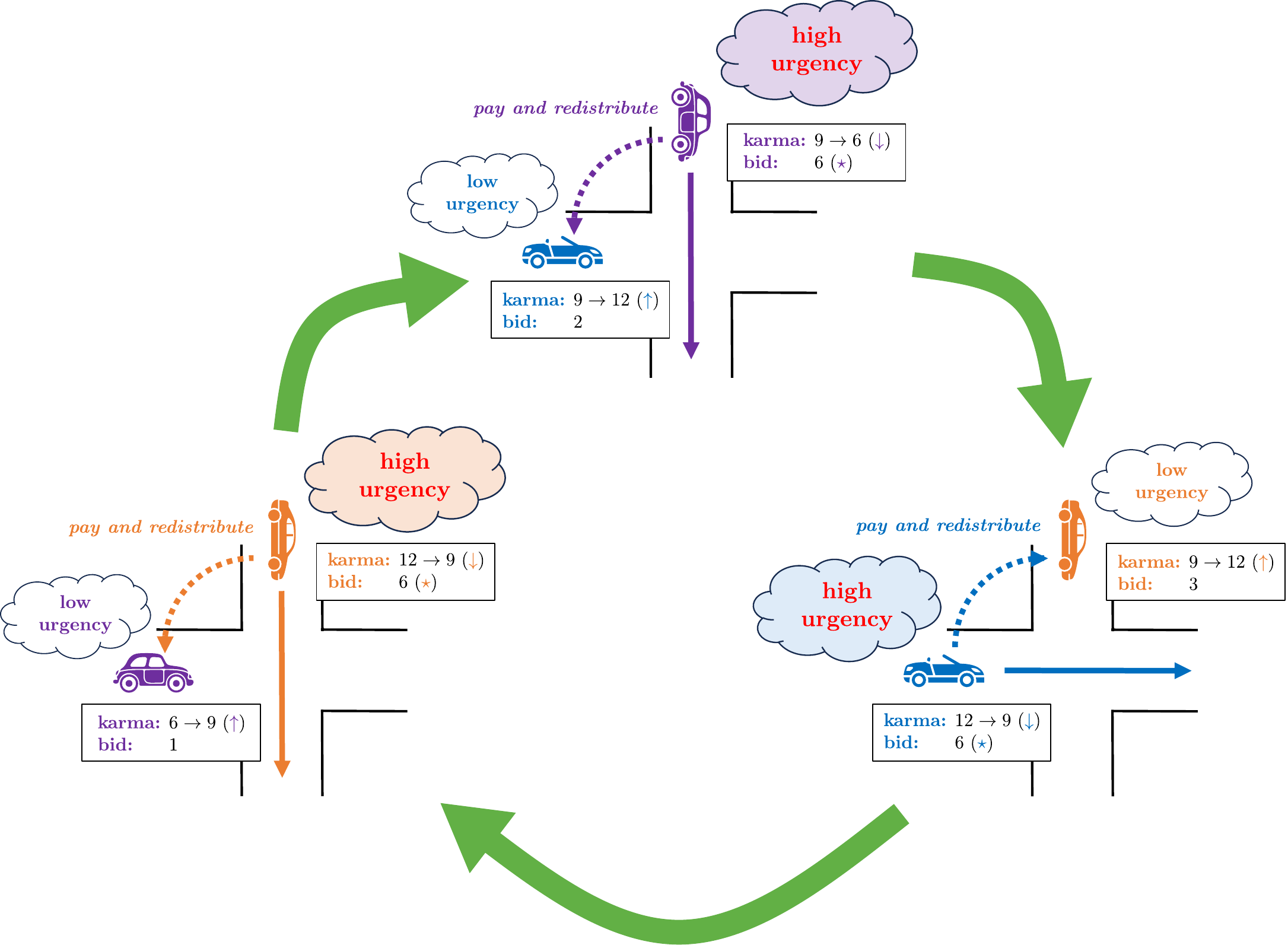}
    \caption{Dynamic resource allocation with karma involving three cars and a repeated sequence of three encounters. Start with the encounter in the top center. The high-urgency purple car has a current karma account of 9 and bids 6, thus outbidding the low-urgency blue car whose karma account is also 9 but bids 2. The bids of the purple car (and, in general, of other cars winning in parallel encounters) are paid and redistributed in the population. As a result, the blue car's karma account goes up by 3 to 12. Let's move along clockwise. Blue now happens to have high urgency and bids 6, thus outbidding and getting priority over orange whose karma account is 9 and bids 3. Now the orange karma goes up by the redistribution share of 3 to 12. Let's move along clockwise. Orange now has high urgency and bids 6, thus outbidding low-urgency purple who has 6 karma left and bids 1. Thus the circle closes, et cetera, et cetera. }
    \label{fig:karma-illustration}
\end{figure}

Efficiency and fairness in determining who gets what and when are the two major objectives in resource allocation situations under scarcity, and many mechanisms and market designs have been proposed~\citep{roth2015gets}.
An important class of allocation problems is when goods are repeatedly and indefinitely allocated amongst a fixed population: for example, farmers require daily access to shared groundwater resources, commuters require regular access to roads, students require frequent access to scarce computing clusters, food banks require daily access to food donations, etc.
In situations like these, it is often the case that one person relative to another gets higher utility from the good in one period but lower utility in another, which we shall refer to as time-varying levels of \emph{urgency}.
A recent innovation to address exactly those kinds of allocation problems goes under the name \emph{karma}~\citep{vishnumurthy2003karma,vuppalapati2023karma,elokda2023self,elokda2025vision}.
The karma mechanism mirrors what Western popular culture associates with the notions of karma and samsara stemming from Indian religions according to which one's deeds in the present (karma) affect the quality of one's future life (phala) and thereafter (samsara) \citep{reichenbach1990law,kyabgon2015karma}.
The mechanism proposed in that literature is implemented via individual accounts of non-tradable credits called karma.
Individuals may bid some amount from their current account at every instance of resource allocation. In the baseline implementation of the mechanism, the highest bidders win and obtain priority for the resource, and must pay their bids, which are then \emph{redistributed in the population}\footnote{This corresponds to the first-price auction implementation of the mechanism. Other auctions, like second-price, have also been discussed.}.
Figure~\ref{fig:karma-illustration} illustrates how the karma mechanism works out to the benefit of everyone at hand of an example involving three infinitely re-occurring road intersection encounters. 

The karma mechanism is simple and appealing, but not yet frequently used in practice. An exception is the `choice system' proposed by \cite{prendergast2022allocation}, which is adopted by U.S. food banks, where karma tokens are called `shares.' Food banks under the choice system bid shares to obtain priority for available food donations, pay their bids upon winning based on a first-price auction mechanism, and the total payment of shares is redistributed at the end of each day\footnote{\cite{prendergast2022allocation} discusses alternatives to first-price auctions and their potential benefits, but mentions that the real-world implementation partners had a strong preference for first-price owed to its transparency and simplicity.}. This real-world example is emblematic of the kind of resource allocations for which karma is suited: the use of monetary transfers is deemed inappropriate or highly undesirable; the allocations are repeated frequently (i.e., daily); and, importantly, there is no finite horizon in sight for when the allocations will cease to occur.
Indeed, the design choice of redistributing the paid shares is critical to address the infinite repetition of the allocation: by forming a `closed economy' in which total shares are preserved over time, a stationary regime can be reached and henceforth repeated indefinitely.

The idea of sacrificing resource consumption today in favor of future consumption in periods of higher urgency is intuitive, and it is not surprising that this same idea underpins several related lines of works, including linking decisions across periods~\citep{jackson2007overcoming,hortala2010inefficiencies,escobar2013efficiency}, trading favors~\citep{mobius2001trading,olszewski2018efficient1,olszewski2018efficient2,leo2017taking}, token or artificial currency-based mechanisms~\citep{johnson2014analyzing,gorokh2021monetary,gorokh2021remarkable,siddartha2023robust}, and trading votes~\citep{casella2021does,casella2005storable,casella2006experimental,hortala2010simple,hortala2012qualitative,casella2019trading,casella2021trading,macias2024storable}.
With respect to these works, which are discussed in Section~\ref{sec:literature}, karma and the aforementioned choice system have the distinguishing feature of forming closed economies that are particularly suited for infinitely repeated resource allocations.

The choice system, which has indeed resulted in significant aggregate gains in terms of food bank participation and fluidity of donations~\citep{prendergast2022allocation}, provides empirical evidence that karma mechanisms can be successful in practice.
On the theory side, recent works have proposed game-theoretic models of karma mechanisms in repeated resource allocation settings~\citep{elokda2023self,elokda2024carma,elokda2025vision}. These works have been motivated by the use of karma as a public policy instrument that is an equitable alternative to classical (monetary) congestion pricing policies, e.g., for allocating priority roads and other public infrastructures.
However, in order to realize the potential benefits of karma-based policies, and employ these policies effectively in human populations, systematic behavioral evidence is needed.
The present paper contributes a first controlled experimental test of karma to this emergent strand of literature.
We conduct an experiment on karma with different bidding schemes and different urgency processes in order to better understand the behavioral uptake of karma by human actors and its efficiency and distributional consequences.

The theoretical predictions of~\cite{elokda2023self,elokda2024carma} are based on the solution concept of the \emph{\gls{SNE}}: a compact, time-invariant predictor of optimal rational behavior, which is guaranteed to exist due to the preservation of karma.
\cite{elokda2023self} develop computation tools for the \gls{SNE} and show that farsighted \gls{SNE} are almost fully efficient with respect to the private urgency of the users.
It focuses on the setting of repeated pairwise matches competing over a single resource, which generalizes the stylized intersection example illustrated in Figure~\ref{fig:karma-illustration} for a continuum of agents from which pairwise matches are randomly drawn.
We note, however, that the same theoretical and computational framework is applicable to more general settings in which more than two agents compete over more than one resource in each time-step~\citep{elokda2024carma,elokda2024travel}.
The theoretical predictions in these works are based on idealized assumptions including rationality, far-sightedness, perfect adoption, infinite population, and that the stationary conditions for which the \gls{SNE} is optimal are reached.
In this paper, we perform a behavioral investigation with real humans who do not necessarily follow such idealized assumptions.
In particular, we recruit subjects from an online sample via \gls{MTurk}, sometimes criticized for more randomness in the resulting choice data, which for us is not a concern as we are interested in evidence that is robust to noisy and irrational behaviors of non-experts in such situations.
We adopt the randomized pairwise matching setting of~\cite{elokda2023self} to focus our investigation on the fundamental decision problem of trading-off present versus future resource consumption with karma.
We want to know whether the inexperienced online subjects find it natural to adopt the mechanism, are able to realize efficiency gains, and whether stationarity is attained in practice.
In our analysis we focus on the overall efficiency effects of karma as well as on distributional consequences and associated fairness properties of the mechanism.
In our treatments, we vary the nature of the dynamic urgency distributions (more frequent and less intense versus less frequent and more intense) and the auction process of the karma mechanism (binary bidding versus full bidding).
We use random allocation as the main benchmark to compare our findings to.
This domain-independent benchmark is representative of common schemes that are unaware of the individual private urgency, including fixed turn-taking schemes.


The main insights of our experiment summarize as follows:
\begin{itemize}
    \item Almost all participants benefit in the karma-based allocation compared with random allocation: this is the case for 90\% of the participants, and the remaining 10\% are mostly dropouts who do not participate in karma bidding actively.
    If we look at active participants only, karma led to an almost Pareto improvement.


    \item The realized aggregate benefits, while greater than in random allocation, fall short of theoretical Nash predictions.
    Analysis of the bidding behaviors reveals that close to Nash-level benefits are attained by participants that deviated from Nash by one karma bid unit on average, while positive benefits are still attained by participants that deviated from Nash by up to 3--4 karma bid units on average.
    Moreover, the main deviation to Nash behavior takes the form of irrational over-bidding in low urgency rounds.

    \item The realized variations in the karma and bid distributions over time are small and comparable in magnitude to the variations that would be attained if all participants followed stationary Nash behavior.
    This suggests that an approximately stationary regime is reached despite of the present deviations to Nash.
    Moreover, the variations are smaller under binary bidding than full bidding, suggesting that the karma auctions are particularly predictable under binary bidding.

    \item The above findings hold robustly across all treatment combinations considered.
    No significant differences are found between treatments, but there is (weak) evidence that benefits are particularly pronounced in situations when high urgency is dynamically more intense and less frequent, and the bidding scheme is designed to be minimal (i.e., binary).
\end{itemize}

These findings provide first formal evidence that karma may be used beneficially and robustly in human interactions.
Our study also points in several directions for further theoretical investigation and serves as a benchmark for future experiments.

\section{Related Mechanisms}
\label{sec:literature}

In this section, we highlight how karma mechanisms differ from several previously proposed mechanisms that share the same intuitive idea of trading off between present or future access to resources.

\paragraph{Linking decisions}
Mechanisms based on linking decisions~\citep{jackson2007overcoming,hortala2010inefficiencies,escobar2013efficiency} rely on correlating each individual's reports over time with publicly known distributions of the private urgency, and punishing those individuals that deviate.
A variant of these linking mechanisms was tested behaviorally by~\cite{engelmann2012mechanisms}, which reports near-optimal efficiency gains by the experimental subjects as theoretically predicted.
The mechanism in~\cite{engelmann2012mechanisms} uses exact knowledge of the private urgency process and issues a tailored amount of fixed-value tokens, i.e., for the considered setting of 50-50\% chance of low or high urgency the mechanism issues 20 tokens to spend over 40 periods.
In contrast, karma mechanisms are able to adapt to any private urgency process, through a bidding scheme that does not fix the value of karma.
In our {experiment}, we subject the same karma mechanism to different treatments varying the private urgency processes.

\paragraph{Trading favors}
Mechanisms based on \emph{trading favors}~\citep{mobius2001trading,olszewski2018efficient1} rely on simple book-keeping of favors owed, but these classical mechanisms are tailored to truthful reporting of one's availability to grant favors with no regard to time-varying private urgency. 
\cite{leo2017taking} addresses time-varying urgency specifically for two individuals taking turns to perform chores, while~\cite{olszewski2018efficient2} addresses time-varying urgency in more general settings using probabilistic exchange of karma-like tokens called `chips.'
However, \cite{olszewski2018efficient2}'s mechanism depends on the individual preference distributions in a complex manner and thus does not scale naturally.

\paragraph{Tokens and artificial currency}
Karma-like instruments have been previously referred to as \emph{vouchers}, \emph{tokens}, \emph{scrips}, or \emph{artificial currency}.
We distinguish between \emph{token-based mechanisms} in which the value of the resource is fixed in tokens~\citep{johnson2014analyzing} (typically one resource unit is worth one token); and \emph{artificial currency-based mechanisms} in which, like karma, the value of the resource is determined in an auction-like mechanism~\citep{gorokh2021monetary,gorokh2021remarkable,siddartha2023robust}.
Token-based mechanisms are not well suited to elicit time-varying private urgency;
whereas most previously proposed artificial currency-based mechanisms are tailored to finite resource repetitions: individuals are issued an initial budget of currency to spend over the finite horizon (with no redistribution or other forms of currency exchange).
The non-preservation of total system currency makes these mechanisms less well-suited to infinite resource repetitions: they would require a periodic central endowment of currency (e.g., every month or year), do not forgive mistakes leading to early depletion of currency, and, importantly, lead to non-stationary settings in which optimal strategies depend explicitly on the time left in the horizon.

\paragraph{Trading votes}
Trading votes across issues or proposals is an intuitively appealing and practically prevalent practice, yet it remains unclear to what extent vote trading improves welfare and how to design vote trading mechanisms optimally, as pointed out in~\cite{casella2021does}'s recent review.
\cite{casella2021does} distinguish between two types of vote trading: those in which votes are traded with other voters~\citep{casella2019trading,casella2021trading}; and those in which votes are traded individually with one's \emph{future self} (referred to as storable~\citep{casella2005storable} or qualitative votes~\citep{hortala2012qualitative}).
The latter type of storable votes, which yields particularly favorable efficiency gains in comparison to the other types of vote trading~\citep{casella2021does}, is closely related to the aforementioned class of (finite-horizon) artificial currency mechanisms: voters are issued an initial budget of votes to cast in a (small) finite number of issues~\citep{casella2005storable,casella2006experimental,hortala2012qualitative}.
One recently proposed mechanism by~\cite{macias2024storable} resembles karma more closely: in this mechanism, votes are `paid' by the majority voters and subsequently redistributed, however, \cite{macias2024storable} studies a two voter only model.
Our study thus complements the literature on storable votes, as we are motivated by resource allocations that are repeated more frequently and typically involve more players than in voting.
In contrast to previous experimental studies on vote trading~\citep{casella2006experimental,hortala2010simple,casella2019trading}, our {experiment} involves significantly more rounds and larger groups.

\paragraph{Summary of related mechanisms}
To summarize, the present state of literature does not offer any natural mechanisms that simultaneously a) can be repeated indefinitely; b) are able to elicit private time-varying preferences; and c) do not require exact public knowledge of potentially private preference dynamics to do so.
Karma mechanisms fill this gap.
The lack of natural alternatives furthermore motivates our choice of adopting random allocation as the main benchmark in this first behavioral investigation of karma.
\section{Experimental Methods}


We conducted a balanced two-by-two factorial experiment with 400 participants in total. Treatments varied in the \emph{dynamic urgency process} of the participants and the \emph{richness of the karma scheme}.
For urgency, we distinguished between a \emph{low stake} process where participants have frequent events with moderate urgency, and a \emph{high stake} process where participants have rare events with high urgency.
For richness, we tested a \emph{binary} scheme where participants can choose from only two bid levels that depend on their karma, and a \emph{full range} scheme where participants have full choice over the bid up to their karma.

\begin{table}[!h]
\centering
\caption{List of notation and parameter values}
\label{tab:parameters}
\begin{tabular}{llcc}
\toprule
Parameter \hspace{1pt} & Description \hspace{1pt} & \hspace{1pt} Low Stake \hspace{1pt} & \hspace{1pt} High Stake \hspace{1pt} \\
\hline \hline
& & & \\[-9pt]
$N$ & Number of participants & \multicolumn{2}{c}{20} \\
$T$ & Number of rounds & \multicolumn{2}{c}{50} \\
$\kinit$ & Initial karma & \multicolumn{2}{c}{9} \\
$\kmax$ & Max karma & \multicolumn{2}{c}{18} \\
$\ulow$ & Low urgency level & \multicolumn{2}{c}{1} \\
$\uhigh$ & High urgency level & 5 & 9 \\
$\mathbb{P}(\uhigh)$ & High urgency probability & 0.5 & 0.25 \\
$\ravg$ & Target score & 90 & 101.25 \\
$\rrand$ & Random score & \multicolumn{2}{c}{37.5} \\
$\payavg$ & Target bonus fee & \multicolumn{2}{c}{\$10} \\
$\payrand$ & Random bonus fee & \multicolumn{2}{c}{\$1} \\
$\payfix$ & Fixed fee & \multicolumn{2}{c}{\$1.5} \\
$\Ttest$ & Number of test rounds & \multicolumn{2}{c}{5} \\
$\Tdecision$ & Decision inactivity timer & \multicolumn{2}{c}{10s (test rounds: 20s)} \\
$\Tresults$ & Results inactivity timer & \multicolumn{2}{c}{\phantom{1}5s (test rounds: 20s)} \\
$\Tinactive$ & Inactivity counter & \multicolumn{2}{c}{6} \\
\bottomrule
\end{tabular}
\end{table}

\subsection{The Game}

The game we study is one that proceeds with $N=20$ participants over $T=50$ rounds.
All participants receive an initial endowment of integer karma $k(0)=k(1)=\kinit=9$ and an initial game score $s(0)=0$.
At each round $t \in \{1,\dots,T\}$, participants are randomly matched in pairs to compete over a shared resource, and they do not receive information about their anonymous opponent.
Each participant is given a private urgency value $u(t) \in \{\ulow, \: \uhigh\}$ that is drawn randomly and independently from a process $\mathbb{P}(u)$ that is identical for all players, and must place a sealed integer bid $b(t) \leq k(t)$.
In each pairwise matching, the higher bidder gets allocated the resource: the score $s(t) = s(t-1) + u(t)$ increases by the urgency and the bid is collected as payment.
The lower bidder does not get allocated the resource: the score $s(t) = s(t-1)$ does not increase and no payment is collected.
Ties are settled randomly.
To keep the total amount of karma constant, at the end of each round, the total collected payment $\ptot$ is uniformly redistributed to all participants.
In order to keep the karma states visited by participants integer and finite, and since the uniform redistribution share is not necessarily integer, the system internally keeps track of floating point karma for each participant, but displays the integer part only to the participant in each time-step.
Moreover, each participant is allowed a maximum karma level $\kmax=18$ that the redistribution respects.
Participants with $\kmax$ do not receive additional redistribution which instead gets issued uniformly to the others.
After the redistribution the game proceeds to the next round $t+1$.
To summarize the information available to participants, they know that all participants start with the same karma and follow the same urgency process, and in each round, they are informed of their own urgency and (integer) karma, but not the urgency and karma of their anonymous opponent.

Table~\ref{tab:parameters} summarizes the notation and parameter values introduced above and hereafter.

\subsection{Treatments}

We follow a two-by-two factorial treatment design, where we vary the \emph{dynamic urgency process} of the participants and the \emph{richness of the karma scheme}.
Table~\ref{tab:treatments} summarizes the treatment configurations schematically.
For each of the four resulting treatments we run 5 independent game groups with 20 participants per group resulting in 100 participants per treatment and 400 participants in total.

\begin{table}[!h]
\centering
\caption{2x2 experimental design.}
\label{tab:treatments}
\begin{tabular}{ll||l|l}
\toprule
& & \multicolumn{2}{c}{\textbf{Richness of scheme}} \\
& & \multicolumn{1}{c|}{Binary} & \multicolumn{1}{c}{Full Range} \\
\hline \hline
\multirow{3}{*}[-5pt]{\rotatebox[origin=c]{90}{\textbf{Urgency process}}} \hspace{1pt} & & & \\[-9pt]
& \multirow{1}{*}[17pt]{Low Stake} \hspace{1pt} & \hspace{2pt} \includegraphics[height=0.1\textwidth]{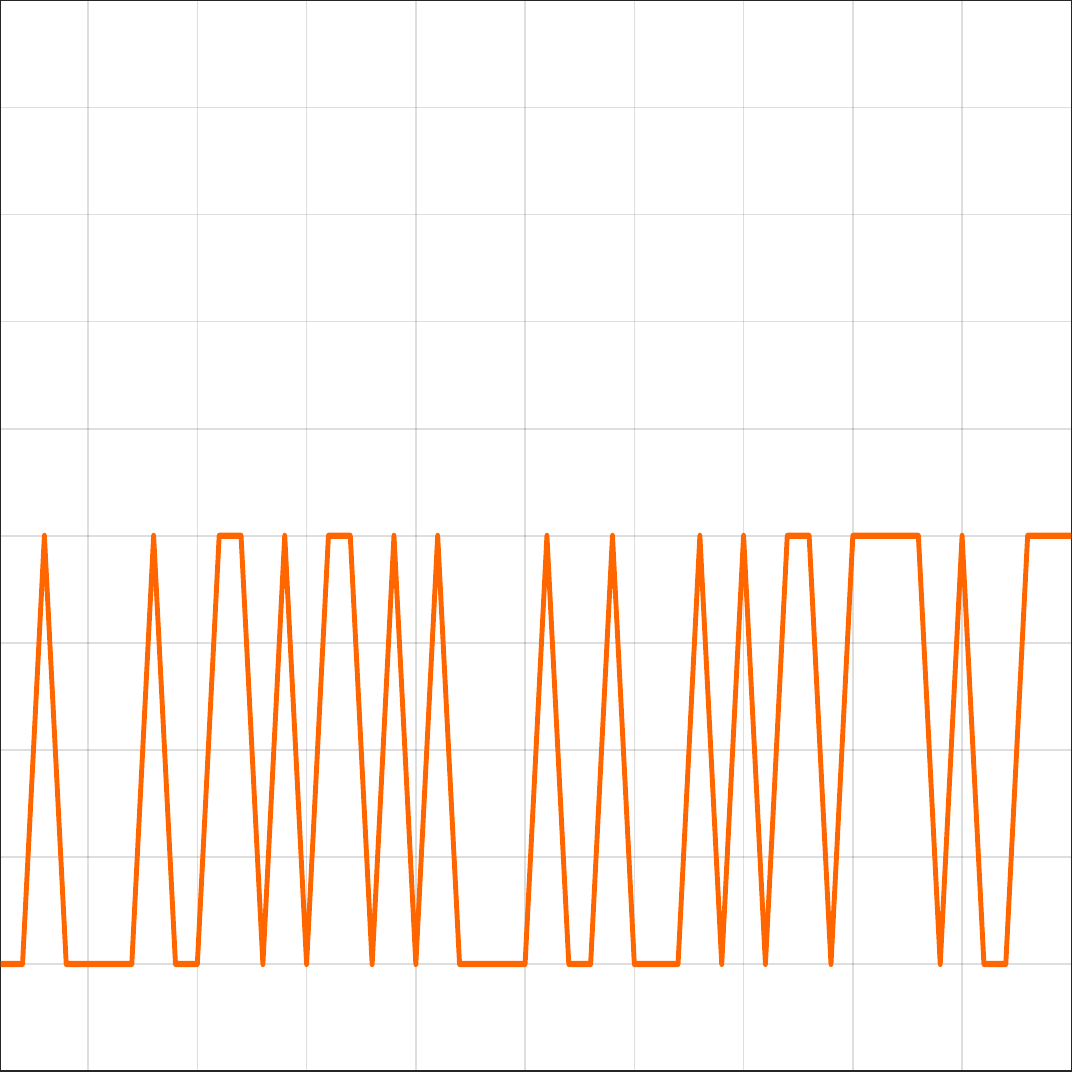} \hspace{5pt} \includegraphics[height=0.1\textwidth]{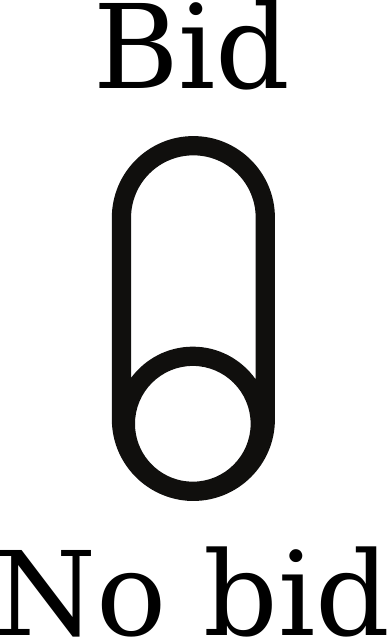} \hspace{2pt} & \hspace{2pt} \includegraphics[height=0.1\textwidth]{figures/low-stake-icon.pdf} \hspace{5pt} \includegraphics[height=0.1\textwidth]{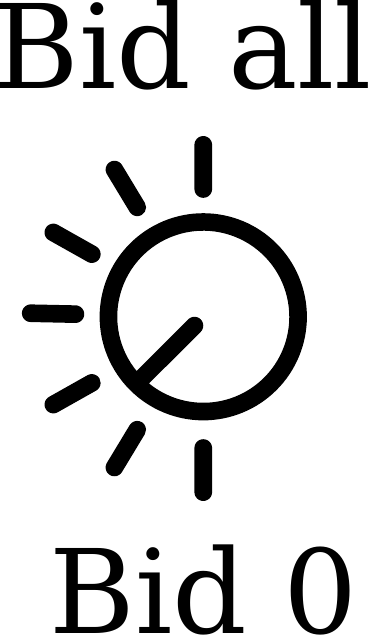} \hspace{2pt} \\ 
\cmidrule{2-4}
& \multirow{1}{*}[18pt]{High Stake} \hspace{1pt} & \hspace{2pt} \includegraphics[height=0.1\textwidth]{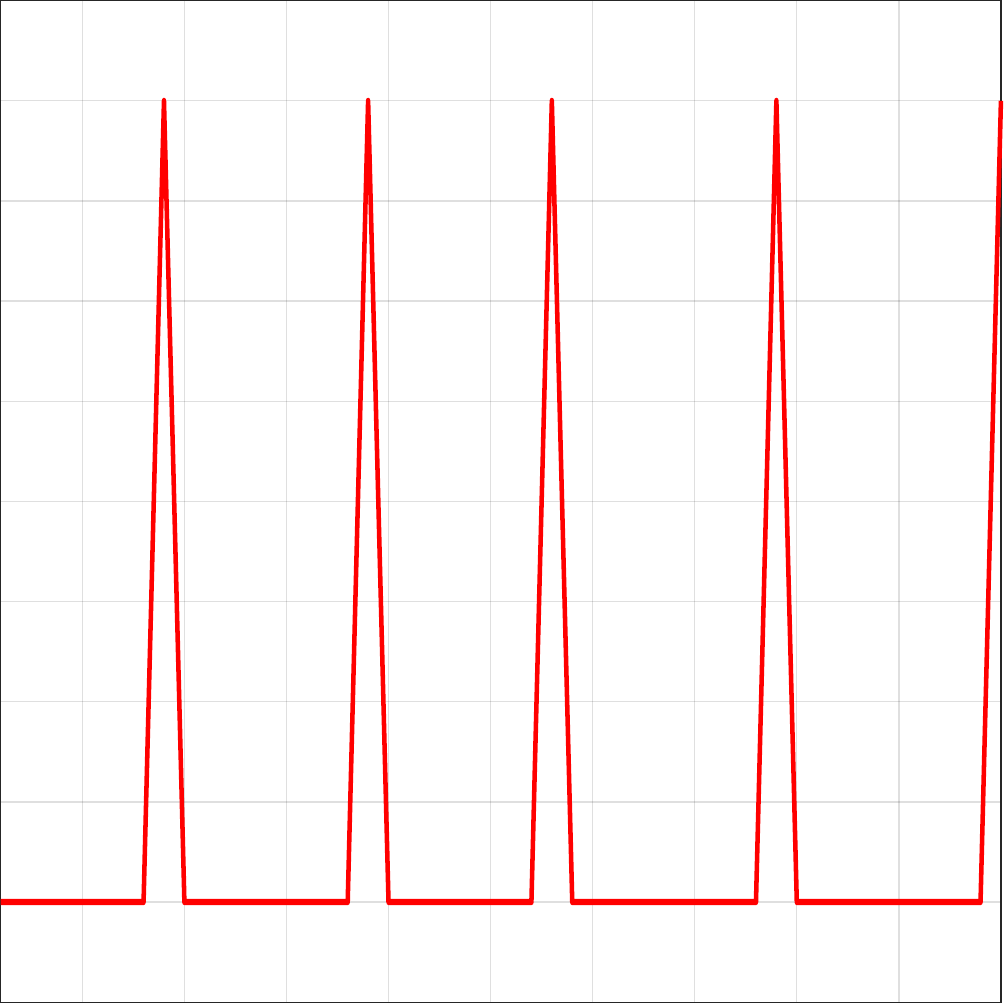} \hspace{5pt} \includegraphics[height=0.1\textwidth]{figures/binary-icon.pdf} \hspace{2pt} & \hspace{2pt} \includegraphics[height=0.1\textwidth]{figures/high-stake-icon.pdf} \hspace{5pt} \includegraphics[height=0.1\textwidth]{figures/full-range-icon.pdf} \hspace{2pt} \\
\bottomrule
\end{tabular}
\end{table}

\subsubsection{Urgency Process}

The urgency processes in both treatment variations of \emph{low stake} and \emph{high stake} have the same low urgency $\ulow=1$, but differ in the magnitude and frequency of the high urgency (low stake: $\uhigh=5$, $\prob(\uhigh)=0.5$; high stake: $\uhigh=9$, $\prob(\uhigh)=0.25$).
 {The motivation for these treatment variations is to investigate behavioral effects under urgency processes of different dynamic nature.}
Both processes have the same urgency on average $\mathbb{E}(u)=3$, and therefore the same expected scores under random allocation.
Low stake represents the case where the high urgency event is frequent but moderate, while in high stake the high urgency event is rare but severe.
 {Notice that due to the different dynamic nature of the processes, there is a greater potential to benefit over random allocation in the high stake process than in the low stake process (cf. Nash efficiency gains in Figure~\ref{fig:mean-efficiency-gains}).}

        
        
        
        
        
        

\subsubsection{Richness of Karma Scheme}

In the treatment variation of \emph{binary}, participants can only choose between two bid levels: $0$ or $\floor{\frac{k}{2}}$ (i.e., half their karma, rounded down to the nearest integer).
In the treatment variation of \emph{full range}, participants can choose any integer bid up to their karma.
 {The motivation of these treatment variations is to investigate the behavioral effects of a reduced action space, i.e., binary, where participants either bid or not.}
 {The particular design choice to restrict bids to $0$ or $\floor{\frac{k}{2}}$ in the binary scheme was guided by theoretical Nash predictions:
using tools from~\cite{elokda2023self}, these binary bid levels were predicted to achieve almost the same efficiency at the Nash equilibrium as in full range bidding, for the urgency processes considered (cf. Nash efficiency gains in Figure~\ref{fig:mean-efficiency-gains}).}
 {Therefore, the simpler binary scheme does not trade-off performance, in theory,} and any differences observed in the outcomes of the two bidding schemes will be due to behavioral effects.

\subsection{Theoretical Predictions}
\label{sec:equilibrium}
A mathematical model for this game was developed in~\cite{elokda2023self}, and we provide here a high-level description of the main elements.
The karma economy is modelled as a \emph{\gls{DPG}}~\citep{elokda2024dynamic} in which it is assumed that there is a continuum of players, each with private individual states (i.e., the urgency and karma) and playing an individual action in each time-step (i.e., the karma bid).
The time horizon is infinite and the players seek to maximize their discounted infinite-horizon payoffs.
They are assumed to follow symmetric, stationary Markovian policies that are functions of their individual states only (and not of the private states of others).
A policy is thus a map from individual urgency and karma to a (potentially randomized) bid.
A macroscopic description of the game, referred to as the \emph{social state}, is given by the \emph{pair} of policy \emph{and} state distribution, the latter specifying the fraction of players occupying different states at any given time-step.
An individual player thus plays a game against the social state from which a sequence of random anonymous opponents and their bids are drawn.
A natural solution concept in this setting is the \emph{\textbf{\acrfull{SNE}}}, which is a special social state satisfying the following two coupled fixed point conditions:
\begin{itemize}
    \item \emph{Optimality:} The \gls{SNE} policy is optimal for the discounted infinite-horizon problem of each player given that all players follow the \gls{SNE} policy and their states are distributed as per the \gls{SNE} state distribution;
    
    \item \emph{Stationarity:} The \gls{SNE} state distribution is \emph{stationary} under the Markovian dynamics generated by the \gls{SNE} policy; thus the optimality of the policy is consistent.
\end{itemize}
A \gls{SNE} is guaranteed to exist in karma \glspl{DPG}~\citep[Theorem~1]{elokda2023self}, and approximates \emph{\glspl{MPE}} of suitably defined finite-player stochastic games for a sufficiently large number of players~\citep[Theorem~2]{adlakha2015equilibria}.

\begin{figure}[!tb]
    \centering
    \begin{subfigure}[b]{0.49\textwidth}
        \centering
        \includegraphics[width=\textwidth]{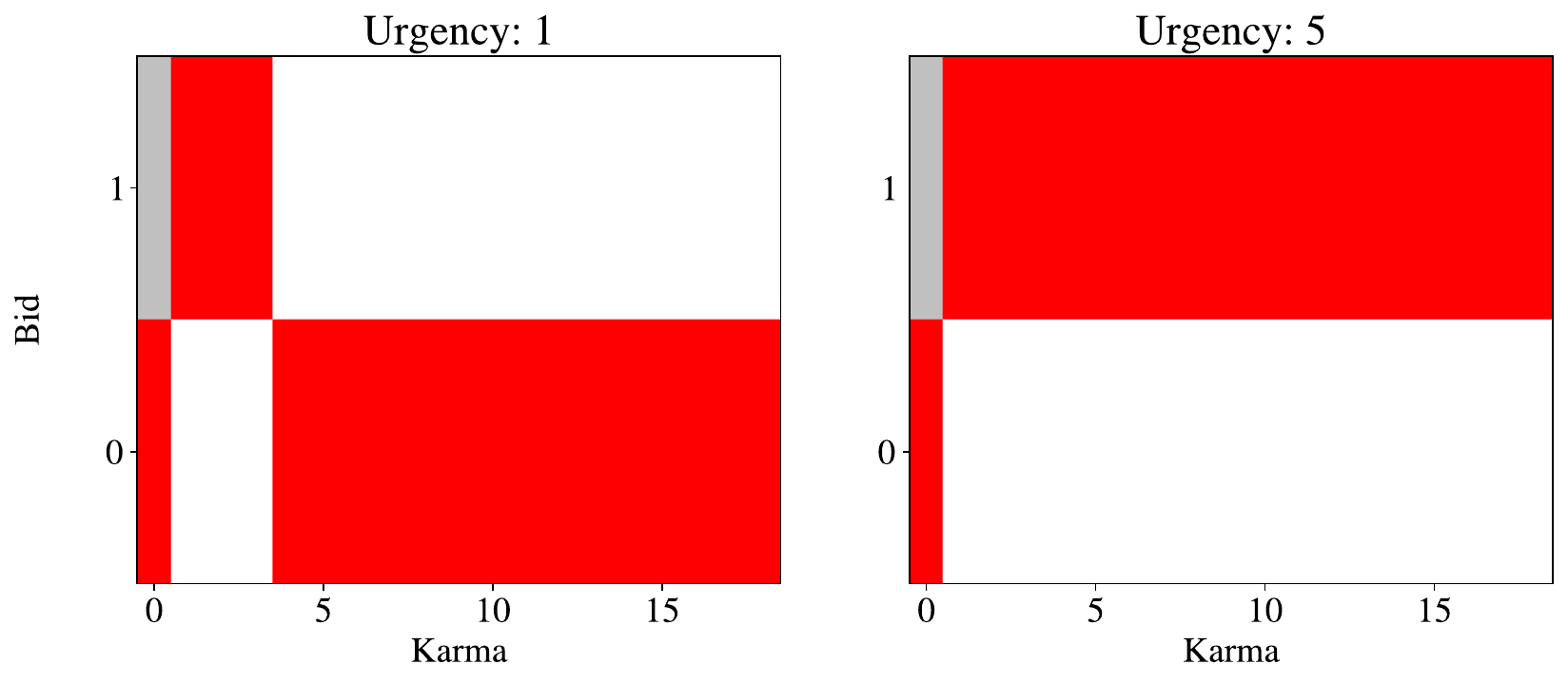}

        \includegraphics[width=0.5\textwidth]{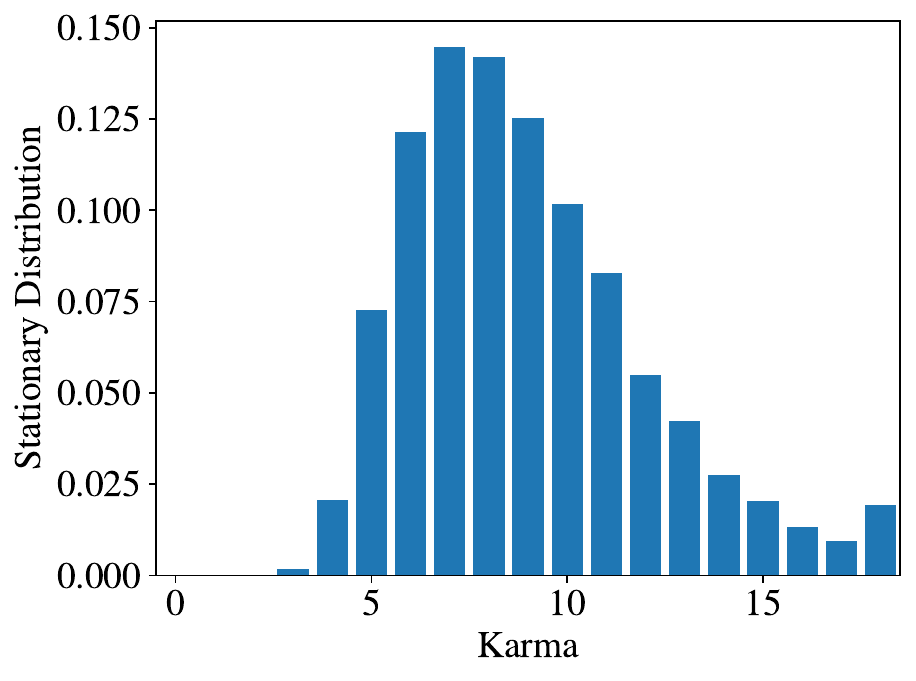}
        \caption{Low Stake, Binary.}
        \label{fig:equilibrium-low-stake-binary}
    \end{subfigure}
    \hfill
    \begin{subfigure}[b]{0.49\textwidth}
        \centering
        \includegraphics[width=\textwidth]{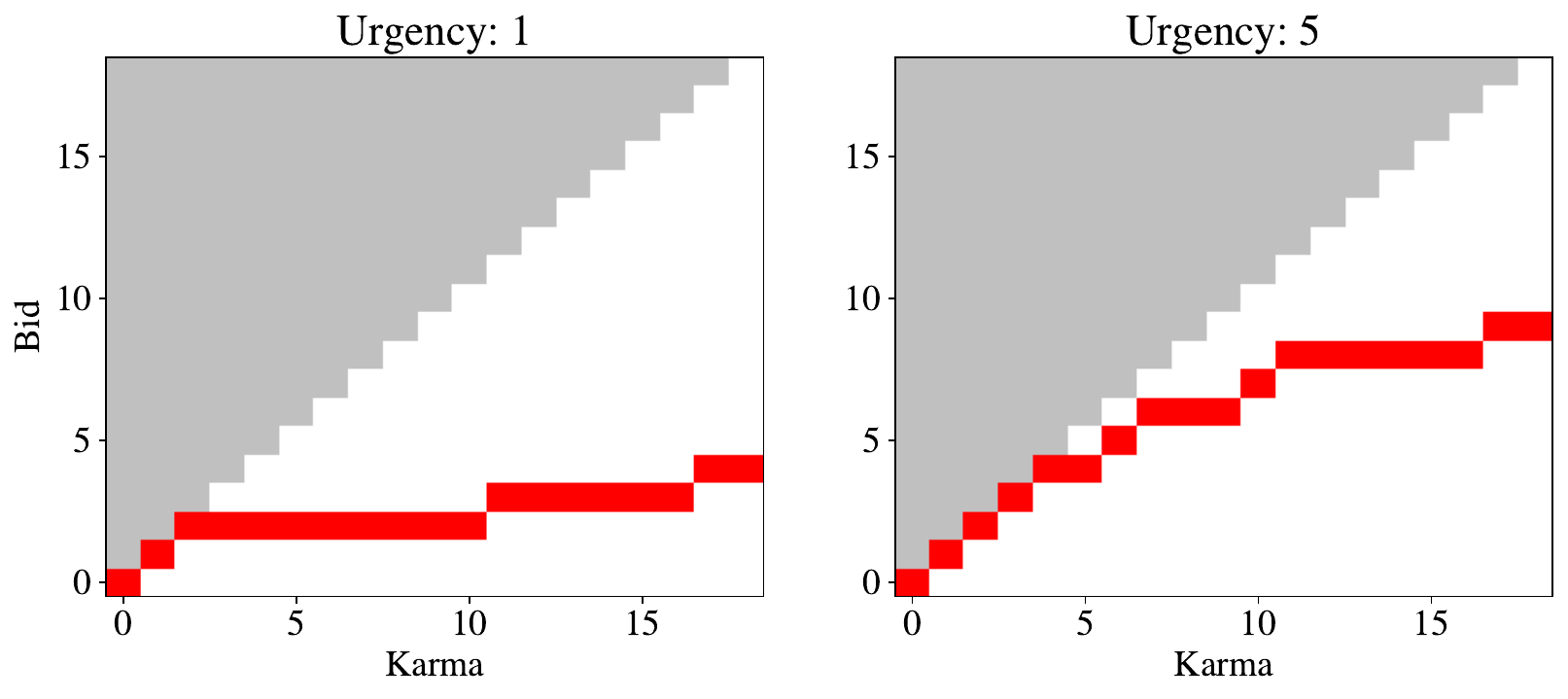}

        \includegraphics[width=0.5\textwidth]{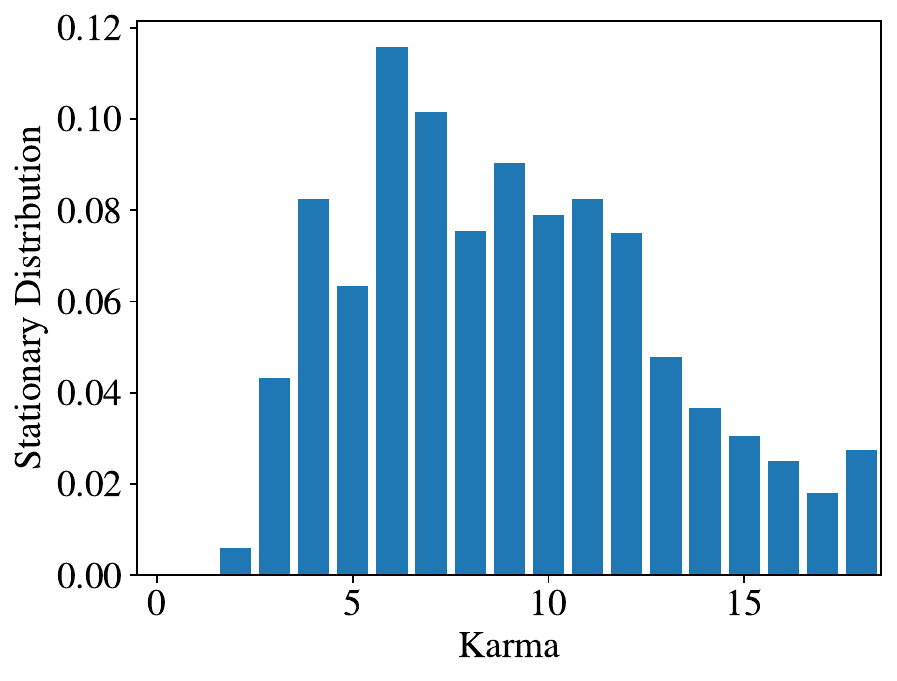}
        \caption{Low Stake, Full Range.}
        \label{fig:equilibrium-low-stake-full-range}
    \end{subfigure}

    \medskip
    
    \begin{subfigure}[b]{0.49\textwidth}
        \centering
        \includegraphics[width=\textwidth]{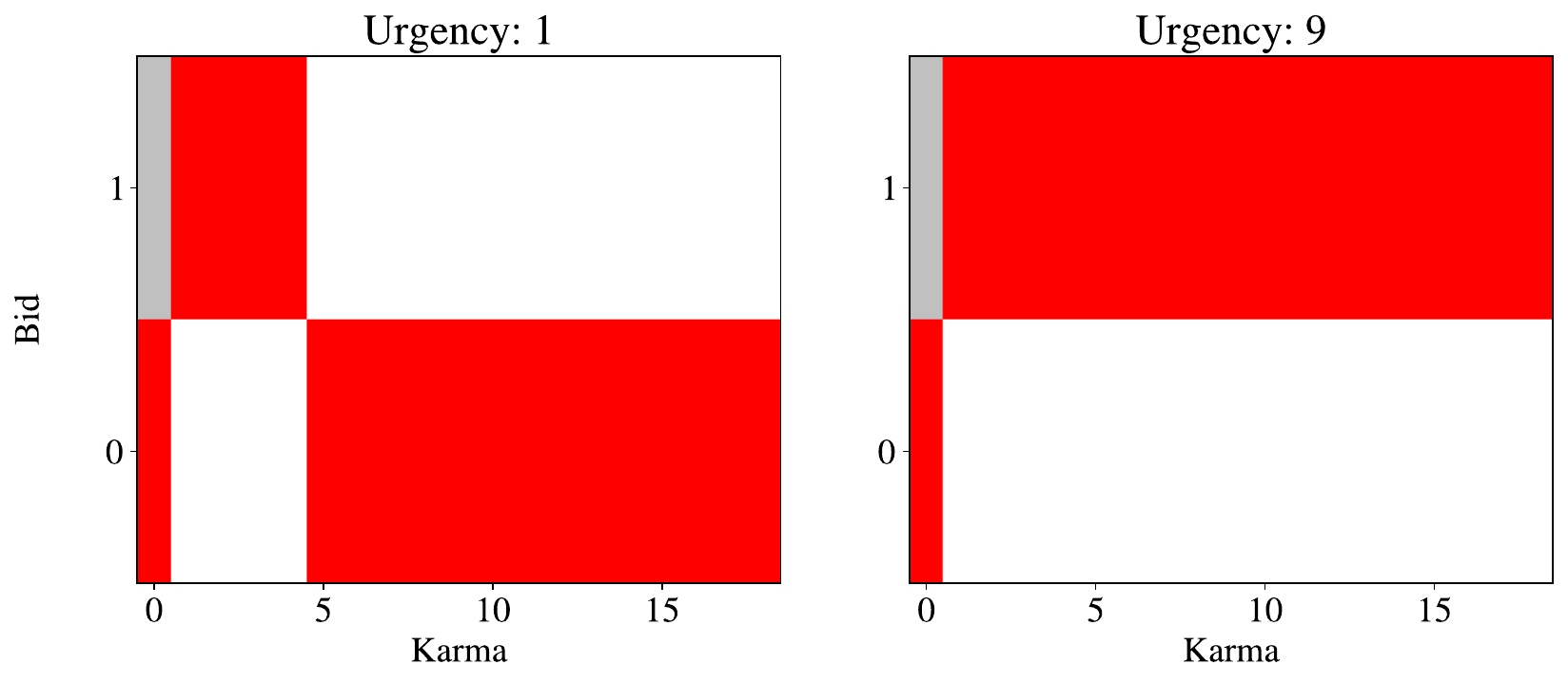}

        \includegraphics[width=0.5\textwidth]{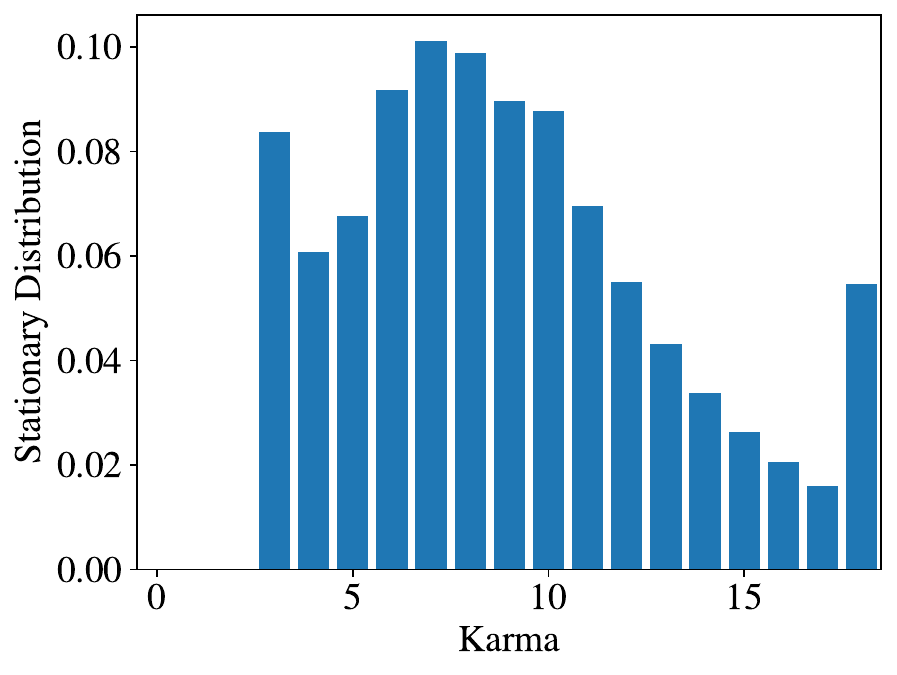}
        \caption{High Stake, Binary.}
        \label{fig:equilibrium-high-stake-binary}
    \end{subfigure}
    \hfill
    \begin{subfigure}[b]{0.49\textwidth}
        \centering
        \includegraphics[width=\textwidth]{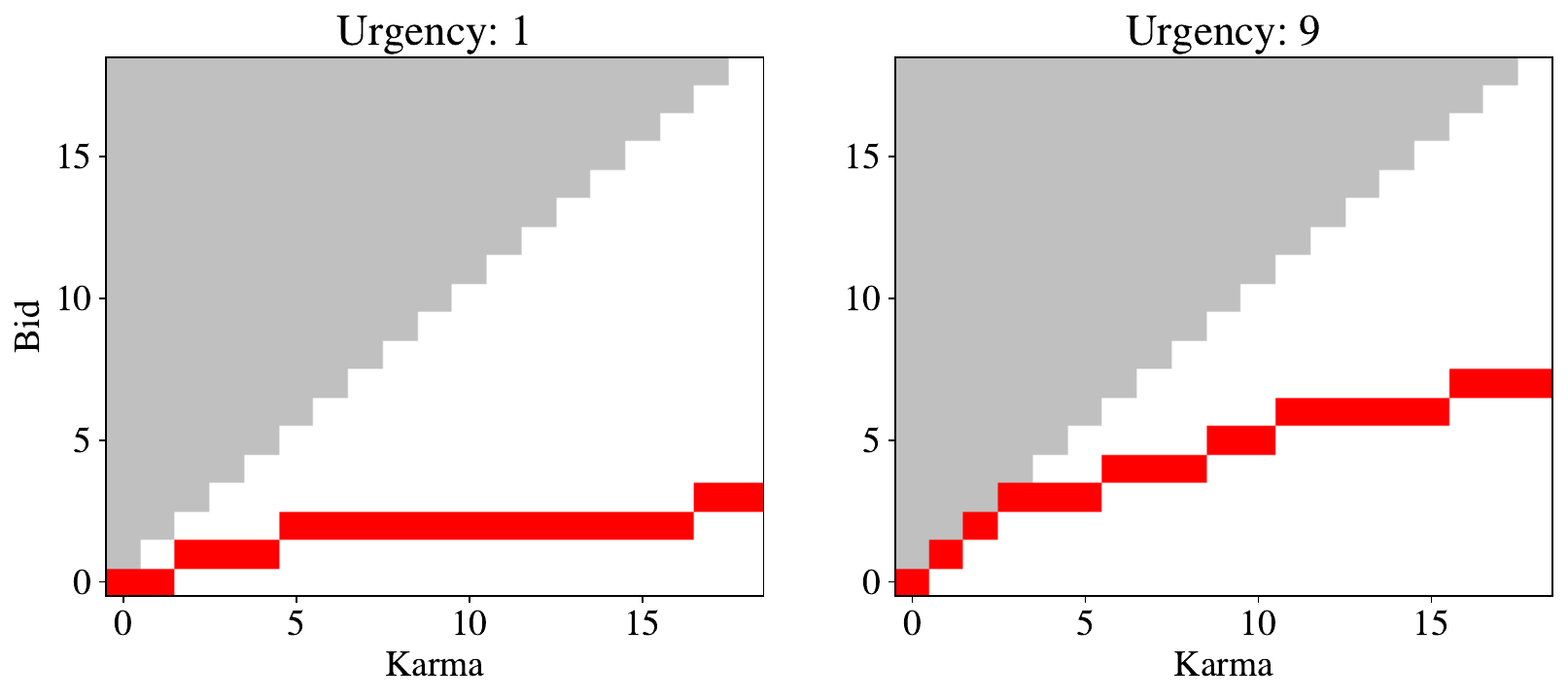}

        \includegraphics[width=0.5\textwidth]{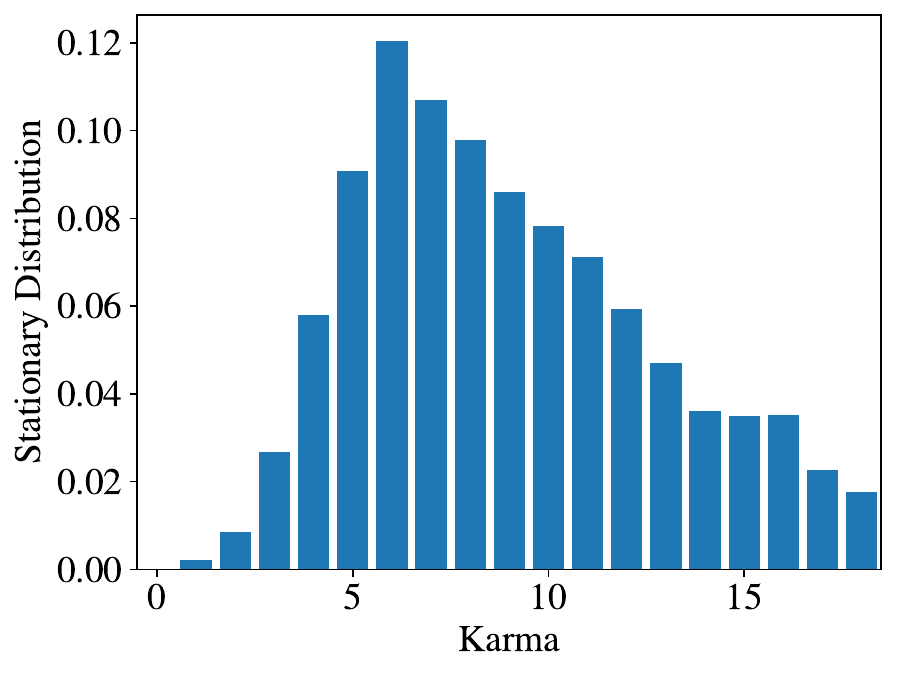}
        \caption{High Stake, Full Range.}
        \label{fig:equilibrium-high-stake-full-range}
    \end{subfigure}
    \caption{Theoretical \acrfull{SNE} for the four treatment combinations and a future discount factor of $0.98$. The \gls{SNE} consists of an optimal symmetric bidding policy (top) and an associated stationary distribution of karma states (bottom).}
    \label{fig:equilibrium}
\end{figure}

The afore-described \gls{DPG} model and its \gls{SNE} solution concept have important advantages for tractability in comparison to finite player and horizon models.
Namely, the continuum of players assumption enables compacting a combinatorial number of joint states into distributions, and the infinite horizon assumption enables looking for compact, indefinitely repeated, stationary policies.
These assumptions are clearly not satisfied in our finite-player and horizon experiment.
Nonetheless, the \gls{SNE} is a natural theoretical benchmark to consider in the absence of a tractable finite-player and horizon model, as motivated also by the fact that it attains the same theoretical near-optimal efficiency when simulated in finite-player and horizon environments.
Figure~\ref{fig:equilibrium} shows what the \gls{SNE} looks like in the different treatments, computed using the tools in~\cite{elokda2023self} for a future discount factor of $0.98$.
All of these \gls{SNE} achieve $>99\%$ of the full possible efficiency, as is evident by the fact that high urgency players are able to outbid low urgency ones the vast majority of times.
The small efficiency loss is due precisely to the occurrence of low karma states in which high urgency players are not able to signal their urgency.


\subsection{Recruitment}
\label{subsec:subjects}
Participants were recruited on \gls{MTurk}, using established IP-based filtering techniques by the \gls{DeSciL} to pre-select participants.
The main pre-selection criteria are:
\begin{itemize}
    \item Participants must be located in the United States;
    \item Participants must have received approval for at least 98\% of their previous \gls{MTurk} tasks;
    \item Participants must not belong to \gls{DeSciL}'s global blacklist of IP ranges. This blacklist includes IP addresses that made multiple identical submissions and displayed other bot-like behaviors in \gls{DeSciL}'s previous \gls{MTurk} projects. 
    \item In order to avoid recruiting the same participant more than once, only one participant is recruited per IP subnet.
\end{itemize}
Despite of these best efforts, it is difficult to completely eliminate noisy and bot-like behaviors of the \gls{MTurk} subject pool considered.
This is not a concern for our experiment, as we seek to analyze the robust performance of karma in such imperfect populations.
Beyond the aforementioned pre-selection criteria, and for the sake of simplicity and anonymity, we did not filter participants based on demographic attributes such as age, gender, education, etc., nor did we collect these attributes; 
cf. \cite{difallah2018demographics,moss2023ethical} for information on general \gls{MTurk} demographics.

\subsection{Implementation}
\label{subsec:implementation}

The game is implemented as a real-time online experiment using oTree~\citep{chen2016otree}, cf.~\cite{decision_science_laboratory_2026_22810650} for the implementation code.
Approval for the experiment was granted by the ETH Zurich ethics committee, and the experiment was pre-registered on the Open Science Framework~\citep{elokda2023karma}.
Two technical pre-tests were conducted and led to a few revisions to the pre-registered implementation plan, cf.~\ref{sec:pretests} for details.

Apart from the technical pre-tests, a total of 20 experiment groups (5 per treatment) were formed over three sessions, each session on a different day, within a two-week window.
The first session included 8 groups and lasted approximately 3 hours; and the other two sessions each included 6 groups and lasted approximately 2 hours.
All sessions were conducted during typical working hours in the United States.
Participants were grouped in block and assigned to treatments according to an alternating, pre-specified order.
This choice attempts to balance the treatments per session and accordingly, we expect no major imbalance in the attributes of participants assigned to different treatments.

The landing page of the experiment gives a context-free introduction to the karma mechanism as a generic resource allocation scheme, cf. Figure~\ref{fig:introduction-page} in~\ref{sec:pages} for the exact appearance and wording.
The page concludes with a hint at a possible application for karma being to establish allocating priority lanes or express roads.
The landing page is followed by a consent form, after which participants must wait for up to 10 minutes to form a group of 20 participants.

Participants are then given detailed instructions, cf. Figure~\ref{fig:instruction-page} in~\ref{sec:pages} for an example, and the opportunity to familiarize with the game over $5$ test rounds that do not contribute to the game score.
The main game then proceeds with $T=50$ rounds.
This relatively large number of rounds is chosen to simulate an infinite repetition, following well-established observations that human subjects do not generally perform backward induction over many rounds and instead treat repetitions as infinite until the final few rounds (cf., e.g., \cite{engelmann2012mechanisms}).
Each round consists of a \emph{decision page}, in which participants are shown their randomly drawn urgency and current (integer) karma balance and must place a bid, and then a \emph{results page} follows which provides feedback on the outcomes of that round.
The bid must be placed within $10$ seconds ($20$ seconds in test rounds), otherwise the participant is signalled as inactive, and the bid defaults to zero.
The results page communicates the outcome of the round in terms of whether priority is granted, the karma payment and redistribution, and the updated karma balance and game score.
In addition, it gives feedback on the opposing bid (but not the opposing urgency and karma).
The participant is given $5$ seconds ($20$ seconds in test rounds) to acknowledge the results page.
Figures~\ref{fig:decision-page}--\ref{fig:results-page} in~\ref{sec:pages} show examples of these pages.

Participants that fail to show activity, either by placing an active bid on the decision page or acknowledging the results page, for more than 6 consecutive times are considered to \emph{drop out} of the experiment.
Dropouts are handled as participants that are not interested in gaining resources: they continue to partake in the pairwise matches and the karma redistribution, but their bids default to zero automatically in each round.
Every five rounds, however, they are prompted for 5 seconds to signal their activity, which allows them to return to normal bidding.
Active participants are not informed of the presence of dropouts as that could bias them to place low bids.
However, a zero opposing bid would be observed in the results page after matching with a dropout.

\subsection{Compensation}

At the end of the game, participants are awarded a monetary payoff consisting of a fixed fee $\payfix=\$1.5$ which compensates for a maximum waiting time of 10 minutes to form a group; and a bonus fee $\paybonus$ that depends on the final score $s(T)$.
Therefore, the bonus fee encourages to be the winning bidder in as many rounds as possible, and especially in high urgency rounds.
The bonus fee is determined according to the following rule:
\begin{itemize}
    \item Persistent dropouts that failed to signal their activity when prompted throughout the experiment are not awarded any payoff, i.e., $\payfix = \paybonus = 0$;

    \item For all other participants, the bonus fee is computed as an affine function of the score, given by
    \begin{align*}
        \paybonus &= \max\left\{(\payavg - \payrand) \: \frac{s(T) - \rrand}{\ravg - \rrand} + \payrand, \: 0\right\}.
    \end{align*}
\end{itemize}

The above affine payment rule linearly interpolates between two payment levels: $\payavg=\$10$ is the payment associated with a target expected average score $\ravg$, which is chosen to compensate the typical participant for a maximum experiment duration of 40 minutes; and $\payrand=\$1$ is the payment associated with an expected score for random bidding $\rrand$, which serves the purpose of disincentivizing random play.
The parameters $\ravg$ and $\rrand$ were tuned based on the results of the technical pre-tests to approximately achieve the desired payments, c.f. Table~\ref{tab:parameters} for their values.
While the bonus payment function was not provided to the participants in mathematical terms, its main features were explained, including that the bonus depends on performance and is expected to be low under random play.
This explanation was provided both on the introductory landing page and again on the detailed instructions page, c.f. Figures~\ref{fig:introduction-page}--\ref{fig:instruction-page} in~\ref{sec:pages} for the exact wording.

Table~\ref{tab:earnings} in~\ref{sec:earnings-statistics} provides statistics on the actual total earnings (fixed plus bonus fee) in the experiment, per treatment and overall.
Dropouts for which $\payfix = \paybonus = 0$ are not included in these statistics, c.f. Section~\ref{sec:dropouts-timeouts} for the frequency of dropouts.
Despite best efforts to a-priori tune the payment function such that the mean total earning is $\$11.5$, the actual overall mean was roughly $\$2$ short at $\$9.61$, and mean earnings were roughly $\$1$ less in the high stake versus the low stake treatments.
Nonetheless, the majority of payments exceeded standard MTurk rates in the realm of $\$7.25$/hr (or $\$4.83$ for a maximum duration of 40 minutes), with the lower quartile lying above this figure in all treatments.

\section{Results}
\label{sec:results}

In order to present our main findings in Sections~\ref{sec:efficiency}--\ref{sec:stationarity}, we first introduce the welfare measures used to quantitatively assess our results in Section~\ref{sec:welfare}; the benchmarks used to compare our findings to in Section~\ref{sec:benchmarks}; and report statistics on the frequency of experimental dropouts and time-outs in Section~\ref{sec:dropouts-timeouts}.
The raw experiment data and analysis code are included in~\cite{elokda_2026_22811110}.

\emph{Remark:} With the best intentions, an analysis plan was included in the pre-registration of our experiment~\citep{elokda2023karma}.
Some but not all of the analyses presented in this section were indeed pre-registered, and there are also further analyses from the pre-registration that are not included.
In~\ref{sec:prereg}, we recall the pre-registered analysis plan, highlight which of the pre-registered analyses are performed in this section, and explain the unfortunate need to replace and/or append the remaining analyses.
This is complemented with several robustness checks of our main findings performed in~\ref{sec:robustness}.

\subsection{Welfare Measures}
\label{sec:welfare}

We perform our analysis on two levels: groups and individuals.
In the group-level analysis, we treat aggregate outcomes of each experimental group as one observation leading to five independent observations per treatment.
In the individual-level analysis, we treat the outcomes of each individual as one observation leading to 100 observations per treatment.
Although these observations are correlated for individuals of the same group, we expect the correlations between any given pair of individuals to be low due to the size and anonymity of group participants.
Analyzing individual observations thus forms a trade-off that allows us to make finer-grained insights than the aggregate group observations, as well as increase the power of our statistical tests in cases where the group-level analysis fails to show significant effects.
A power calculation is performed in \ref{sec:power} which indicates the effect sizes that can be detected by each level of analysis.

Our central welfare measure in both levels of analysis is the \emph{efficiency gain}, which we define next.
On the individual level, for a participant $i$, let $\left(u_i(t)\right)_{t \in \{1,\dots,T\}}$ be the vector of \emph{realized} urgency in the experiment, $S_i = s_i(T)$ the total score at the end of the experiment (which recall does not include the test rounds), and $\Rrand_i = \frac{1}{2} \sum_{t=1}^T u_i(t)$ the expected total score under random allocation, \emph{given the urgency realization}.
Then the efficiency gain of participant $i$ is defined as
\begin{align}
\label{eq:efficiency-gain}
E_i = \frac{S_i - \Rrand_i}{\Rrand_i},
\end{align}
and expresses the relative improvement with respect to the expected random score given the urgency realization.
The definition is conditioned on the urgency realization in order to control for randomness in the urgency processes.
On this basis, in the individual-level analysis we will assess overall efficiency based on the \emph{mean efficiency gain} among participants, and fairness based on whether \emph{Pareto improvements} in efficiency gains are achieved.

In the group level analysis, for a group $G$, the efficiency gain is aggregated over the participants of the group $i \in G$ as follows:
\begin{align}
\label{eq:efficiency-gain-group}
E_G = \frac{\sum_{i \in G} \left(S_i - \Rrand_i\right)}{\sum_{i \in G} \Rrand_i}.
\end{align}

Note that both of our efficiency gain definitions use the scores attained over the whole experiment length (excluding test rounds), which potentially includes undesirable end-game effects.
This choice was made for simplicity, since end-game effects play a minor role for the chosen experiment length of 50, as confirmed with two-sided Wilcoxon signed rank tests of the individual-level efficiency gains in first versus second half of the experiment finding no statistically significant differences overall
($n=400$; first half median $11.11\%$; second half median $12.01\%$; two-sided Wilcoxon test $W=38\,850$, $p=0.7111$;
see Tables~\ref{tab:wilcoxon-first-second-half}--\ref{tab:wilcoxon-first-second-half-no-dropouts} in~\ref{sec:sup-tests} for detailed results per treatment).

\subsection{Benchmarks}
\label{sec:benchmarks}

As reviewed in Section~\ref{sec:literature}, there are no well-established mechanisms for indefinitely repeated allocation settings without monetary transfers or central knowledge of urgency processes.
We thus employ \emph{random allocation (labelled ``Random'')} as the main benchmark in our analysis.
Random is a natural, domain-independent representative of Pareto efficient allocation rules that do not take the private urgency into account, either explicitly or because users would have an incentive to misreport their urgency.
For example, it represents the \emph{random serial dictatorship} mechanism popularly employed in single-shot allocations~\citep{abdulkadirouglu1998random}.
Moreover, a mechanism which allocates resources simply based on player reports would yield random allocation, in theory, since they will have an incentive to report highest urgency always.

In addition to Random, we consider the following auxiliary theoretical benchmarks:
\begin{itemize}
    \item \emph{Turn-taking (labelled ``Turns''):} The resource is allocated to the player that received priority the least amount of times in the past.
    In case of a tie on how often both players received priority previously, the resource is allocated randomly.

    \item \emph{Karma \acrfull{SNE} (labelled ``Nash''):} This is the theoretical benchmark of stationary Nash play as detailed in Section~\ref{sec:equilibrium} and portrayed in Figure~\ref{fig:equilibrium}.
    Recall that the \gls{SNE} considered is for a future discount factor of $0.98$, and achieves near-optimal first-best efficiency in all treatments.
    Thus, this benchmark represents the ideal outcomes that can be attained under karma or any other mechanism.

    \item \emph{Linking mechanism with incorrect prior (labelled ``Linking''):} While the linking mechanism introduced by~\cite{jackson2007overcoming} and studied experimentally by~\cite{engelmann2012mechanisms} serves a different setting than karma mechanisms, namely, finitely repeated allocations in which urgency processes are centrally known, we include as a further theoretical benchmark a linking mechanism in which the central entity \emph{mis-estimates} the private urgency process.
    Namely, we simulate a linking mechanism designed for the high stake urgency process in the low stake treatments, and vice-versa in the high stake treatments\footnote{In particular, in low stake treatments, participants are issued 12 instead of 25 high urgency tokens to consume in 50 rounds; and in the high stake treatments, they are issued 25 instead of 12.}.
    Players bid `as truthfully as possible'~\citep{jackson2007overcoming}, i.e., they start by reporting high urgency truthfully until they either run out of high urgency tokens or the remaining tokens equal the remaining rounds.
    Note that if the central entity uses the correct urgency process, the linking mechanism would achieve near-optimal efficiency similar to the Nash benchmark of the karma mechanism which, however, is not specifically tailored to any urgency process.
\end{itemize}

In the forthcoming analysis, all of the above benchmarks were simulated 100 independent times to obtain confidence intervals.

\subsection{Dropouts and Time-outs}
\label{sec:dropouts-timeouts}

Moreover, our forthcoming analysis includes the observations corresponding to those participants that dropped out of the experiment.
This choice is consistent with our pre-registration in which we did not anticipate the exclusion of any observations~\citep[Section~~2.3]{elokda2023karma}, motivated by the fact that a) the outcomes of active participants are correlated to those of dropouts, and therefore it is not straightforward to exclude the latter; and b) it is the more conservative choice since dropouts perform particularly poorly, and results are quantitatively slightly more favorable if they are excluded.
However, this choice does not affect our main qualitative insights, as our robustness checks in~\ref{sec:robustness-with-without-dropouts} reveals.
This is likely due to the fact that dropouts occupied less than $7\%$ of the population in all treatments, as reported in Table~\ref{tab:dropouts-timeouts}.
This table also reports the frequency of rounds in which active participants failed to show activity in time on either the decision or the results page.
With the time-out frequency lying per treatment in the range of $9.3\%$--$11.5\%$, Table~\ref{tab:dropouts-timeouts} suggests that the vast majority of data analyzed is based on active bidding decisions made on time by active participants.

\begin{table}[!h]
\centering
\caption{Frequency of participants that dropped out of the experiment (labelled ``'Dropouts''), and frequency of rounds in which active participants failed to place a bid or acknowledge the results in time (labelled ``Time-outs''), for all treatments.}
\label{tab:dropouts-timeouts}
\begin{tabular}{ll||l|c|c}
\toprule
& & \multicolumn{3}{c}{\textbf{Richness of scheme}} \\
& & \multicolumn{1}{c|}{Binary} & \multicolumn{1}{c|}{Full Range} & \multicolumn{1}{c}{\textbf{Combined}} \\
\hline \hline
\multirow{4}{*}[-4pt]{\rotatebox[origin=c]{90}{\begin{tabular}{{@{}c@{}}}
     \textbf{Urgency} \textbf{process}
\end{tabular}}} \hspace{1pt} & & & \\[-9pt]
& Low Stake \hspace{1pt} & \hspace{1pt} \begin{tabular}{@{}lc@{}} Dropouts: & 7 / 100 \\ Time-outs: & \phantom{0}534 / 4\,650 \end{tabular} \hspace{1pt} &
\hspace{1pt} \begin{tabular}{@{}c@{}} 5 / 100 \\ 458 / 4\,750 \end{tabular} \hspace{1pt} &
\hspace{1pt} \begin{tabular}{@{}c@{}} 12 / 200 \\ 992 / 9\,400 \end{tabular} \hspace{1pt} \\
\cmidrule{2-5}
& High Stake \hspace{1pt} & \hspace{1pt} \begin{tabular}{@{}lc@{}} Dropouts: & 7 / 100 \\ Time-outs: & \phantom{0}506 / 4\,650 \end{tabular} \hspace{1pt} &
\hspace{1pt} \begin{tabular}{@{}c@{}} 4 / 100 \\ 445 / 4\,800 \end{tabular} \hspace{1pt} &
\hspace{1pt} \begin{tabular}{@{}c@{}} 11 / 200 \\ 951 / 9\,450 \end{tabular} \hspace{1pt} \\
\cmidrule{2-5}
& \textbf{Combined} \hspace{1pt} & \hspace{1pt} \begin{tabular}{@{}lc@{}} Dropouts: & 14 / 200 \\ Time-outs: & 1\,040 / 9\,300 \end{tabular} \hspace{1pt} & \hspace{1pt} \begin{tabular}{@{}c@{}} 9 / 200 \\ 903 / 9\,550 \end{tabular} \hspace{1pt} & \hspace{1pt} \begin{tabular}{@{}c@{}} 23 / 400 \\ 1\,943 / 18\,850 \end{tabular} \hspace{1pt} \\
\bottomrule
\end{tabular}
\end{table}

\subsection{Efficiency Results}
\label{sec:efficiency}

Efficiency went up in all treatments.
Table~\ref{tab:efficiency-groups} reports all group-level efficiency gains, which are positive with the exception of one group (G2 of the high stake-full range treatment).
Moreover, the efficiency gains are statistically significantly positive in all treatments, as elaborated in Table~\ref{tab:wilcoxon-treatment-groups} and Table~\ref{tab:wilcoxon-treatment-individuals} reporting one-sided Wilcoxon signed rank tests of the group-level and individual-level efficiency gains, respectively.
In the group-level tests (Table~\ref{tab:wilcoxon-treatment-groups}), the high stake-full range treatment falls short of $5\%$ significance with a $p$-value of $6.25\%$, due to the occurrence of one negative group-level efficiency gain in this treatment.
Due to the limited power of the group-level test, we turn to the individual-level test of this treatment (Table~\ref{tab:wilcoxon-treatment-individuals}), which confirms the presence of significantly positive efficiency gains with a $p$-value of $0.04\%$.
Moreover, both excluding dropouts from the analysis, as well as performing Welch's $t$-tests instead of Wilcoxon signed rank, show significantly positive efficiency gains in all treatments, cf. Tables~\ref{tab:wilcoxon-treatment-groups-no-dropouts}--\ref{tab:ttest-treatment-individuals} in~\ref{sec:robustness}.

\begin{table}[!h]
\centering
\caption{Group-level efficiency gains, cf. Equation~\eqref{eq:efficiency-gain-group}, for each experimental group. A total of 20 independent groups, five per treatment (corresponding to groups G1--G5), were formed.}
\label{tab:efficiency-groups}
\begin{tabular}{ll||l|c}
\toprule
& & \multicolumn{2}{c}{\textbf{Richness of scheme}} \\
& & \multicolumn{1}{c|}{Binary} & \multicolumn{1}{c}{Full Range} \\
\hline \hline
\multirow{3}{*}[-20pt]{\rotatebox[origin=c]{90}{\begin{tabular}{{@{}c@{}}}
     \textbf{Urgency} \textbf{process}
\end{tabular}}} \hspace{1pt} & & & \\[-9pt]
& Low Stake \hspace{1pt} & \hspace{1pt} \begin{tabular}{@{}lc@{}} G1: & 2.05\% \\ G2: & 5.18\% \\ G3: & 3.44\% \\ G4: & 5.15\% \\ G5: & 9.74\% \end{tabular} \hspace{1pt} &
\hspace{1pt} \begin{tabular}{@{}lc@{}} G1: & 3.17\% \\ G2: & 0.81\% \\ G3: & 8.02\% \\ G4: & 9.14\% \\ G5: & 5.68\% \end{tabular} \hspace{1pt} \\
\cmidrule{2-4}
& High Stake \hspace{1pt} & \hspace{1pt} \begin{tabular}{@{}lc@{}}  G1: & 7.77\% \\ G2: & 3.33\% \\ G3: & 8.23\% \\ G4: & 17.12\% \\ G5: & 17.26\% \end{tabular} \hspace{1pt} &
\hspace{1pt} \begin{tabular}{@{}lc@{}} G1: & 10.05\% \\ G2: & -0.87\% \\ G3: & 5.61\% \\ G4: & 12.94\% \\ G5: & 13.53\% \end{tabular} \hspace{1pt} \\
\bottomrule
\end{tabular}
\end{table}

\begin{table}[!h]
\centering
\caption{One-sided Wilcoxon signed rank tests with test statistics $W$ and associated $p$-values testing if group-level efficiency gains, cf. Equation~\eqref{eq:efficiency-gain-group}, are positive in all treatments.
Tests passing the $5\%$ significance level are highlighted green.
All treatments have $n=5$ observations.}
\label{tab:wilcoxon-treatment-groups}
\begin{tabular}{ll||c|c|c}
\toprule
& & \multicolumn{3}{c}{\textbf{Richness of scheme}} \\
& & \multicolumn{1}{c|}{Binary} & \multicolumn{1}{c|}{Full Range} & \multicolumn{1}{c}{\textbf{Combined}} \\
\hline \hline
\multirow{4}{*}[-5pt]{\rotatebox[origin=c]{90}{\textbf{Urgency process}}} \hspace{1pt} & & & \\[-9pt]
& Low Stake \hspace{1pt} & \begin{tabular}{@{}rc@{}} $W$: & \cellcolor{green!25}$15.0$ \\ $p$: & \cellcolor{green!25}$0.0313$ \end{tabular} &
\hspace{1pt} \begin{tabular}{@{}c@{}} \cellcolor{green!25}$15.0$ \\ \cellcolor{green!25}$0.0313$ \end{tabular} \hspace{1pt} &
\hspace{1pt} \begin{tabular}{@{}c@{}} \cellcolor{green!25}$55.0$ \\ \cellcolor{green!25}$0.0010$ \end{tabular} \hspace{1pt} \\
\cmidrule{2-5}
& High Stake \hspace{1pt} & \hspace{1pt} \begin{tabular}{@{}rc@{}} $W$: & \cellcolor{green!25}$15.0$ \\ $p$: & \cellcolor{green!25}$0.0313$ \end{tabular} \hspace{1pt} &
\hspace{1pt} \begin{tabular}{@{}c@{}} $14.0$ \\ $0.0625$ \end{tabular} \hspace{1pt} &
\hspace{1pt} \begin{tabular}{@{}c@{}} \cellcolor{green!25}$54.0$ \\ \cellcolor{green!25}$0.0020$ \end{tabular} \hspace{1pt} \\
\cmidrule{2-5}
& \textbf{Combined} \hspace{1pt} & \hspace{1pt} \begin{tabular}{@{}rc@{}} $W$: & \cellcolor{green!25}$55.0$ \\ $p$: & \cellcolor{green!25}$0.0010$ \end{tabular} \hspace{1pt} & \hspace{1pt} \begin{tabular}{@{}c@{}} \cellcolor{green!25}$53.0$ \\ \cellcolor{green!25}$0.0029$ \end{tabular} \hspace{1pt} & \hspace{1pt} \begin{tabular}{@{}c@{}} \cellcolor{green!25}$208.0$ \\ \cellcolor{green!25}$<0.0001$ \end{tabular} \hspace{1pt} \\
\bottomrule
\end{tabular}
\end{table}

\begin{table}[!h]
\centering
\caption{One-sided Wilcoxon signed rank tests with test statistics $W$ and associated $p$-values testing if individual-level efficiency gains, cf. Equation~\eqref{eq:efficiency-gain}, are positive for all treatments.
Tests passing the $5\%$ significance level are highlighted green.
All treatments have $n=100$ observations.}
\label{tab:wilcoxon-treatment-individuals}
\begin{tabular}{ll||c|c|c}
\toprule
& & \multicolumn{3}{c}{\textbf{Richness of scheme}} \\
& & \multicolumn{1}{c|}{Binary} & \multicolumn{1}{c|}{Full Range} & \multicolumn{1}{c}{\textbf{Combined}} \\
\hline \hline
\multirow{4}{*}[-5pt]{\rotatebox[origin=c]{90}{\textbf{Urgency process}}} \hspace{1pt} & & & \\[-9pt]
& Low Stake \hspace{1pt} & \begin{tabular}{@{}rc@{}} $W$: & \cellcolor{green!25}$3\,461.0$ \\ $p$: & \cellcolor{green!25}$0.0006$ \end{tabular} &
\hspace{1pt} \begin{tabular}{@{}c@{}} \cellcolor{green!25}$3\,733.0$ \\ \cellcolor{green!25}$<0.0001$ \end{tabular} \hspace{1pt} &
\hspace{1pt} \begin{tabular}{@{}c@{}} \cellcolor{green!25}$14\,349.0$ \\ \cellcolor{green!25}$<0.0001$ \end{tabular} \hspace{1pt} \\
\cmidrule{2-5}
& High Stake \hspace{1pt} & \hspace{1pt} \begin{tabular}{@{}rc@{}} $W$: & \cellcolor{green!25}$3\,778.0$ \\ $p$: & \cellcolor{green!25}$<0.0001$ \end{tabular} \hspace{1pt} &
\hspace{1pt} \begin{tabular}{@{}c@{}} \cellcolor{green!25}$3\,494.0$ \\ \cellcolor{green!25}$0.0004$ \end{tabular} \hspace{1pt} &
\hspace{1pt} \begin{tabular}{@{}c@{}} \cellcolor{green!25}$14\,484.0$ \\ \cellcolor{green!25}$<0.0001$ \end{tabular} \hspace{1pt} \\
\cmidrule{2-5}
& \textbf{Combined} \hspace{1pt} & \hspace{1pt} \begin{tabular}{@{}rc@{}} $W$: & \cellcolor{green!25}$14\,490.0$ \\ $p$: & \cellcolor{green!25}$<0.0001$ \end{tabular} \hspace{1pt} & \hspace{1pt} \begin{tabular}{@{}c@{}} \cellcolor{green!25}$14\,289.0$ \\ \cellcolor{green!25}$<0.0001$ \end{tabular} \hspace{1pt} & \hspace{1pt} \begin{tabular}{@{}c@{}} \cellcolor{green!25}$57\,435.0$ \\ \cellcolor{green!25}$<0.0001$ \end{tabular} \hspace{1pt} \\
\bottomrule
\end{tabular}
\end{table}

\begin{figure}[!tb]
    \centering
    \begin{subfigure}[b]{0.35\textwidth}
        \centering
        \includegraphics[width=\textwidth]{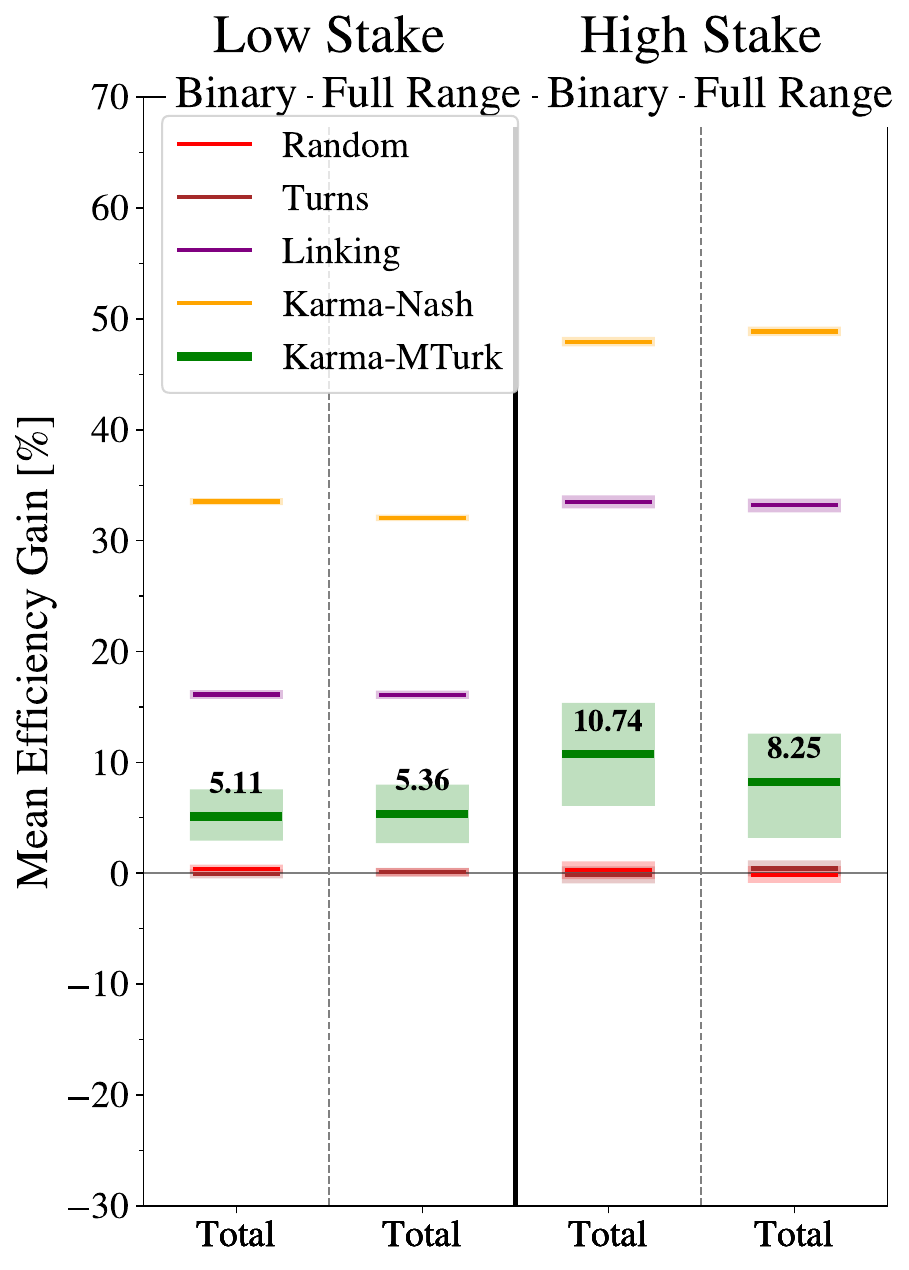}
        \caption{Group level.}
        \label{fig:mean-efficiency-gains-groups}
    \end{subfigure}
    \begin{subfigure}[b]{0.64\textwidth}
        \centering
        \includegraphics[width=\textwidth]{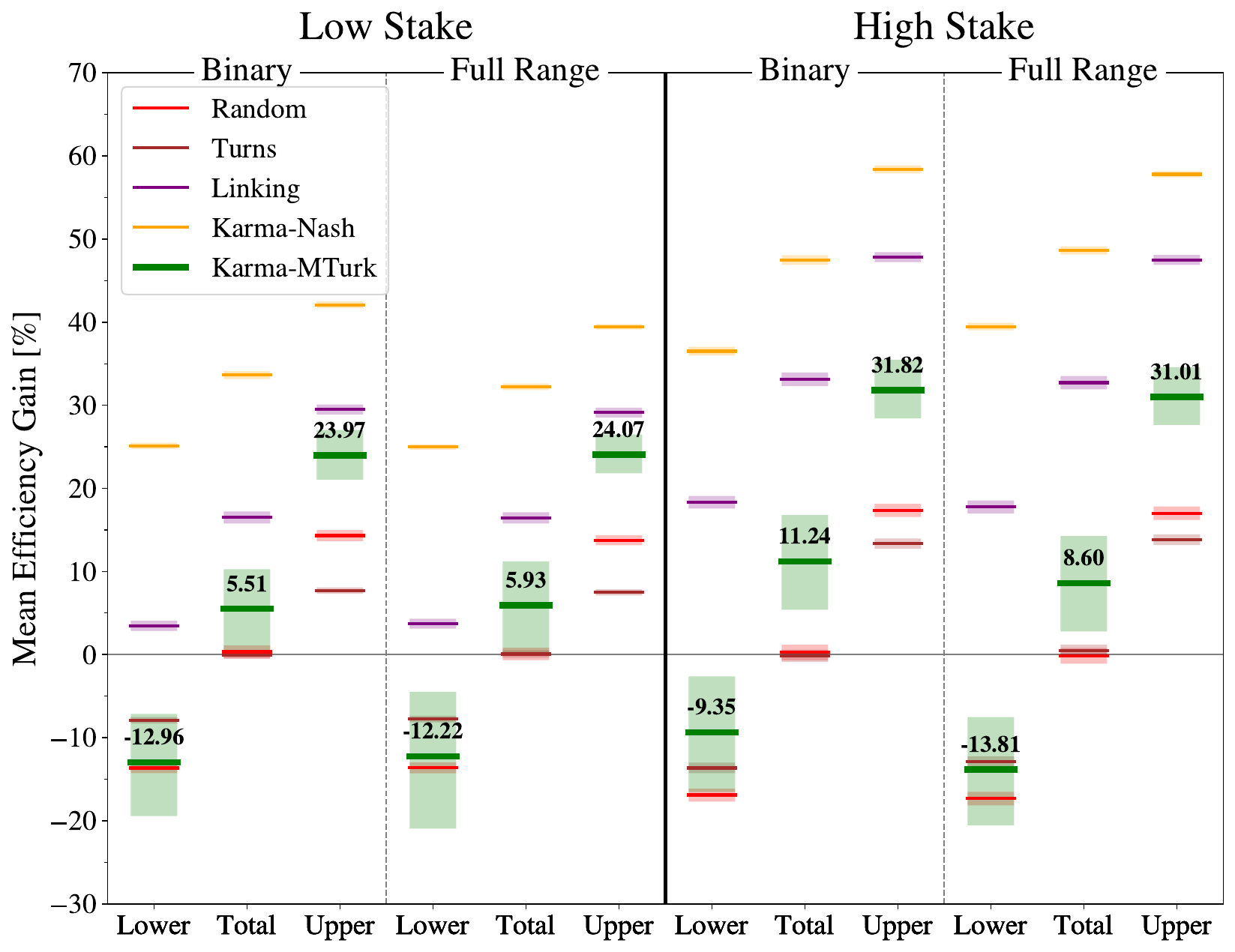}
        \caption{Individual level.}
        \label{fig:mean-efficiency-gains-individuals}
    \end{subfigure}
    
    \caption{Mean efficiency gains in the four treatment combinations attained by the experimental subjects (labelled ``Karma-MTurk'') as well as the simulated theoretical benchmarks introduced in Section~\ref{sec:benchmarks}.
    The means are shown for both (a) the group-level efficiency gains, cf. Equation~\eqref{eq:efficiency-gain-group}; and (b) the individual-level efficiency gains, cf. Equation~\eqref{eq:efficiency-gain}.
    The shaded areas indicate $95\%$ confidence interval estimates using $10\,000$ bootstraps ((a): bootstrapping from 5 group observations for Karma-MTurk and from 100 simulations for the theoretical benchmarks; (b): bootstrapping from 100 individual observations for Karma-MTurk and from $2\,000$ simulated individual observations for the theoretical benchmarks, and note that since there are correlations in the observations of same-group / simulation individuals, the confidence intervals in (b) are heuristic estimates for visualization purposes).
    For the individual-level efficiency gains (b), also shown are the means in the lower and upper halves of the population (labelled ``Lower'' and ``Upper'', respectively), while the total means are labelled ``Total''.
    }
    \label{fig:mean-efficiency-gains}
\end{figure}

This result is corroborated by Figure~\ref{fig:mean-efficiency-gains} which compares the mean efficiency gains attained by the experimental \gls{MTurk} subjects to the simulated theoretical benchmarks introduced in Section~\ref{sec:benchmarks}.
The means were computed with respect to both the group-level efficiency gains (Equation~\eqref{eq:efficiency-gain-group}), cf. Figure~\ref{fig:mean-efficiency-gains-groups}, and the individual-level efficiency gains (Equation~\eqref{eq:efficiency-gain}), cf. Figure~\ref{fig:mean-efficiency-gains-individuals}, and both levels of analysis show similar results in terms of total mean efficiency gains.
Namely, the \gls{MTurk} subjects outperform both the random and turn-taking benchmarks, but fall short of the theoretical karma \gls{SNE} and the linking mechanism (even with incorrect prior).
Random and turn-taking perform similarly since the turn-taking allocation is essentially random with respect to urgency; while the linking mechanism with incorrect prior attains approximately half of the full possible efficiency gains\footnote{The outcome of the linking mechanism is consistent with the considered urgency process mismatch: in the low stake treatments, players are endowed with half of the ex-ante expected correct amount of high urgency tokens; and in the high stake treatments, they are endowed with double the correct amount.}, the latter being almost fully attained by the karma \gls{SNE} in all treatments.
While these results may suggest that it could be more favorable to adopt a linking mechanism, even with an incorrectly assumed urgency process, over a karma mechanism in practice, it is important to note that the performance of the linking mechanism depends on the level of mismatch, and it is possible to construct 
linking mechanisms that perform worse than the \gls{MTurk} subjects even in theory (in addition to the fact that the linking mechanism is not naturally suitable for infinite repetition).

The individual-level analysis in Figure~\ref{fig:mean-efficiency-gains-individuals} additionally provides finer-grained insights on the mean efficiency gains in the bottom and top halves of the population.
The efficiency gains are especially pronounced in the upper half of the population (mean ranging from $23.97\%$ to $31.82\%$); whereas the lower half of the population performs similarly to the lower half under random allocation.
This suggests that overall, karma provides an opportunity for participants to achieve pronounced benefits, without strongly harming those that are less strategic.
An exception is dropouts who perform worse than random and contribute to lowering means in the lower half and overall since with persistent zero bids, they are expected to not receive any resources.
As shown in our robustness checks, cf. Figure~\ref{fig:mean-efficiency-gains-non-dropouts} in~\ref{sec:robustness-mean-median}, excluding dropouts indeed leads to pronouncedly higher mean efficiency gains, both in the lower half of the population and overall, with the lower half under karma more clearly dominating that under random.
We also test the robustness of Figure~\ref{fig:mean-efficiency-gains} against the choice of plotting median efficiency gains instead of means, cf. Figures~\ref{fig:mean-median-efficiency-gains-groups-with-without-dropouts}--\ref{fig:mean-median-efficiency-gains-with-without-dropouts} in~\ref{sec:robustness-mean-median}, which show similar qualitative insights as means, with the exception that medians are less affected by the choice of including or excluding dropouts in the individual-level analysis.

We next compare the efficiency gains between treatments, for which we employ two-sided \gls{MWW} tests to detect statistically significant differences.
The results of these tests are reported in Tables~\ref{tab:mann-whitney-inter-treatment-groups}--\ref{tab:mann-whitney-inter-treatment-individuals-no-dropouts} in~\ref{sec:sup-tests}, respectively for the group-level efficiency gains (Table~\ref{tab:mann-whitney-inter-treatment-groups}), the individual-level gains (Table~\ref{tab:mann-whitney-inter-treatment-individuals}), and the individual-level gains excluding dropouts (Table~\ref{tab:mann-whitney-inter-treatment-individuals-no-dropouts}).
In summary, no statistically significant differences are detected between most treatment pairs.
The main exception is high stake treatments, which combined achieve weakly significantly higher gains than combined low stake treatments in all levels of analyses considered (e.g., group-level, cf. Table~\ref{tab:mann-whitney-inter-treatment-groups}: combined high stake: median $9.14\%$, $n=10$; combined low stake: median $5.16\%$, $n=10$; \gls{MWW} test $U=74$, $p=0.0757$).
This is consistent with the dynamic nature of the two urgency processes, whereby in the high stake process it is ex-ante feasible for all participants to achieve higher efficiency gains compared to the low stake process, cf. difference in Nash efficiency gains between low stake and high stake treatments.
The individual-level analyses additionally reveal that the high stake, binary bidding treatment is particularly favorable.
This treatment achieves significantly higher individual-level gains than the low stake-binary treatment irrespective if dropouts are included or not (e.g., individual-level, dropouts included, cf. Table~\ref{tab:mann-whitney-inter-treatment-individuals}: high stake, binary: median $15.34\%$, $n=100$; low stake, binary: median $7.38\%$, $n=100$; \gls{MWW} test $U=5\,806$, $p=0.0490$).
Excluding dropouts, high stake-binary additionally achieves weakly significantly higher individual-level gains than low stake-full range (individual-level, dropouts excluded, cf. Table~\ref{tab:mann-whitney-inter-treatment-individuals-no-dropouts}: high stake, binary: median $16.92\%$, $n=93$; low stake, full range: median $13.58\%$, $n=95$; \gls{MWW} test $U=5\,061$, $p=0.0848$).
Overall, we find no statistically significant differences between binary and full range bidding (e.g., group-level, cf. Table~\ref{tab:mann-whitney-inter-treatment-groups}: combined binary: median $6.47\%$, $n=10$; combined full range: median $6.85\%$, $n=10$; \gls{MWW} test $U=52$, $p=0.9097$).
This is also in line with the design of the ex-ante feasible efficiency gains under Nash play.

\subsection{Fairness Results}
\label{sec:fairness}

\begin{figure}[!tb]
    \centering
    \includegraphics[width=0.8\textwidth]{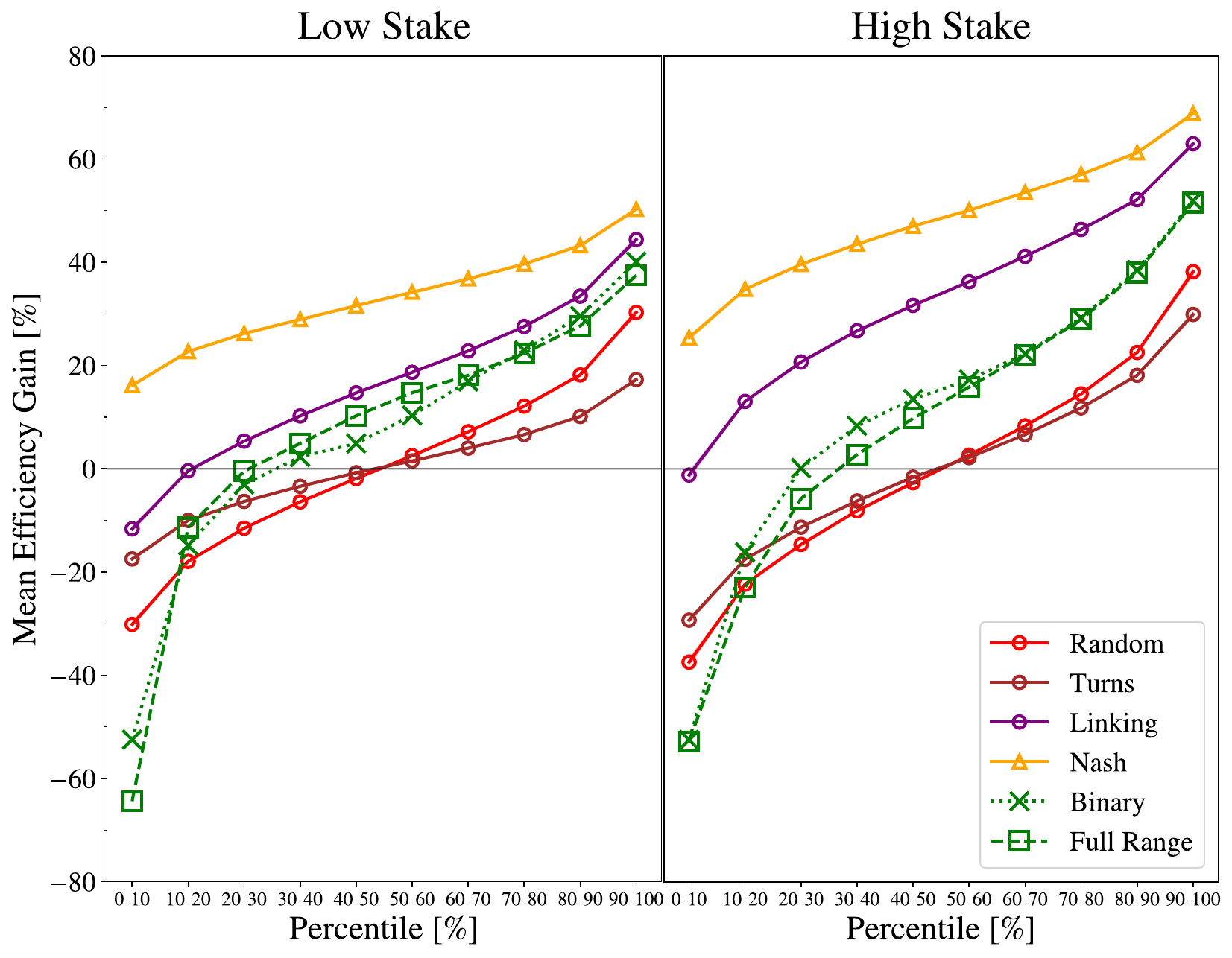}
    \caption{Mean individual-level efficiency gains, cf. Equation~\eqref{eq:efficiency-gain}, per population decile for the four treatment combinations, as attained by the experimental subjects (labelled ``Binary'' and ``Full Range'' for the respective treatments) as well as the simulated theoretical benchmarks introduced in Section~\ref{sec:benchmarks}.
    For the theoretical benchmarks, 95\% confidence interval estimates using $10\,000$ bootstraps from $400$ simulated individual observations in each decile are included, and the confidence intervals narrowly coincide on top of the means}.
    \label{fig:mean-efficiency-gain-per-percentile}
\end{figure}

Figure~\ref{fig:mean-efficiency-gains-individuals} shows that in addition to achieving overall efficiency gains, the karma allocation is more efficient than random allocation in the upper/more fortunate half of the population, and not much less efficient than random in the lower/less fortunate half.
To provide finer-grained insight on the fairness of karma, Figure~\ref{fig:mean-efficiency-gain-per-percentile} further shows the mean individual-level efficiency gain \emph{per population decile} for the four treatments.
It is important to note that while the individual-level efficiency gain (Equation~\eqref{eq:efficiency-gain}) is defined with respect to the ex-ante expected score under random allocation given the urgency realization, not the whole population realizes this score ex-post.
To control for the fact that there will be more or less fortunate individuals due to randomness under random allocation, Figure~\ref{fig:mean-efficiency-gain-per-percentile} also shows the ex-post mean efficiency gains per population decile for random allocation (as well as the other theoretical benchmarks introduced in Section~\ref{sec:benchmarks}).
The key feature to observe is that in all treatments, $90\%$ of the experimental subjects achieve higher efficiency in the karma treatments than if the allocations were random.
Only the lowest decile is worse-off in karma than random.
Recall that as per Table~\ref{tab:dropouts-timeouts}, this decile is dominated by dropouts whose bids defaulted to zero due to inactivity, leading to particularly low scores.
If dropouts are excluded, cf. Figure~\ref{fig:mean-efficiency-gain-per-percentile-non-dropouts} in~\ref{sec:robustness-mean-median}, a Pareto improvement over random is observed across all deciles in almost all treatments.
Comparing instead to the turn-taking benchmark, which results in lower variability in ex-post efficiency gains than random, we find that $80\%$ (and $90\%$ if dropouts are excluded) of the experimental subjects achieve higher efficiency in the karma treatments than if simple, urgency-unaware turn-taking was followed.
These insights are qualitatively similar if we plot medians instead of means, cf. Figure~\ref{fig:mean-median-efficiency-gain-per-percentile-with-without-dropouts} in~\ref{sec:robustness-mean-median}.
While these results hold for all treatments, we note that the treatment combination of high stake-binary is particularly favorable in terms of achieving the largest gap to random allocation across deciles, cf. right panel of Figure~\ref{fig:mean-efficiency-gain-per-percentile}.

To corroborate these findings, we ran two-sided Wilcoxon signed rank tests of the individual-level efficiency gains per decile, comparing the karma treatments to the mean individual-level efficiency gain of the respective decile in the simulated random benchmark, cf. Tables~\ref{tab:mann-whitney-deciles}--\ref{tab:mann-whitney-deciles-no-dropouts} in~\ref{sec:sup-tests}.
With dropouts included in the analysis (Table~\ref{tab:mann-whitney-deciles}), the tests show statistically significant differences in favor of random in the bottom decile (combined karma: median $-52.13\%$, $n=40$; versus mean random: $-33.79\%$; Wilcoxon test $W=26$, $p<0.0001$); either significant differences in favor of karma or no significant differences in the second-lowest decile (combined karma: median $-13.46\%$, $n=40$; versus mean random: $-20.13\%$; Wilcoxon test $W=217.0$, $p=0.0086$); and significant differences in favor of karma in all higher percentiles (e.g., third-lowest decile: combined karma: median $-2.20\%$, $n=43$; versus mean random: $-13.10\%$; Wilcoxon test $W=0.0$, $p<0.0001$).
With dropouts excluded from the analysis (Table~\ref{tab:mann-whitney-deciles-no-dropouts}), even the bottom decile shows no significant differences between karma and random (combined karma: median $-30.33\%$, $n=40$; versus mean random: $-33.79\%$; Wilcoxon test $W=317$, $p=0.2163$); while all higher deciles show significant differences in favor of karma.

\subsection{Analysis of Bidding Behaviors}
\label{subsec:bidding}

\begin{figure}[!tb]
    \centering
    \begin{subfigure}[b]{0.49\textwidth}
        \centering
        \includegraphics[width=\textwidth]{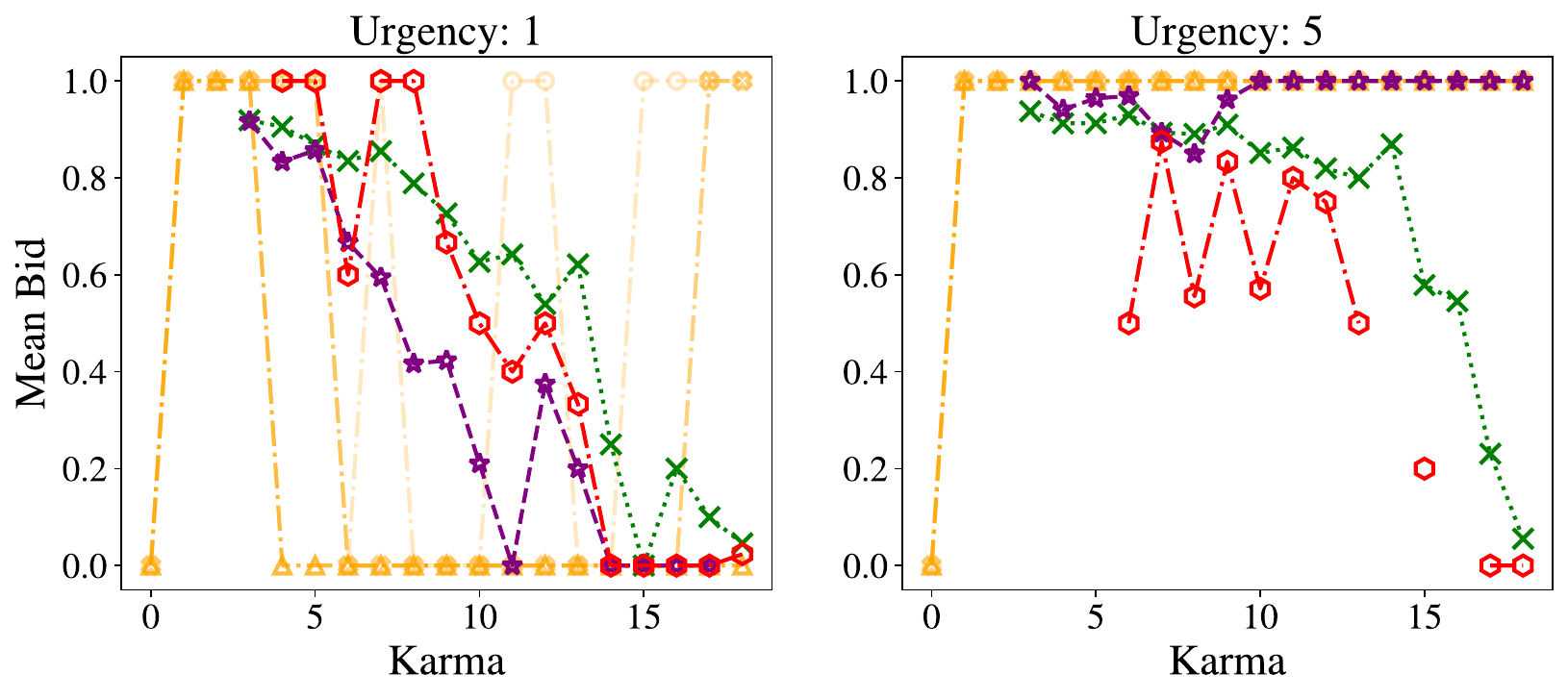}
        \caption{Low Stake, Binary.}
        \label{fig:mean-bid-low-stake-binary}
    \end{subfigure}
    \hfill
    \begin{subfigure}[b]{0.49\textwidth}
        \centering
        \includegraphics[width=\textwidth]{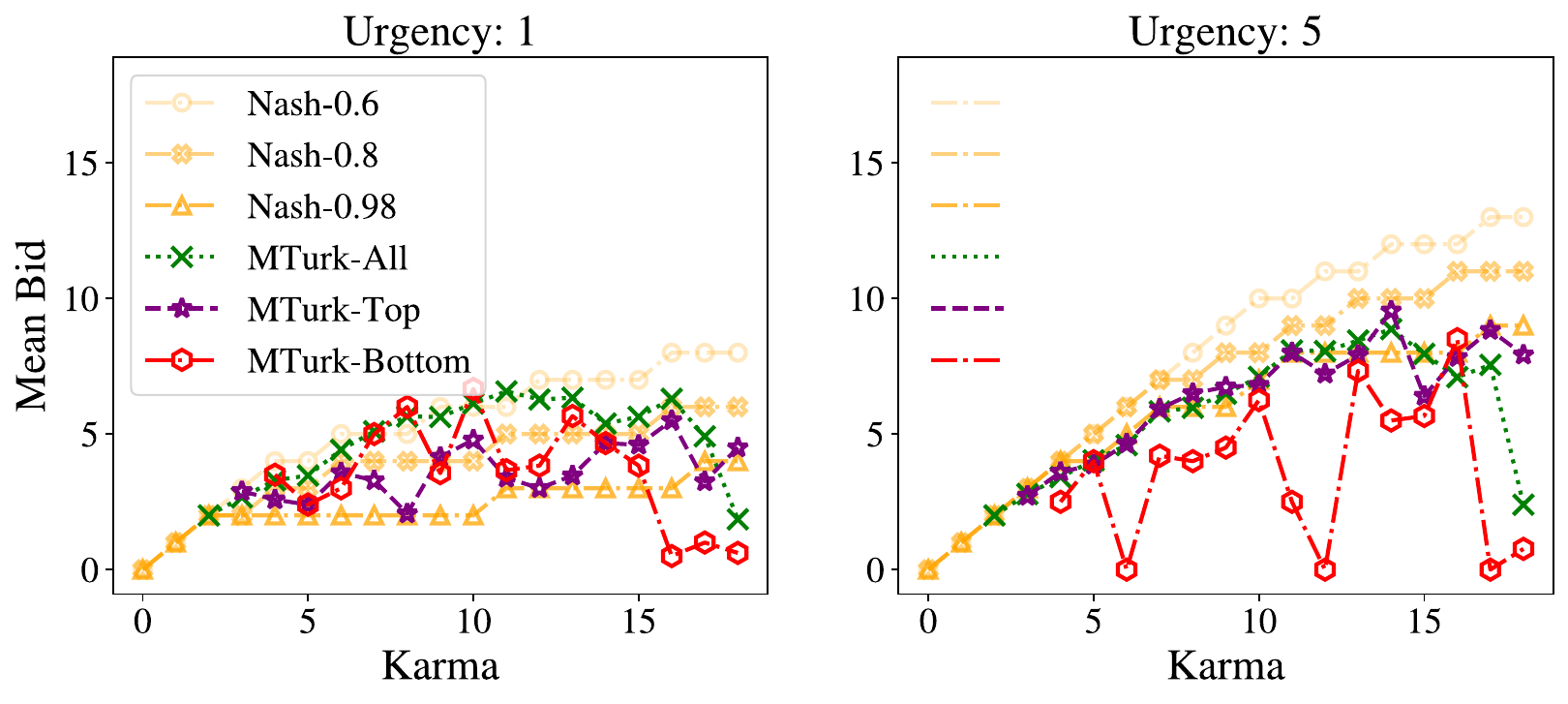}
        \caption{Low Stake, Full Range.}
        \label{fig:mean-bid-low-stake-full-range}
    \end{subfigure}

    \medskip
    
    \begin{subfigure}[b]{0.49\textwidth}
        \centering
        \includegraphics[width=\textwidth]{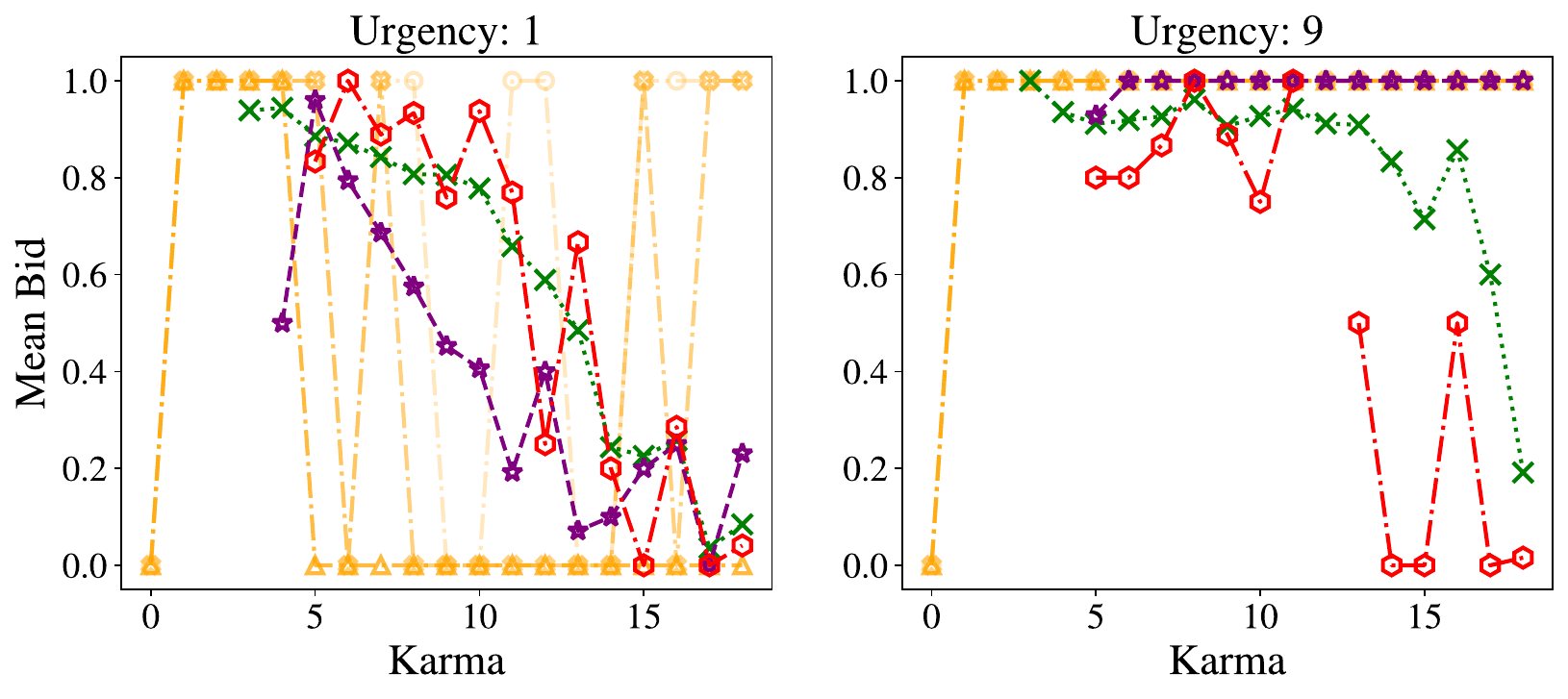}
        \caption{High Stake, Binary.}
        \label{fig:mean-bid-high-stake-binary}
    \end{subfigure}
    \hfill
    \begin{subfigure}[b]{0.49\textwidth}
        \centering
        \includegraphics[width=\textwidth]{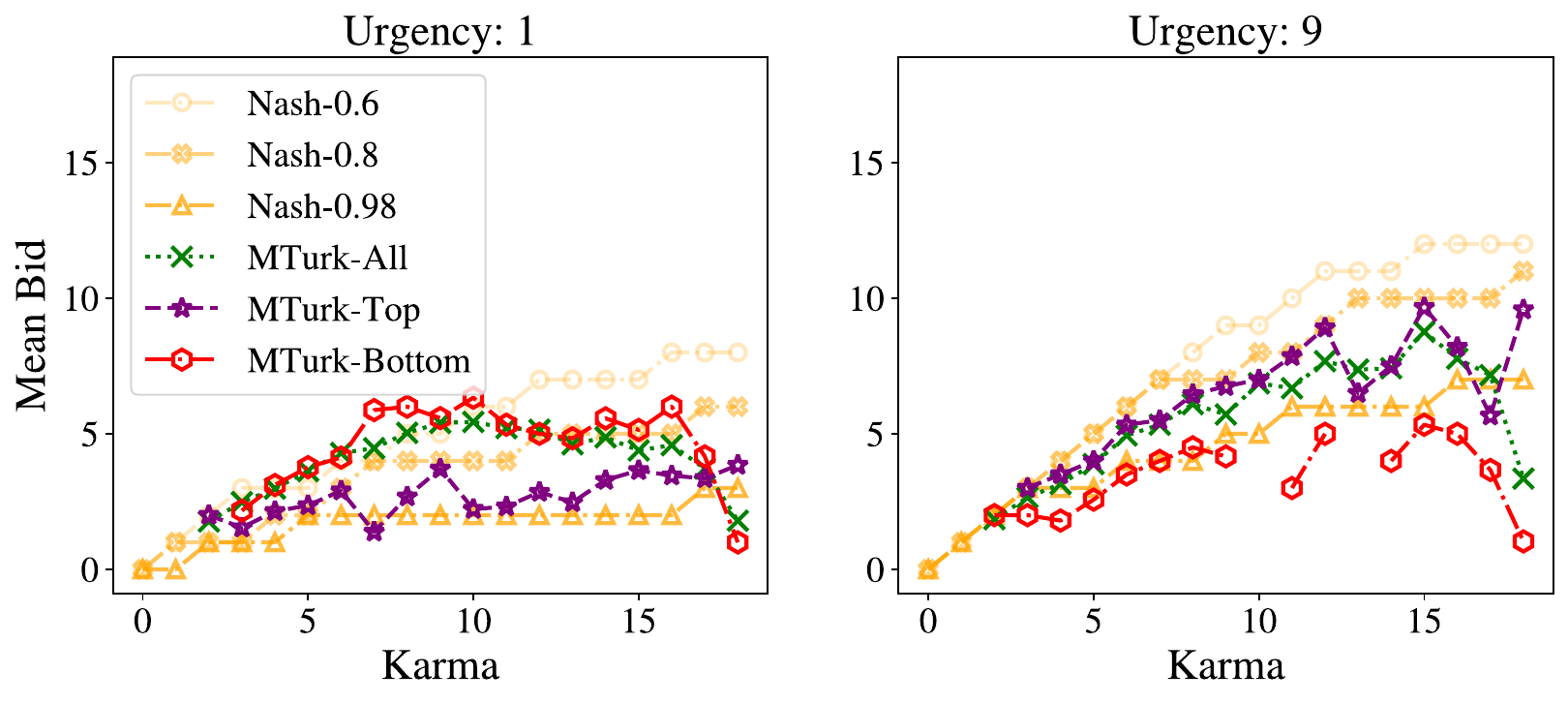}
        \caption{High Stake, Full Range.}
        \label{fig:mean-bid-high-stake-full-range}
    \end{subfigure}
    \caption{Mean bid per urgency and karma for the four treatment combinations.
    Each sub-figure shows the mean bid per urgency (left vs. right panel) and karma ($x$-axis), computed using the aggregate bid data of all participants (labelled ``\gls{MTurk}-All''); the bid data corresponding to the top-performing decile of participants only (labelled ``\gls{MTurk}-Top''); and the bid data corresponding to the bottom-performing decile of participants only (labelled ``\gls{MTurk}-Bottom'').
    The theoretical \gls{SNE} bidding policies associated to future discount factors $0.6$, $0.8$, and $0.98$ are also shown, in increasing intensity of color per discount factor.
    }
    \label{fig:mean-bid-policies}
\end{figure}

In order to provide insight on the attained efficiency gains, and the observed gap to Nash predictions, Figure~\ref{fig:mean-bid-policies} visualizes the mean bidding behaviors for the four treatments, aggregated separately over all participants; the top-performing decile of participants; and the bottom-performing decile of participants, and contrasted to the theoretical \gls{SNE} policies for three levels of future discounting ($0.6$, $0.8$, and $0.98$; and the policies for discount factor $0.98$ are the same as shown in Figure~\ref{fig:equilibrium} and used thus far in the theoretical Nash benchmark).
These mean bids, as well as the forthcoming analysis, are based on the bid data in the main experiment rounds (excluding test rounds).
Moreover, it is important to note that an unfortunate technical bug in the logging of bids has led to an unrecoverable loss of a few bid data-points.
Namely, if a participant actively selected a bid on the decision page, but failed to acknowledge the results page, the selected bid was used in that round correctly but logged as zero incorrectly.
Using the difference in karma between rounds (which was logged correctly), it was possible to recover the incorrectly logged bid if the participant won the round, but not if the participant lost the round.
Therefore, all losing, zero bids for which either the decision or the results page timed out were considered `invalid' and not included in the forthcoming analysis.
The frequency of occurrence of these excluded bids equals that of time-outs reported in Table~\ref{tab:dropouts-timeouts}, which recall ranges per treatment between $9.3\%$--$11.5\%$.

The main finding of Figure~\ref{fig:mean-bid-policies} is that overall, there was a tendency to over-bid in the low urgency state\footnote{Note that due to the aforementioned bug which excluded some of the losing bids, the means in Figure~\ref{fig:mean-bid-policies} could be slightly biased towards higher winning bids.
However, given the relatively small fraction of excluded bids, we do not expect such a bias to be pronounced.}.
The aggregate bidding behaviors are not well-explained by a single value of future discounting:
overall mean bids fit better to far-sighted behavior (discount factors $0.8$--$0.98$) in high urgency, but to short-sighted behavior (discount factor $0.6$) in low urgency.
In contrast, the mean bids of top-performing participants are well-fitted to far-sighted behavior in both urgency levels, which explains the high efficiency gains achieved by these participants.
On the other end of the spectrum, the mean bids of bottom-performing participants display little understanding of the concept of urgency by these participants who showed similar mean bids in both urgency states.
Moreover, it is evident that the bottom decile included most dropouts, as can be seen from the presence of low mean bids in high karma states, corresponding to the persistent zero bids placed on behalf of dropouts who subsequently saturated their karma at the maximum level.

\begin{figure}[!tb]
    \centering
    \begin{subfigure}[b]{0.49\textwidth}
        \centering
        \includegraphics[width=\textwidth]{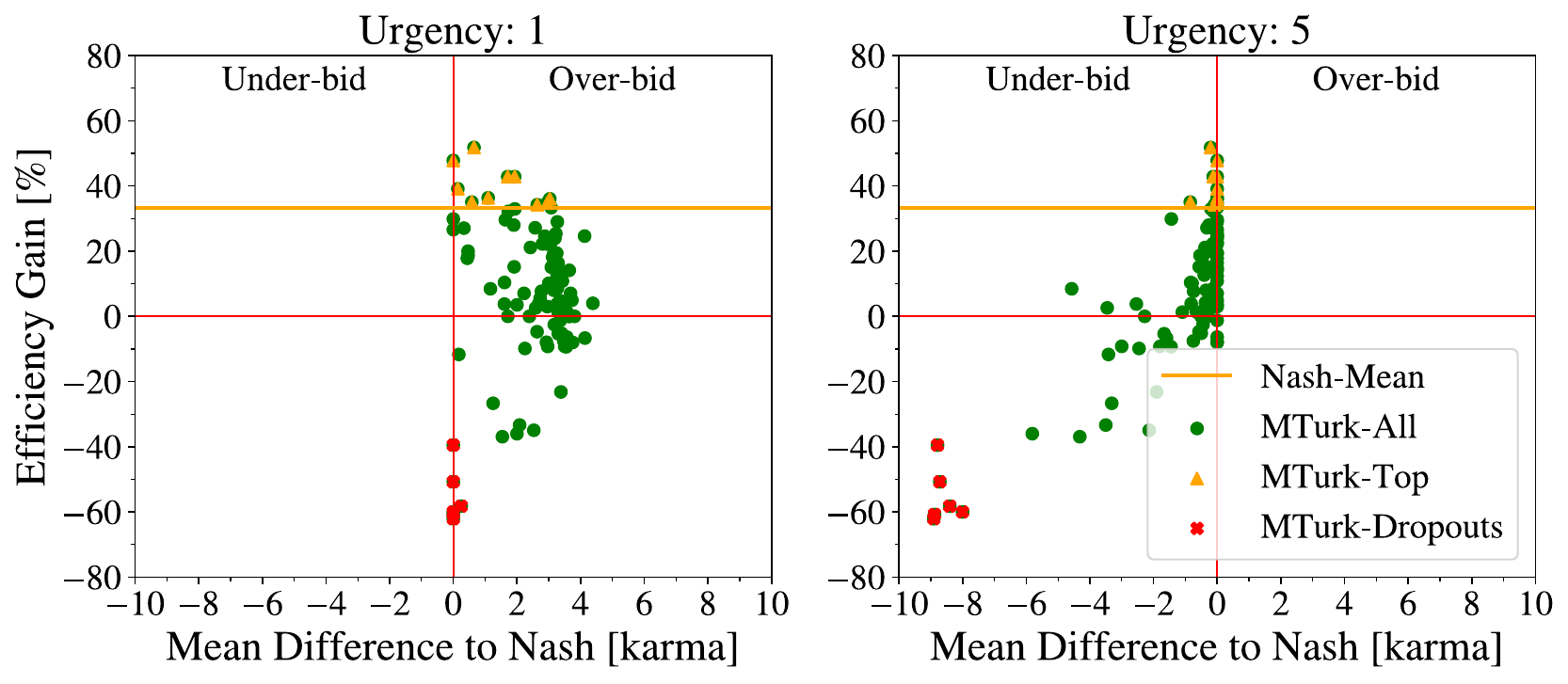}
        \caption{Low Stake, Binary.}
        \label{fig:mean-equilibrium-diff-low-stake-binary}
    \end{subfigure}
    \hfill
    \begin{subfigure}[b]{0.49\textwidth}
        \centering
        \includegraphics[width=\textwidth]{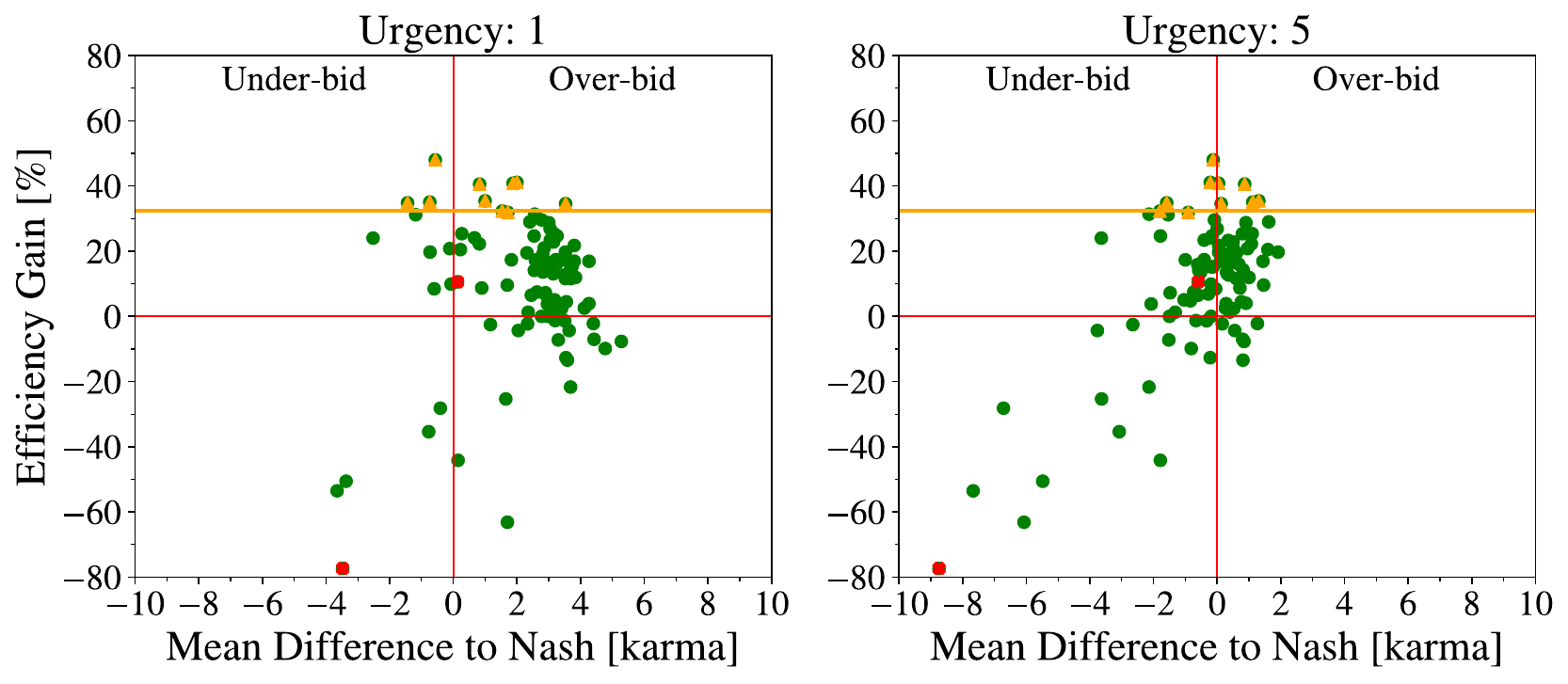}
        \caption{Low Stake, Full Range.}
        \label{fig:mean-equilibrium-diff-low-stake-full-range}
    \end{subfigure}

    \medskip
    
    \begin{subfigure}[b]{0.49\textwidth}
        \centering
        \includegraphics[width=\textwidth]{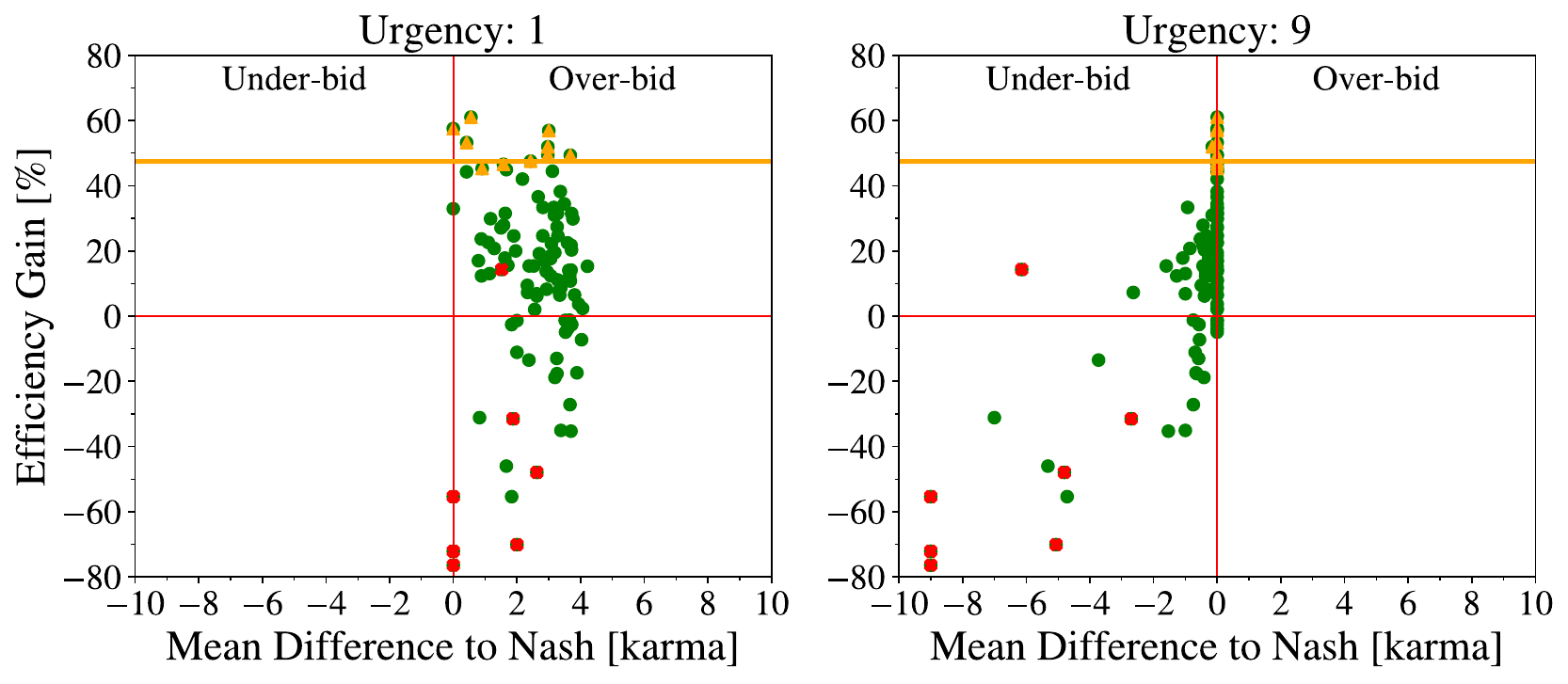}
        \caption{High Stake, Binary.}
        \label{fig:mean-equilibrium-diff-high-stake-binary}
    \end{subfigure}
    \hfill
    \begin{subfigure}[b]{0.49\textwidth}
        \centering
        \includegraphics[width=\textwidth]{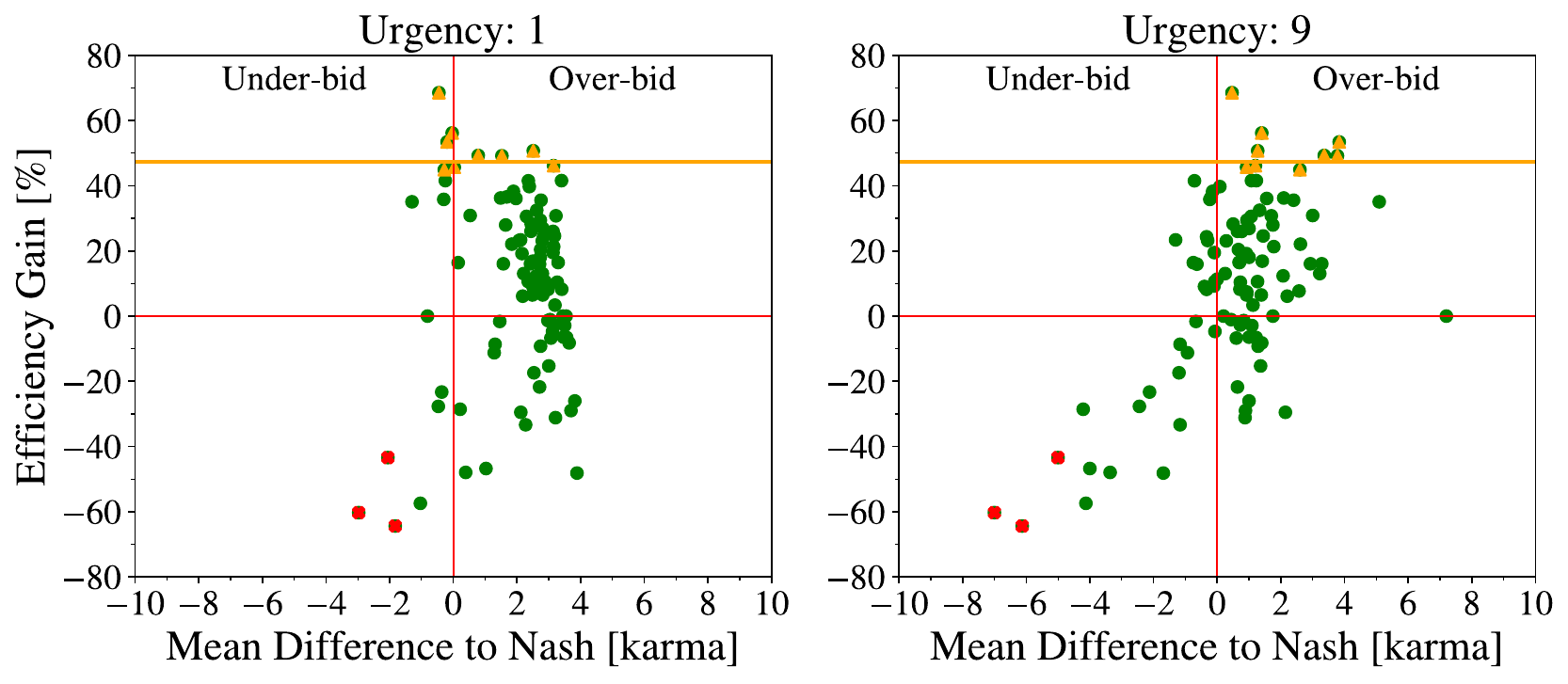}
        \caption{High Stake, Full Range.}
        \label{fig:mean-equilibrium-diff-high-stake-full-range}
    \end{subfigure}
    \caption{Individual-level efficiency gain, cf. Equation~\eqref{eq:efficiency-gain}, vs. mean signed difference between the bids placed by the experimental \gls{MTurk} subjects and the theoretically optimal \gls{SNE} bids (associated to future discount factor $0.98$), per urgency, for the four treatment combinations (labelled ``Mturk-All'').
    The top-10 performing participants are labelled ``MTurk-Top'', and dropouts are labelled ``MTurk-Dropouts''.
    The mean efficiency gain achieved by simulating the theoretical Nash benchmark is also shown (labelled ``Nash-Mean'').
    In the binary treatments, the differences are taken with respect to the resulting bids from the binary choice, i.e., zero or half of the present karma.
    }
    \label{fig:mean-equilibrium-diff}
\end{figure}

To provide further insight on how over/under-bidding during low/high urgency affected performance, Figure~\ref{fig:mean-equilibrium-diff} shows a scatter of the \emph{mean signed difference to Nash-$0.98$ bids} per participant and urgency level, versus the attained individual-level efficiency gains, for the four treatments.
As expected, \emph{under-bidding in high urgency} negatively affects performance, cf. third quadrant of the high urgency panels, which includes the majority of participants achieving negative efficiency gains.
On the other hand, the effect of \emph{over-bidding} in either of the urgency levels is inconclusive.
Consistently in all treatments, the single-top performing participant bid close to Nash.
However, several top performing participants also tended to over-bid in low urgency, and a few of these participants managed to over-bid in both low and high urgency, cf. Figure~\ref{fig:mean-equilibrium-diff-high-stake-full-range}.

\begin{figure}[!tb]
    \centering
    \includegraphics[width=0.8\textwidth]{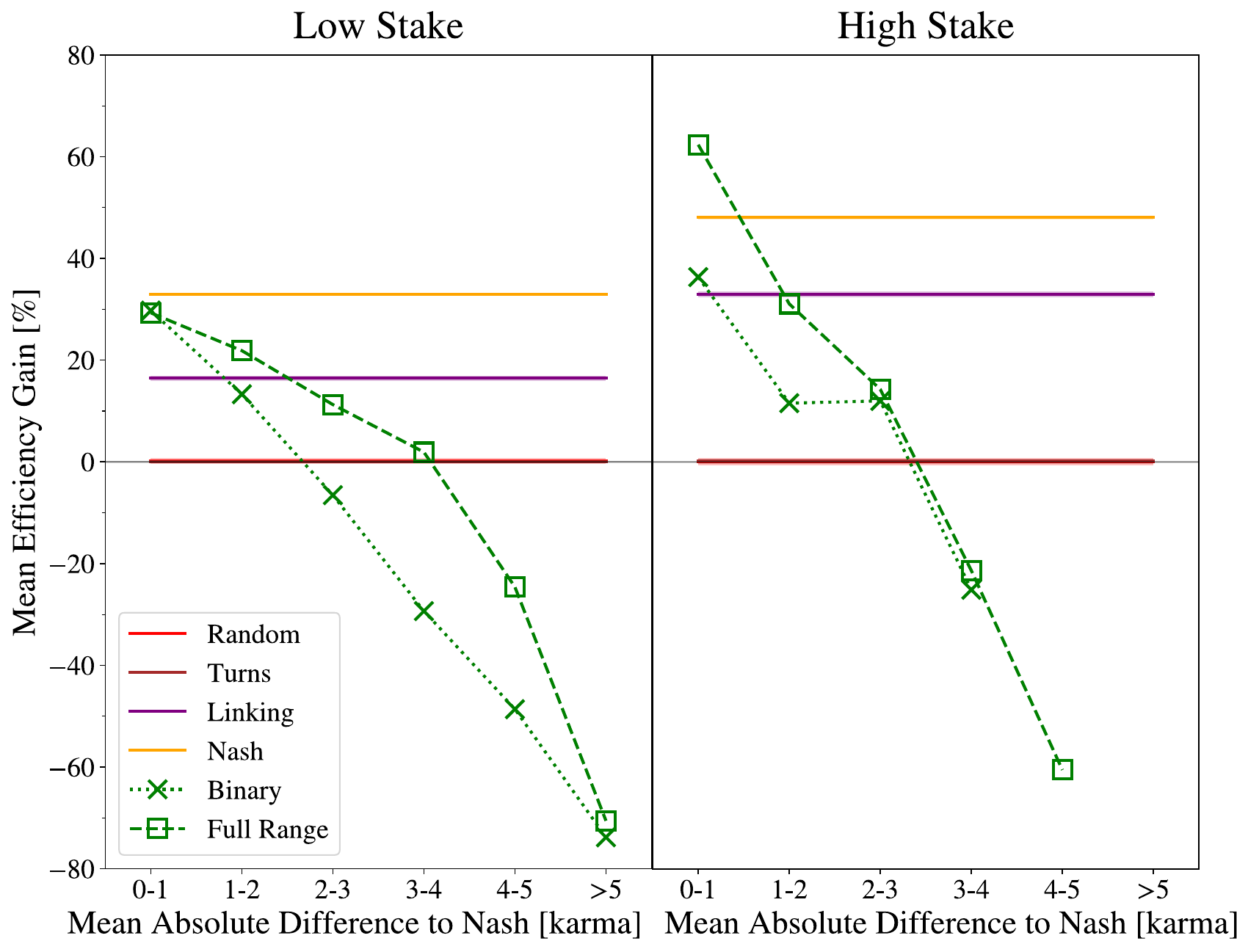}
    \caption{Mean individual-level efficiency gains, cf. Equation~\eqref{eq:efficiency-gain}, among participants for which the mean absolute difference between bids placed and theoretically optimal \gls{SNE} bids (associated to future discount factor $0.98$) lies within the range indicated on the $x$-axis, for the four treatment combinations (labelled ``Binary'' and ``Full Range'' for the respective treatments).
    The mean individual-level efficiency gains of the simulated theoretical benchmarks introduced in Section~\ref{sec:benchmarks} are also shown, along with 95\% confidence intervals estimates using $10\,000$ bootstraps from $4\,000$ simulated individual observations.
    In the binary treatments, the differences are taken with respect to the resulting bids from the binary choice, i.e., zero or half of the present karma.}
    \label{fig:mean-efficiency-gain-per-mean-equilibrium-abs-diff}
\end{figure}

The effect of deviations to theoretically optimal Nash bidding is further investigated in Figure~\ref{fig:mean-efficiency-gain-per-mean-equilibrium-abs-diff}, in which participants are binned based on their \emph{mean absolute difference to Nash-$0.98$ bids}, and the mean individual-level efficiency gain is plotted per mean absolute difference bin, for the four treatments.
Overall, efficiency gains are monotonically decreasing in the mean absolute difference to Nash.
This serves to validate the theoretical \gls{SNE} solution concept, since bidding close to the \gls{SNE} achieves highest efficiency gains despite there being a mismatch between the theoretical assumptions and the experimental setting.
Moreover, Figure~\ref{fig:mean-efficiency-gain-per-mean-equilibrium-abs-diff} suggests that there is some robustness against matching the Nash bidding precisely, especially in the full-range treatments, in which efficiency gains decline more gradually with mean absolute difference to Nash than in the binary treatments.
This is expected since the binary bid levels are already pre-selected to achieve high gains under Nash play; thus a deviation from these pre-selected levels is more severe.
Almost full Nash-level efficiency is attained by participants who deviated from Nash by one karma bid unit on average, roughly half of that efficiency is attained by participants who deviated by two bid units on average, while there are positive efficiency gains in most treatments attained by participants who deviated by three bid units on average.
These results are corroborated by two-sided Wilcoxon signed rank tests of whether individual-level efficiency gains per mean absolute difference to Nash bin are statistically significantly different from zero, c.f. Table~\ref{tab:wilcoxon-diff-to-Nash} in~\ref{sec:sup-tests}.
The test results mirror the trends in Figure~\ref{fig:mean-efficiency-gain-per-mean-equilibrium-abs-diff}:
in low stake-binary, gains are significantly positive for participants that deviated from Nash by up to two karma bid units on average,
and in all other treatments, gains are significantly positive for deviations of up to three bid units on average.

\subsection{Analysis of Stationarity}
\label{sec:stationarity}

An important feature of the karma mechanism is that it forms a closed economy in which karma is preserved over time. In theory, this feature enables reaching a stationary regime, in which a predictable, time-invariant optimal behavior can be repeated indefinitely, cf. Section~\ref{sec:equilibrium}.
In this section, we investigate whether such a stationary regime is also attained in our experiment despite of the mismatch to the theory.
This would signify that overall, experimental subjects who do not necessarily follow optimal Nash behavior nonetheless find simple stationary bidding rules to follow, the game becomes predictable to play, and there are no major cycles or periodic behaviors present.
In the forthcoming analysis, we quantify stationarity (or the lack thereof) by measuring variations in the karma and bid distributions over the course of the main experiment rounds.
We measure the variation in distributions with the \emph{Wasserstein-1 distance}, also known as the \emph{Earth mover's distance}, between realized karma/bid distributions of successive rounds.
The choice of the Wasserstein-1 distance is interpretable: for example, a distance of one corresponds to shifting the distribution by one karma unit on average.
Since perfect stationarity, i.e., zero variation, is unlikely to be attained in our finite population experiment, we compare realized variations with those attained when simulating the theoretical \gls{SNE}.

\begin{figure}[!tb]
    \centering
    \begin{subfigure}[b]{0.49\textwidth}
        \centering
        \includegraphics[width=\textwidth]{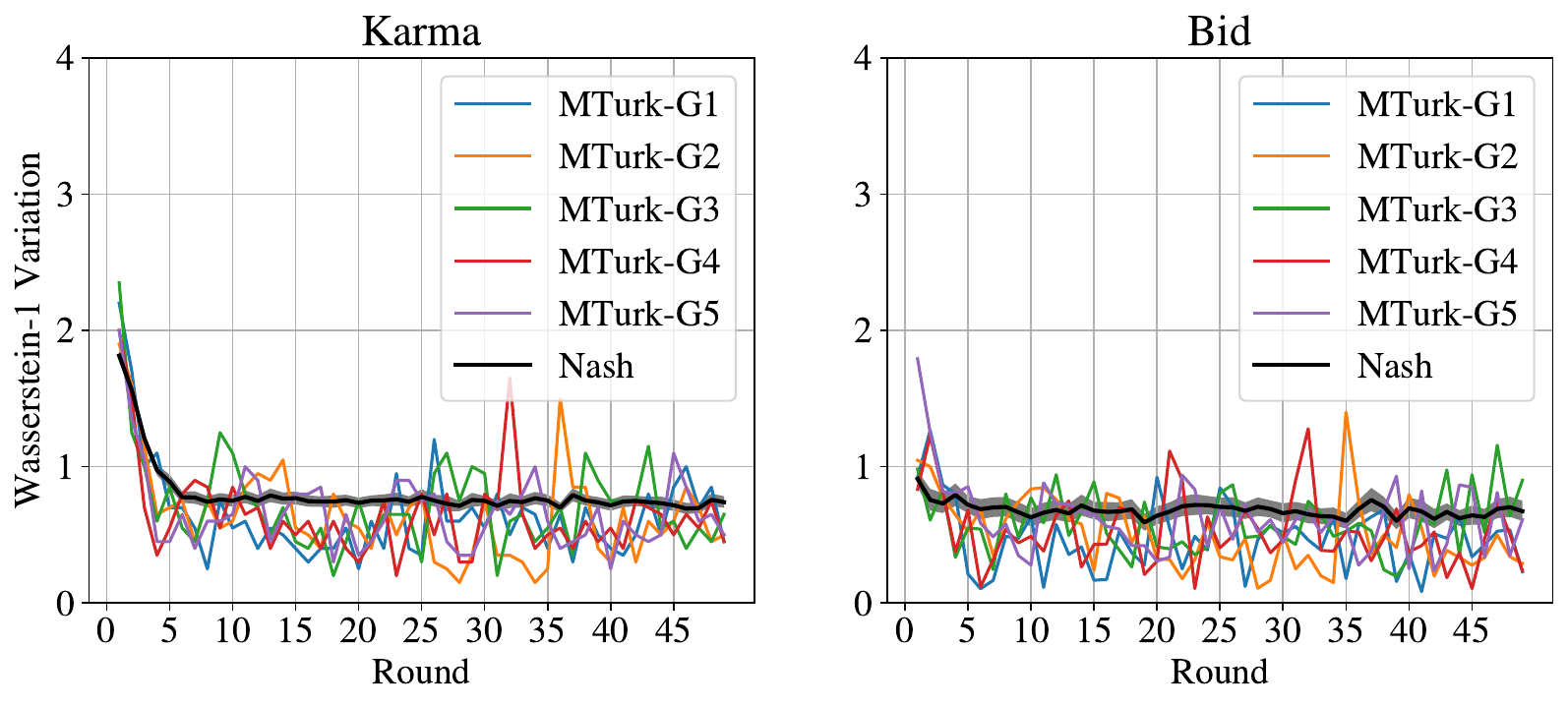}
        \caption{Low Stake, Binary.}
        \label{fig:karma-bid-distribution-diff-low-stake-binary}
    \end{subfigure}
    \hfill
    \begin{subfigure}[b]{0.49\textwidth}
        \centering
        \includegraphics[width=\textwidth]{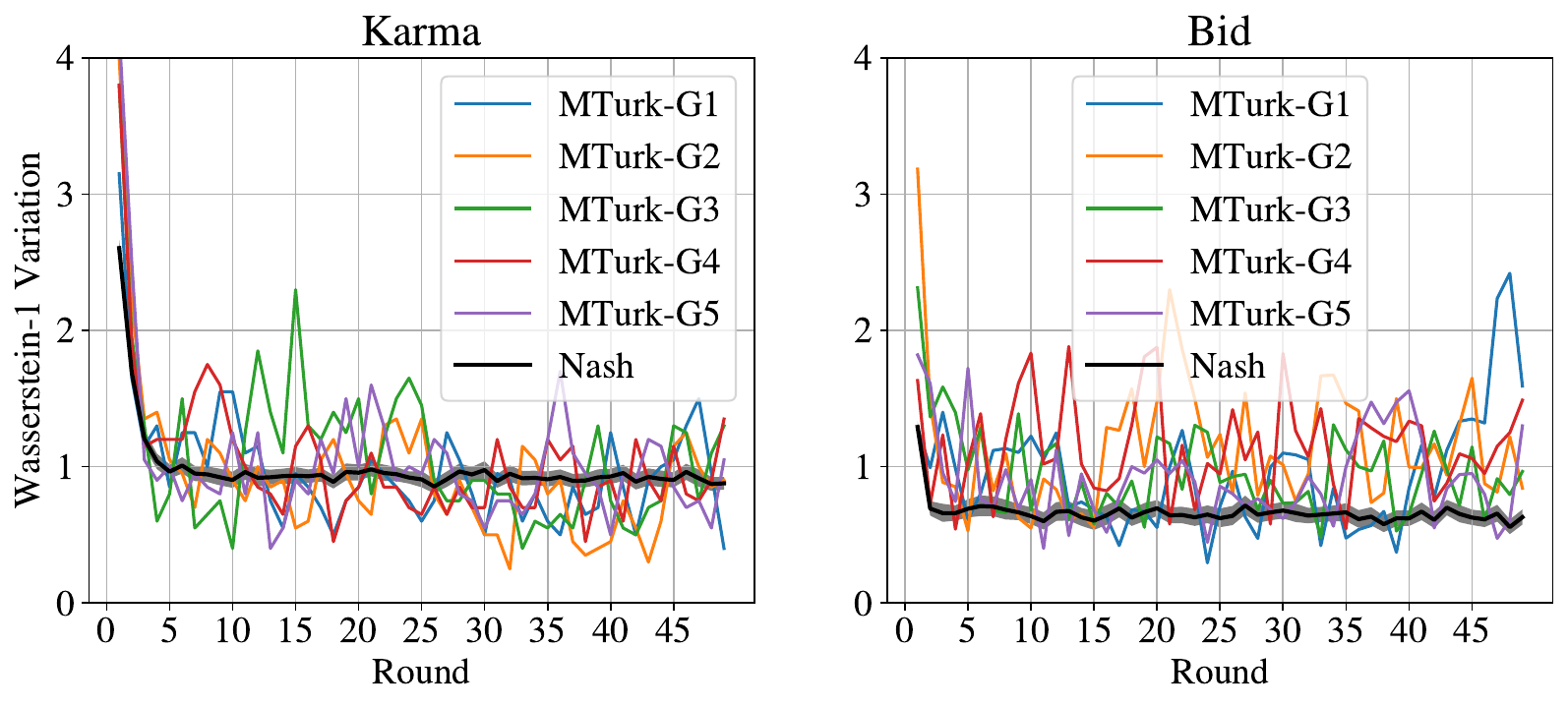}
        \caption{Low Stake, Full Range.}
        \label{fig:karma-bid-distribution-diff-low-stake-full-range}
    \end{subfigure}

    \medskip
    
    \begin{subfigure}[b]{0.49\textwidth}
        \centering
        \includegraphics[width=\textwidth]{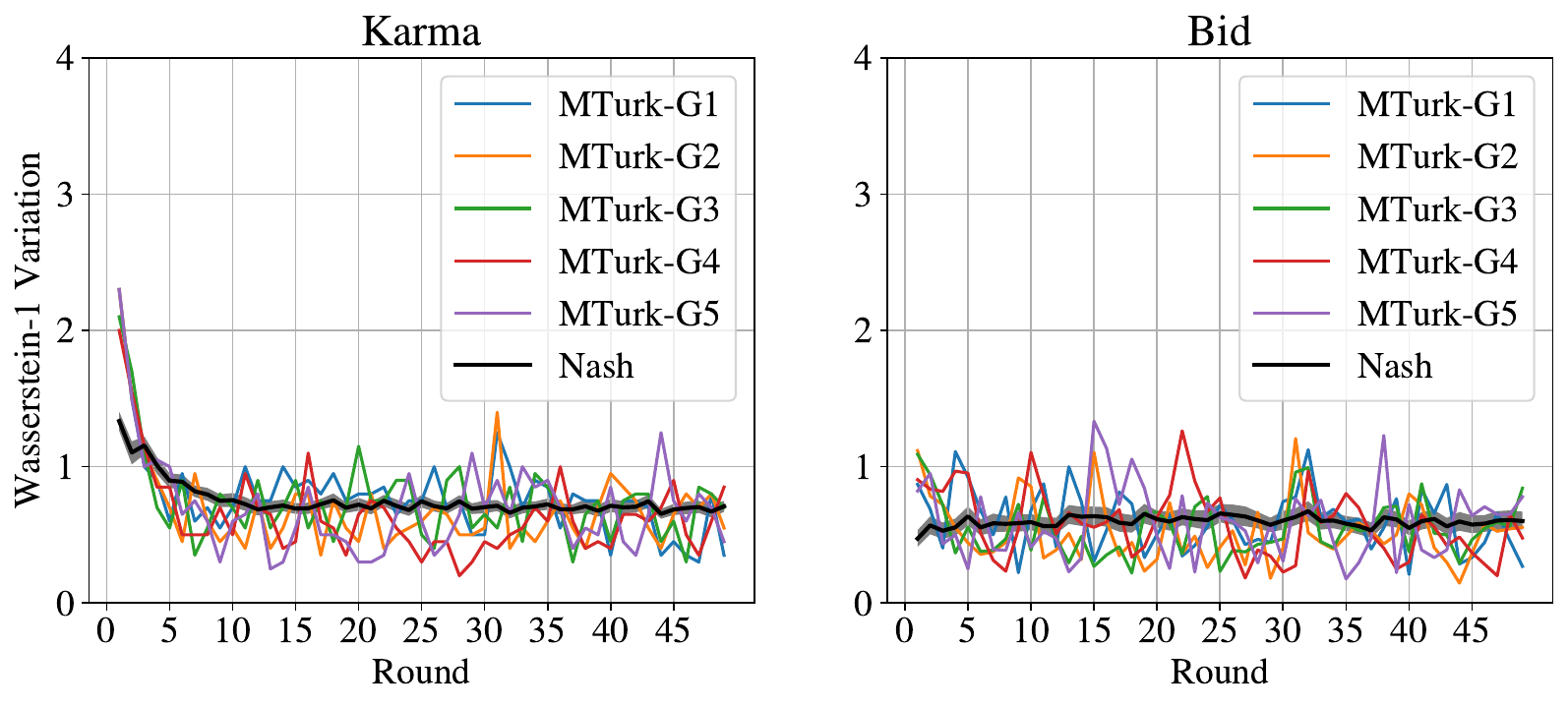}
        \caption{High Stake, Binary.}
        \label{fig:karma-bid-distribution-diff-high-stake-binary}
    \end{subfigure}
    \hfill
    \begin{subfigure}[b]{0.49\textwidth}
        \centering
        \includegraphics[width=\textwidth]{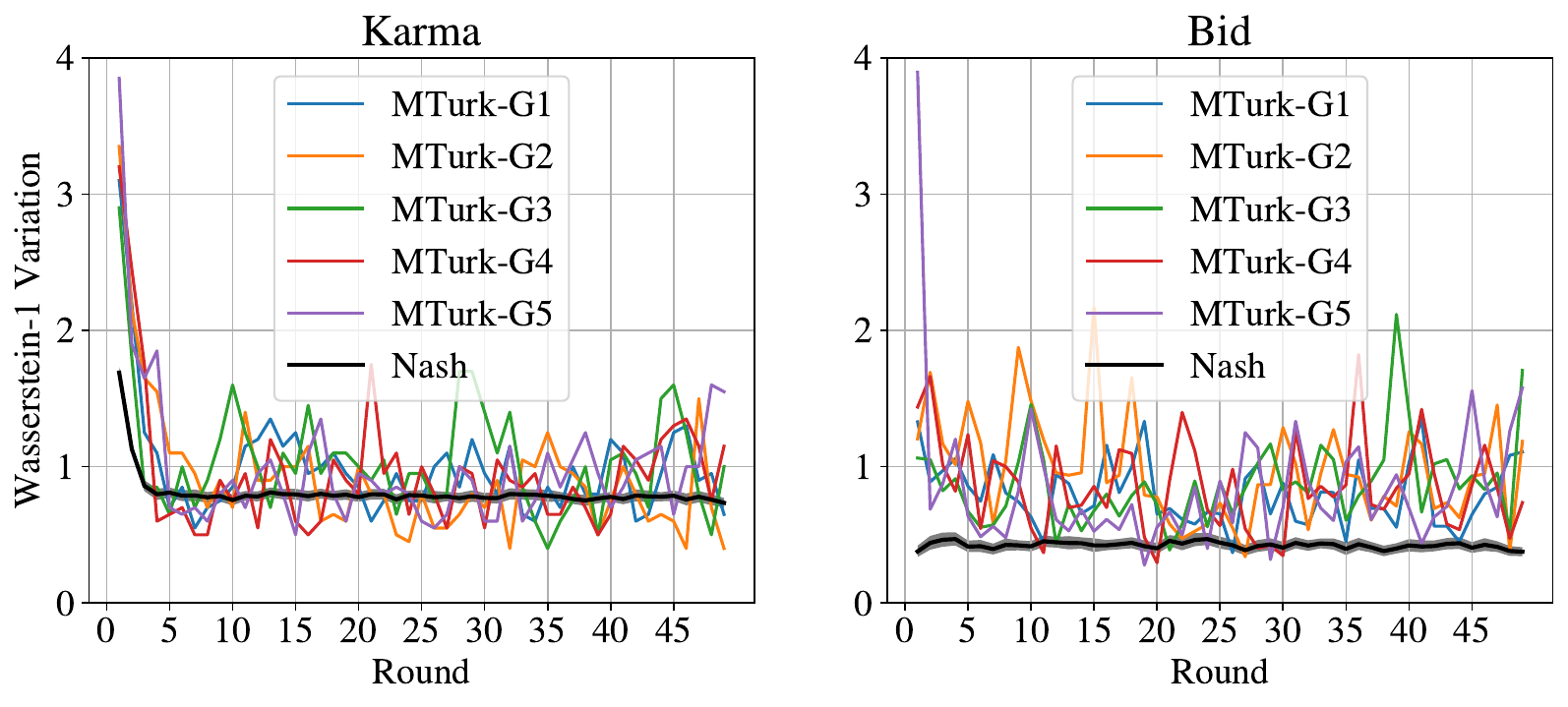}
        \caption{High Stake, Full Range.}
        \label{fig:karma-bid-distribution-diff-high-stake-full-range}
    \end{subfigure}
    \caption{Variation in the distributions of karma (left panel) and bids (right panel) over the course of the main experiment for the four treatment combinations, as measured using the Wasserstein-1 distance between realized distributions of successive rounds.
    For each treatment, the variations are shown for each of the five experiment groups (labelled ``\gls{MTurk}-G1--G5''), and contrasted to the variations attained in the Nash simulations (labelled ``Nash''; the shaded area is the 95\% confidence interval estimate using $10\,000$ bootstraps).
    }
    \label{fig:karma-bid-distribution-diff}
\end{figure}

The results of this stationarity analysis are shown in Figure~\ref{fig:karma-bid-distribution-diff}. The main finding is that in all treatment combinations, after a short initial transient of less than five rounds, the realized variations in the karma and bid distributions are close in magnitude to, if not smaller than, those attained under stationary Nash play.
Two exceptions are the \emph{bid distributions in the full range treatments}, cf. right panels of Figures~\ref{fig:karma-bid-distribution-diff-low-stake-full-range}, \ref{fig:karma-bid-distribution-diff-high-stake-full-range}, in which the experimental variations lie at a visibly higher level than the stationary Nash variations.
Nonetheless, these variations are still relatively small, oscillating around a Wasserstein-1 distance of one.
These findings are corroborated by two-sided Wilcoxon signed rank tests of the realized Wasserstein-1 variations in experimental distributions versus the mean variation attained in the simulated Nash benchmark, cf. Table~\ref{tab:mann-whitney-stationarity} in~\ref{sec:sup-tests}.
These tests show that for the binary treatments, there are statistically significant differences suggesting that experimental distribution variations are \emph{lower} than under simulated Nash; and for the full range treatments, there are either statistically significant differences suggesting that experimental distribution variations are \emph{higher} than under simulated Nash, or no significant differences.
Overall, these findings suggest that an approximately stationary regime is reached, and even more so in the binary treatments.
\section{Discussion}

In sum, we interpret the main results of our study to be that the aggregate efficiency gains of a karma scheme compared with random allocation are pronounced and statistically significant in all treatments.
This constitutes the first set of behavioral evidence that a formal karma mechanism indeed can work to the benefit of the population.
Moreover, in all treatments, (almost) all participants manage to benefit from the karma scheme in comparison to random allocation, with the exception of the lowest decile, which largely consists of dropouts from the experiment who did not actively participate in the bidding.

To us these results are promising because our experimental subjects were recruited from a population of totally untrained and inexperienced users, from an online sample (on \gls{MTurk}). 
Their behavior, even though more efficient than random, is not as efficient as is theoretically feasible under Nash equilibrium play. We view the efficiency gains of our experiment as some kind of behavioral lower bound on the effects.
Analysis of the experimental bidding behaviors reveals a consistent tendency to over-bid in low urgency rounds, however, this observation does not explain the gap to Nash efficiency entirely, driven by what appears to be irrational and noisy behavior.
A natural follow-up question is, therefore, whether karma could be capable of achieving higher efficiency gains than realized in our online experiment if the human population consisted of participants that were better trained.
As a first step in this direction, we conducted an auxiliary experiment under the low stake-full range treatment with a group of `expert' subjects consisting of graduate students in an applied game theory class.
In this expert sample, indeed, the achieved efficiency gains are close to Nash levels (\gls{MTurk}: median $12.86\%$, $n=100$; experts: median $36.65\%$, $n=28$, two-sided \gls{MWW} test $U=667.0$, $p<0.0001$).
This finding supports the interpretation of the efficiency gains from the online experiment as behaviorally \emph{robust lower bounds} on the performance of the karma scheme given the relatively low training and commitment of the subject pool considered.


In addition to providing lower bounds on aggregate gains, our experiment also gives insight on how close individuals have to get to optimal Nash bids in order to attain similarly high gains, and how far they can afford to be before the karma mechanism starts hurting them.
We find a small robustness margin for achieving near-optimal gains (up to one bid unit deviation to Nash on average), and a relatively large margin for achieving positive gains (up to 3--4 bid unit deviations on average).
This serves to validate the optimality and robustness of the theoretical \gls{SNE} concept in light of the mismatch between theory and experiment.

Another important consideration regarding implementation of karma with human participants is whether a simpler scheme (binary) or a richer scheme (full range) is favourable.
That fact that binary led to less variations in the distribution of bids and thus more predictable auction outcomes, meanwhile we found no significant differences in terms of realized efficiency gains compared to full range, provides preliminary evidence that the simpler binary scheme is advantageous--at least for applications similar to the ones we studied.
Such applications feature, as we have investigated in our experiment, urgency processes with binary levels, for which a binary scheme is arguably also particularly natural.
Theoretically, Nash equilibrium under binary bidding will lose in efficiency with more than two urgency levels, and it remains an open question to test what the trade-off is between behavioral simplicity of a binary (or otherwise limited) bidding scheme and the theoretical benefits that come with richer schemes.
Some additional reasons to believe simplicity is beneficial are based on theories of decision fatigue~\citep{baumeister2003psychology,pignatiello2020decision} and simplicity in mechanism design~\citep{pycia2023theory}.
A richer analysis of this complexity-performance trade-off provides fruitful avenues for future investigations, both in theory and in the behavioral lab.

Moreover, we would like to comment on the treatment combination of high stake urgency process under binary bidding scheme, for which there is (weak) evidence that both mean efficiency gains and the distribution of gains are higher than in other treatments.
Having rare but important urgency realizations seemed to make it easier for our subjects to decide whether to bid or not, and there were no subtleties regarding how much to bid as a function of the history of play, etc. given the binary nature of the bidding scheme.
In future experiments, we shall investigate what kind of automated bidding tools could be most beneficial.
One may envision a setting where the human decides whether or not to engage in bidding, but the actual bid is determined by a bot who bids optimally.

Finally, we would like to mention a few avenues for future work that we consider particularly promising. Here, we adopted the relatively simple setting of random pairwise matching, without providing specific context, as a first step towards understanding if humans are able to trade-off present versus future resource consumption effectively with karma.
As a next step, before we can hopefully unleash karma mechanisms in the real-world, we would like to run further contextualized experiments tailored to specific applications including multi-issue voting, priority allocation in traffic and other public goods, flexible EV charging and other forms of demand side management such as in smart grids, renewable energy integration and curtailment, cloud data management, allocation of computing clusters, et cetera, et cetera.
One natural big question is whether karma mechanisms should be completely compartmentalized, or whether karma should be used across multi-resource allocation domains. We consider experiments a natural point of departure to identify whether compartmentalizing or combining issues is beneficial depending on the underlying (e.g., complements versus substitutes) nature of the resources in question.

\section*{Acknowledgments}
We would like to acknowledge Stefan Wehrli and Oliver Bragger from the \acrfull{DeSciL} for their great help in coding and running the experiment.
We would also like to thank the anonymous reviewers and the Associate Editor for all their diligent and helpful comments.
Research supported by NCCR Automation, a National Centre of
Competence in Research, funded by the Swiss National Science
Foundation (grant number 51NF40\_80545).

\bibliographystyle{elsarticle-harv} 
\bibliography{main}

\newpage

\appendix

\section{Implementation Details}
\label{sec:det-implementation}

\subsection{Experiment Pages}
\label{sec:pages}

\begin{figure}[!htb]
    \centering
    \includegraphics[width=\textwidth]{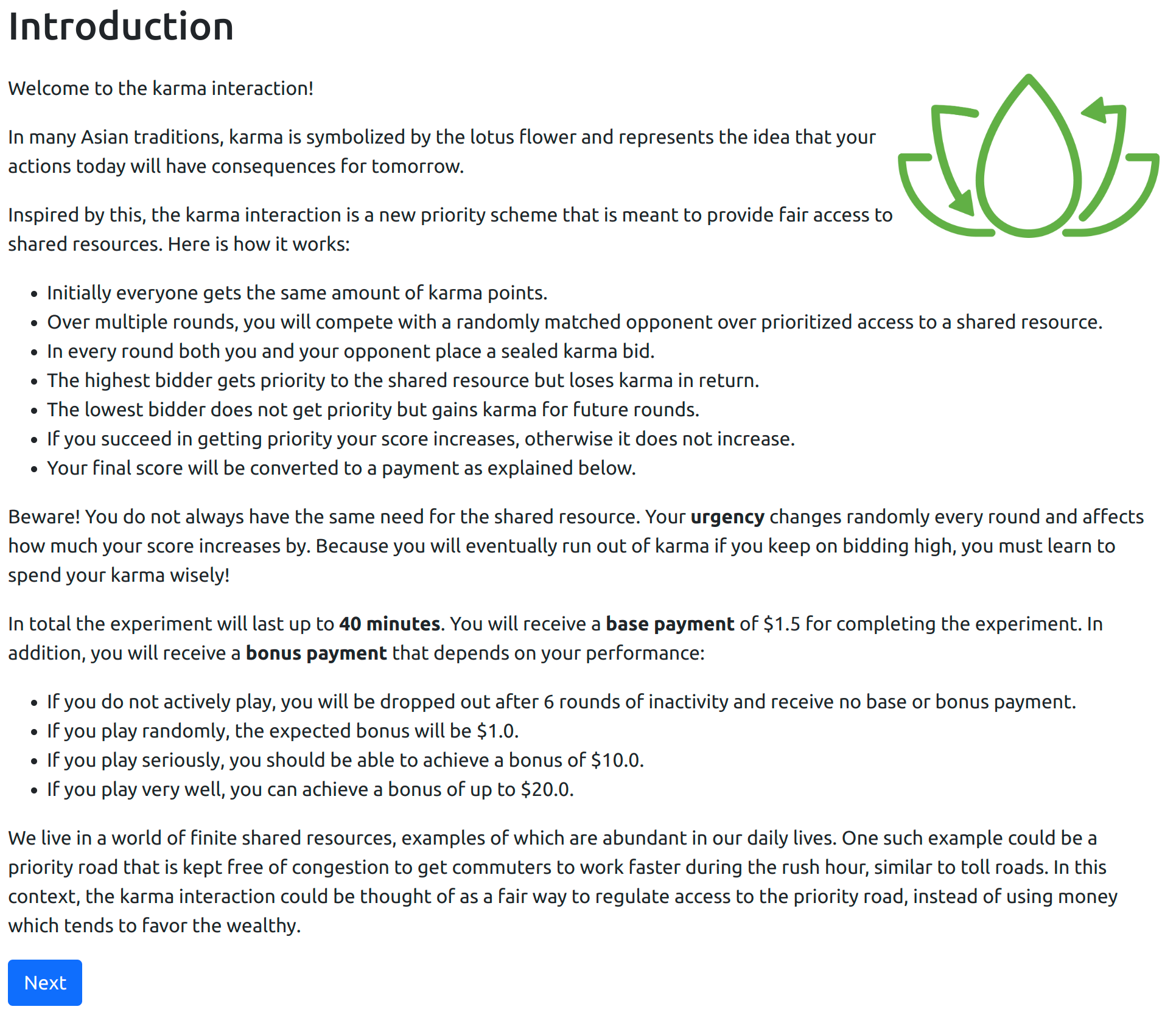}
    \caption{Landing page.}
    \label{fig:introduction-page}
\end{figure}

\begin{figure}[!htb]
    \centering
    \includegraphics[width=\textwidth]{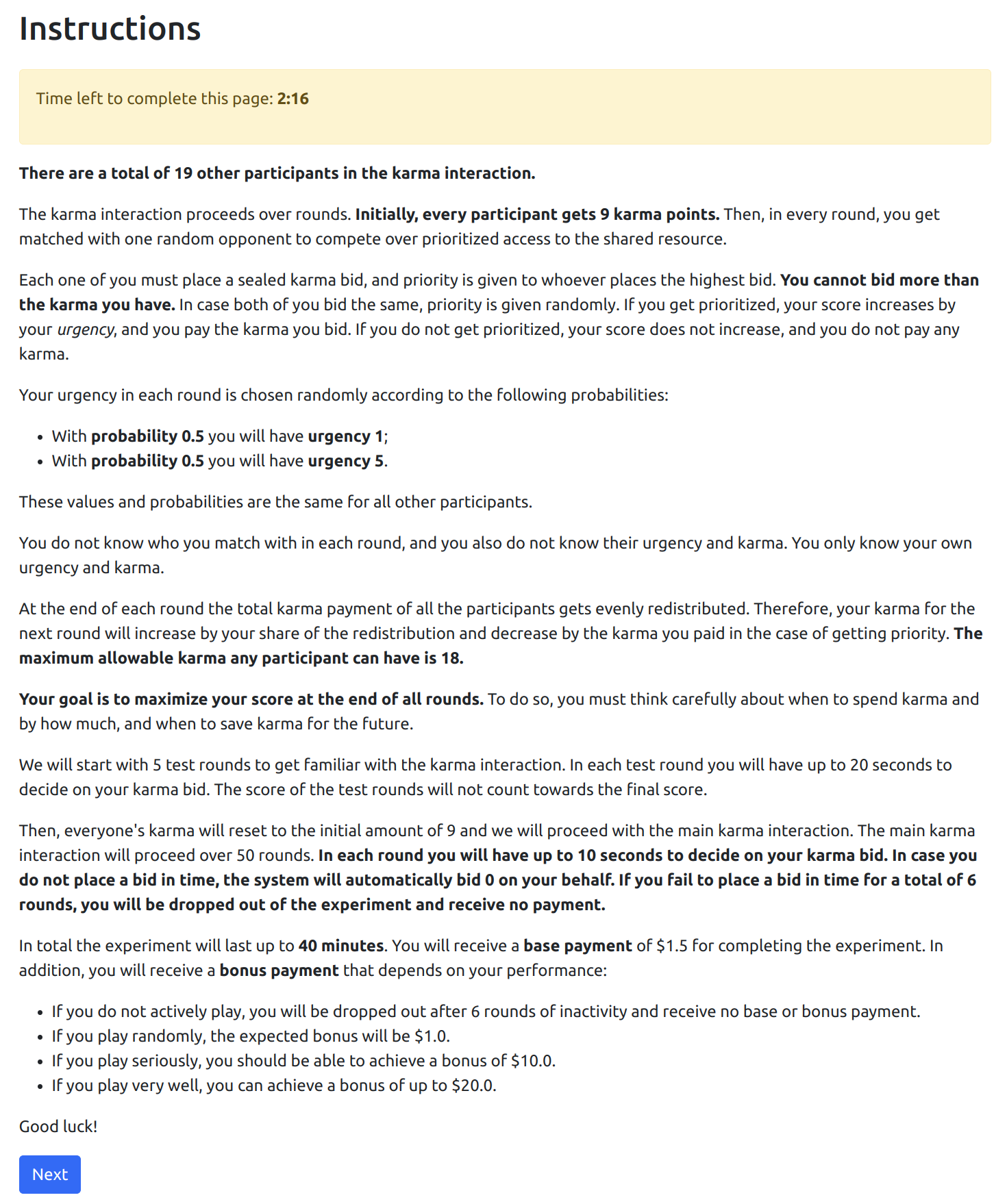}
    \caption{Detailed instructions page (corresponding to the low stake, full range treatment).}
    \label{fig:instruction-page}
\end{figure}

\begin{figure}[!htb]
    \centering
    \begin{subfigure}[b]{\textwidth}
        \centering
        \includegraphics[width=0.85\textwidth]{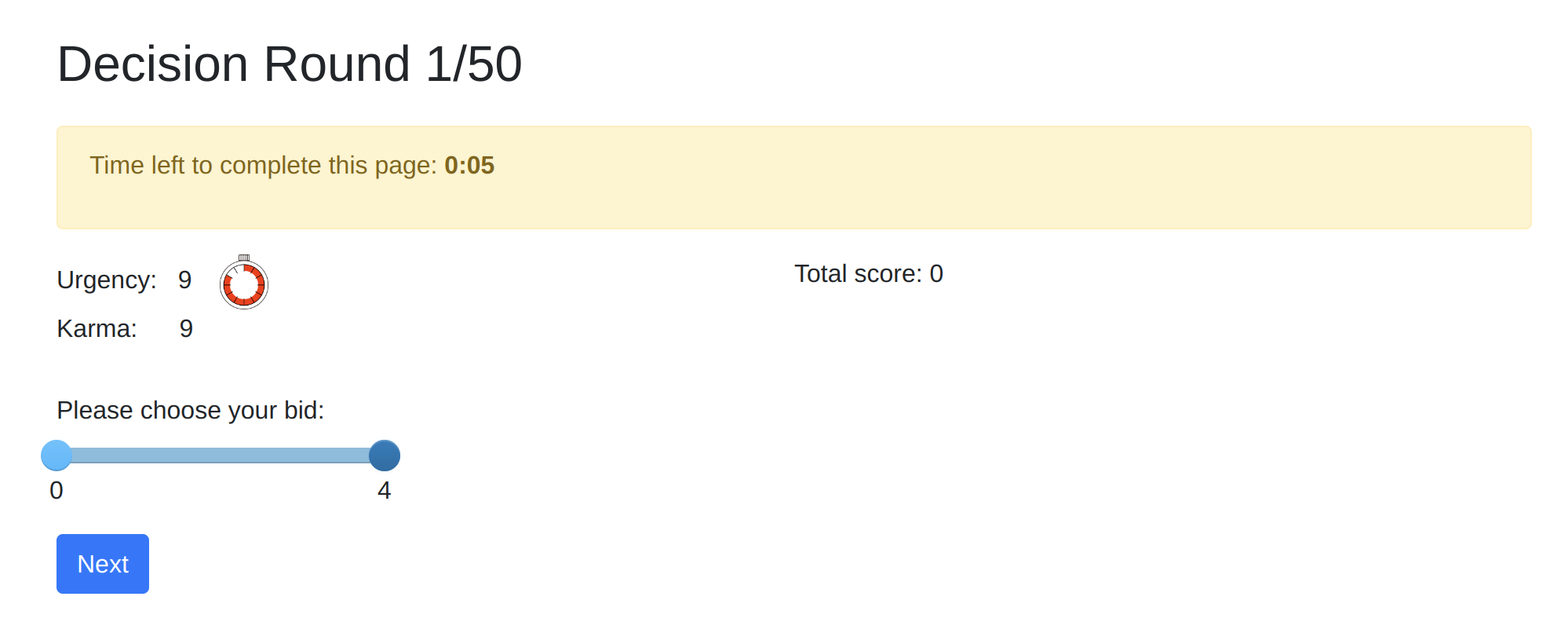}
        \caption{High stake, binary treatment.}
        \label{fig:decision-high-stake-binary}
    \end{subfigure}

    \medskip
    
    \begin{subfigure}[htb!]{\textwidth}
        \centering
        \includegraphics[width=0.85\textwidth]{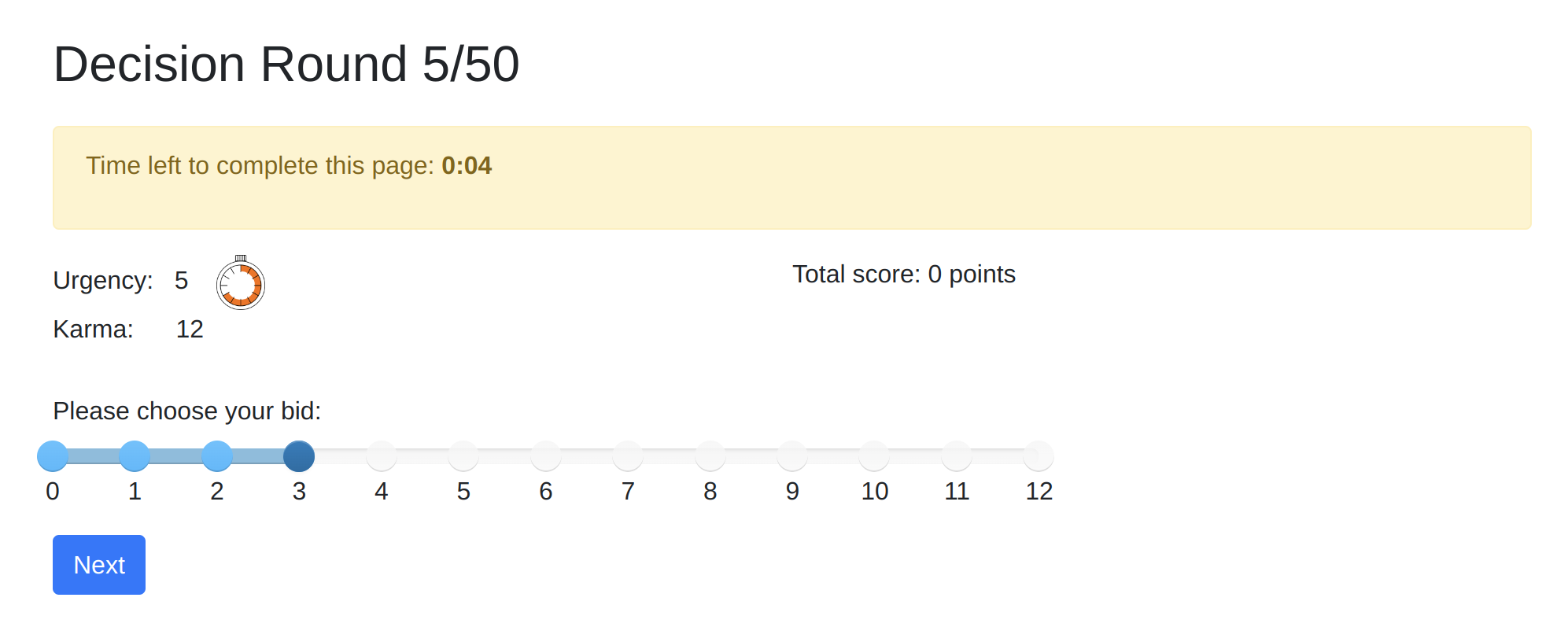}
        \caption{Low stake, full range treatment.}
        \label{fig:decision-low-stake-full-range}
    \end{subfigure}
    \caption{Decision page examples.}
    \label{fig:decision-page}
\end{figure}

\begin{figure}[!htb]
    \centering
    \includegraphics[width=0.85\textwidth]{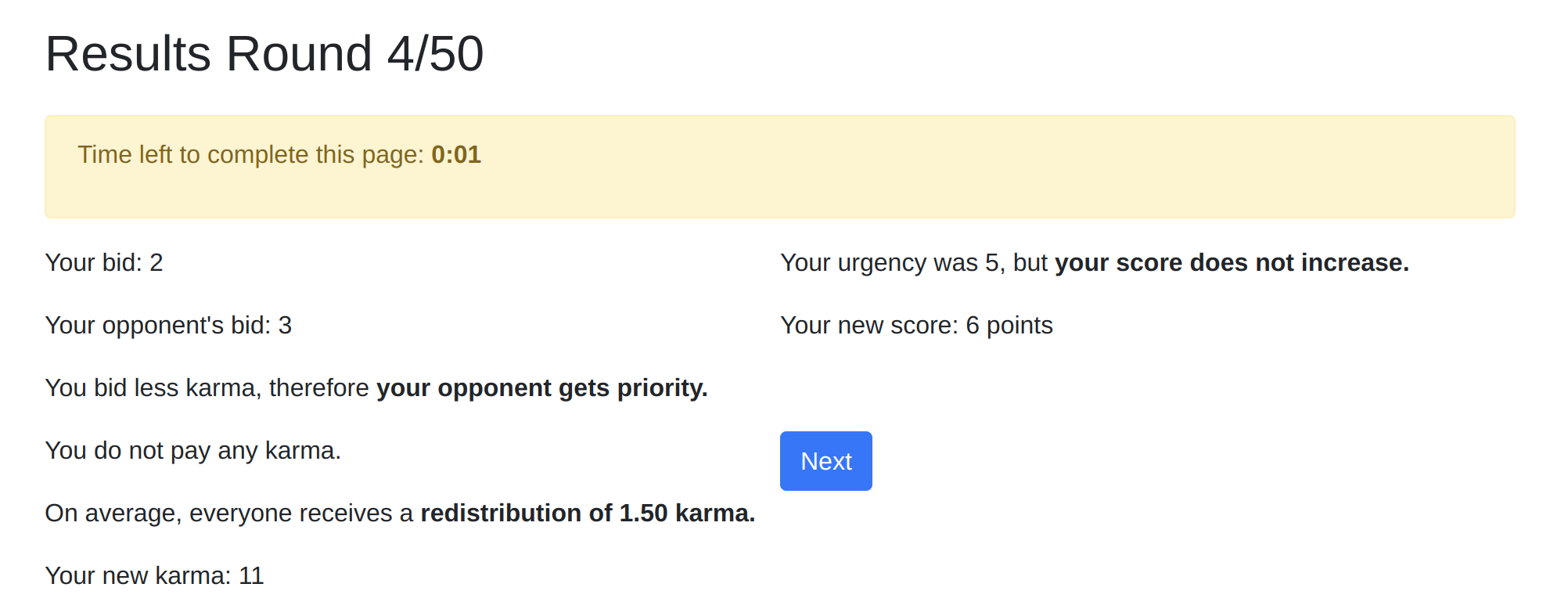}
    \caption{Results page example.}
    \label{fig:results-page}
\end{figure}

Figure~\ref{fig:introduction-page} shows the introductory landing page of the online experiment; Figure~\ref{fig:instruction-page} shows an example of the detailed instructions page; Figure~\ref{fig:decision-page} shows examples of the decision page; and
Figure~\ref{fig:results-page} shows an example of the results page.

\subsection{Technical Pre-tests}
\label{sec:pretests}

Two technical pre-tests were conducted, each with two groups.
Originally, we designed and pre-registered the experiment to have 30 rather than 20 participants per group, and run for 100 rather than 50 main rounds.
The initial pre-test showed that this plan was technically too ambitious, as we encountered technical issues related to time-synchronization of participants and server overload that led to significant unanticipated delays.
From the two groups only one managed to complete the experiment in a significantly longer time than anticipated (1h20min instead of 40min).
Due to these delays and the excessive length of the experiment, many participants dropped out before completion.
Moreover, a preliminary analysis showed no visible difference in bidding behaviors in the first versus the second half of the experiment.
For these reasons, we had to make a pragmatic compromise to decrease the number of participants per group to 20 and the number of main rounds to 50.
This change was tested in the second pre-test which ran successfully in the planned time of 40min once synchronization and server delays were accounted for.
\section{Pre-registered analysis plan}
\label{sec:prereg}

The pre-registration of our experiment included an analysis plan in addition to the design and implementation plan~\citep{elokda2023karma}.
The analysis presented in the main paper constitutes a partial deviation from this pre-registered analysis. 
Here, for completeness, we state the hypotheses that were originally pre-registered, discuss the hypotheses that were addressed by our main analysis, and our motivations for deviating from the analysis of the remaining hypotheses.

The pre-registration reads~\citep[Section~2.1]{elokda2023karma}:\\
\emph{``We wish to test the following hypotheses.
Let us define efficiency as the mean of participant final rewards and fairness as the negated standard deviation of final rewards.
\begin{enumerate}
    \item The karma allocation is more efficient than a random allocation.
    \item The karma allocation is more fair than a random allocation.
    \item The efficiency of the karma allocation is within 10\% of the most efficient allocation.
    \item The karma allocation is fairer than an efficiency-maximizing but history-unaware allocation.
    \item There is a positive correlation between the participants' bid and urgency.
    \item A full bidding scheme is more efficient than a binary bidding scheme.
    \item A binary bidding scheme is more fair than a full bidding scheme.
    \item Participants with high urgency spread achieve higher rewards than those with low urgency spread.''
\end{enumerate}
}

The group-level efficiency gain, cf. Equation~\eqref{eq:efficiency-gain-group}, is an affine variant of the mean of participant final rewards aimed at controlling for the random realization of urgency.
Accordingly, hypothesis (1) is strongly supported by the Wilcoxon signed rank tests of karma efficiency gains reported in Table~\ref{tab:wilcoxon-treatment-groups}; hypothesis (6) is rejected by the inter-treatment \gls{MWW} tests reported in Tables~\ref{tab:mann-whitney-inter-treatment-groups}--\ref{tab:mann-whitney-inter-treatment-individuals-no-dropouts}; and these tests also support hypothesis (8).
Hypothesis (3) is also clearly rejected by investigating the raw group-level efficiency gains reported in Table~\ref{tab:efficiency-groups} which all fall below 10\% of the full possible efficiency.

As regards hypothesis (5), Figure~\ref{fig:mean-bid-policies} shows an overall tendency of higher mean bids in high urgency over low urgency.
We furthermore performed two-sided \gls{MWW} tests of the bid observations in low versus high urgency which showed strongly significantly higher bids in high urgency for all treatments and overall (overall results: low urgency: median $3$, $n=11\,236$; high urgency: median $4$, $n=6\,706$; \gls{MWW} test $U=31\,553\,792.5$, $p<0.0001$).
For the binary treatments, these tests were conducted using the resulting bids from the binary choice, i.e., zero or half of the present karma.
This supports hypothesis (5).

Therefore, the main deviation to the pre-registered analysis plan regards the fairness-related hypotheses (2), (4) and (7).
We departed from this analysis because in hindsight we realized that the negated standard deviation is not a good measure of fairness.
Namely, it fails to satisfy the basic Pareto principle and leads to uninterpretable results, e.g., a scheme in which everyone performs equally poorly is deemed `fairer' than one where everyone performs better with some variation.
Therefore, the fairness analysis performed in Section~\ref{sec:fairness} focuses on the notion of Pareto improvements instead.
If one accepts this difference in fairness definitions, Figure~\ref{fig:mean-efficiency-gain-per-percentile} and Tables~\ref{tab:mann-whitney-deciles}--\ref{tab:mann-whitney-deciles-no-dropouts} provide strong support for hypothesis (2); partial support for hypothesis (7), in particular for the high stake treatments; but reject hypothesis (4) since an efficiency-maximizing scheme performs similarly to the theoretical karma Nash (even if it is history-unaware) and leads to Pareto improvements over the experimental karma subjects.

The pre-registration furthermore states~\citep[Section~2.3]{elokda2023karma}:\\
\emph{``As these are the first karma experiments, we have no benchmarks for expected effect sizes and variances, hence we are unable to run a formal sample size calculator but instead are forced to rely on rules of thumb for sample size and power in order to perform non-parametric tests a la Mann-Whitney-Wilcoxon.''}\\
Consistently with this statement, non-parametric Wilcoxon signed rank and \gls{MWW} tests are the main statistical test methods used in our analysis.
\newpage
\section{Detailed Results}
\label{sec:sup-results}

\subsection{Overview of Earnings}
\label{sec:earnings-statistics}

Table~\ref{tab:earnings} shows an overview of the actual total earnings (fixed plus bonus fee) in the experiment.

\begin{table}[!h]
\centering
\caption{Overview of total earnings $\payfix + \paybonus$ for all treatments. Dropouts for which $\payfix = \paybonus = 0$ are not included in the statistics. IQR refers to the interquartile range.}
\label{tab:earnings}
\begin{tabular}{ll||c|c|c}
\toprule
& & \multicolumn{3}{c}{\textbf{Richness of scheme}} \\
& & \multicolumn{1}{c|}{Binary} & \multicolumn{1}{c|}{Full Range} & \multicolumn{1}{c}{\textbf{Combined}} \\
\hline \hline
\multirow{4}{*}[-50pt]{\rotatebox[origin=c]{90}{\textbf{Urgency process}}} \hspace{1pt} & & & \\[-9pt]
& Low Stake \hspace{1pt} & \begin{tabular}{@{}lc@{}} Mean: & $\$10.03$ \\ Median: & $\$9.96$ \\ Range: & $\$3.10-\$17.67$ \\ IQR: & 
$\$8.59-\$11.84$ \end{tabular} &
\hspace{1pt} \begin{tabular}{@{}c@{}} $\$10.14$ \\ $\$10.30$ \\ $\$1.73-\$16.64$ \\ $\$8.93-\$11.93$ \end{tabular} \hspace{1pt} &
\hspace{1pt} \begin{tabular}{@{}c@{}} $\$10.09$ \\ $\$9.96$ \\ $\$1.73-\$17.67$ \\ $\$8.71-\$11.84$ \end{tabular} \hspace{1pt} \\
\cmidrule{2-5}
& High Stake \hspace{1pt} & \hspace{1pt} \begin{tabular}{@{}lc@{}} Mean: & $\$9.34$ \\ Median: & $\$8.92$ \\ Range: & $\$1.50-\$18.10$ \\ IQR: & $\$7.65-\$11.18$ \end{tabular} \hspace{1pt} &
\hspace{1pt} \begin{tabular}{@{}c@{}} $\$8.93$ \\ $\$8.99$ \\ $\$2.57-\$18.38$ \\ $\$6.63-\$11.01$ \end{tabular} \hspace{1pt} &
\hspace{1pt} \begin{tabular}{@{}c@{}} $\$9.13$ \\ $\$8.92$ \\ $\$1.50-\$18.38$ \\ $\$7.23-\$11.18$ \end{tabular} \hspace{1pt} \\
\cmidrule{2-5}
& \textbf{Combined} \hspace{1pt} & \hspace{1pt} \begin{tabular}{@{}lc@{}} Mean: & $\$9.69$ \\ Median: & $\$9.49$ \\ Range: & $\$1.50-\$18.10$ \\ IQR: & 
$\$7.91-\$11.58$ \end{tabular} \hspace{1pt} & \hspace{1pt} \begin{tabular}{@{}c@{}} $\$9.53$ \\ $\$9.79$ \\ $\$1.73-\$18.38$ \\ $\$7.45-\$11.71$ \end{tabular} \hspace{1pt} & \hspace{1pt} \begin{tabular}{@{}c@{}} $\$9.61$ \\ $\$9.61$ \\ $\$1.50-\$18.38$ \\ $\$7.79-\$11.67$ \end{tabular} \hspace{1pt} \\
\bottomrule
\end{tabular}
\end{table}

\subsection{Robustness Checks}
\label{sec:robustness}

\subsubsection{Including vs. Excluding Dropouts}
\label{sec:robustness-with-without-dropouts}
In order to test the robustness of our results against excluding dropouts from the analysis, we omit observations corresponding to dropouts in the forthcoming analysis.
Letting $D^G$ be the set of dropouts in group $G$, the group-level efficiency gain without dropouts is defined analogously to Equation~\eqref{eq:efficiency-gain-group} as
\begin{align}
\label{eq:efficiency-gain-group-no-dropouts}
E_G^{\textup{no-dropout}} = \frac{\sum_{i \in G \setminus D^G} \left(S_i - \Rrand_i\right)}{\sum_{i \in G\setminus D^G} \Rrand_i}.
\end{align}

Tables~\ref{tab:efficiency-groups-no-dropouts}--\ref{tab:wilcoxon-treatment-individuals-no-dropouts} mirror the results of Tables~\ref{tab:efficiency-groups}--\ref{tab:wilcoxon-treatment-individuals} but with dropouts excluded.
Excluding dropouts reinforces the main insight that efficiency gains are statistically significantly positive in all treatments, as it leads to a Pareto improvement of group-level efficiency gains over including dropouts, with all groups in all treatments achieving positive gains when dropouts are excluded.

\begin{table}[!h]
\centering
\caption{Analogue of Table~\ref{tab:efficiency-groups} of the group-level efficiency gains, but without including dropouts, cf. Equation~\eqref{eq:efficiency-gain-group-no-dropouts}.}
\label{tab:efficiency-groups-no-dropouts}
\begin{tabular}{ll||l|c}
\toprule
& & \multicolumn{2}{c}{\textbf{Richness of scheme}} \\
& & \multicolumn{1}{c|}{Binary} & \multicolumn{1}{c}{Full Range} \\
\hline \hline
\multirow{3}{*}[-32pt]{\rotatebox[origin=c]{90}{\begin{tabular}{{@{}c@{}}}
     \textbf{Urgency process}
\end{tabular}}} \hspace{1pt} & & & \\[-9pt]
& Low Stake \hspace{1pt} & \hspace{1pt} \begin{tabular}{@{}lc@{}} G1: & 17.22\% \\ G2: & 5.18\% \\ G3: & 8.61\% \\ G4: & 8.64\% \\ G5: & 13.52\% \end{tabular} \hspace{1pt} &
\hspace{1pt} \begin{tabular}{@{}lc@{}} G1: & 7.36\% \\ G2: & 6.73\% \\ G3: & 14.06\% \\ G4: & 9.14\% \\ G5: & 11.40\% \end{tabular} \hspace{1pt} \\
\cmidrule{2-4}
& High Stake \hspace{1pt} & \hspace{1pt} \begin{tabular}{@{}lc@{}}  G1: & 18.29\% \\ G2: & 5.04\% \\ G3: & 13.35\% \\ G4: & 17.12\% \\ G5: & 23.64\% \end{tabular} \hspace{1pt} &
\hspace{1pt} \begin{tabular}{@{}lc@{}} G1: & 10.05\% \\ G2: & 4.61\% \\ G3: & 5.61\% \\ G4: & 19.17\% \\ G5: & 17.97\% \end{tabular} \hspace{1pt} \\
\bottomrule
\end{tabular}
\end{table}

\begin{table}[!h]
\centering
\caption{Analogue of Table~\ref{tab:wilcoxon-treatment-groups} of the one-sided Wilcoxon signed rank tests of group-level efficiency gains, but without including dropouts, cf. Equation~\eqref{eq:efficiency-gain-group-no-dropouts}.}
\label{tab:wilcoxon-treatment-groups-no-dropouts}
\begin{tabular}{ll||c|c|c}
\toprule
& & \multicolumn{3}{c}{\textbf{Richness of scheme}} \\
& & \multicolumn{1}{c|}{Binary} & \multicolumn{1}{c|}{Full Range} & \multicolumn{1}{c}{\textbf{Combined}} \\
\hline \hline
\multirow{4}{*}[-5pt]{\rotatebox[origin=c]{90}{\textbf{Urgency process}}} \hspace{1pt} & & & \\[-9pt]
& Low Stake \hspace{1pt} & \begin{tabular}{@{}rc@{}} $W$: & \cellcolor{green!25}$15.0$ \\ $p$: & \cellcolor{green!25}$0.0313$ \end{tabular} &
\hspace{1pt} \begin{tabular}{@{}c@{}} \cellcolor{green!25}$15.0$ \\ \cellcolor{green!25}$0.0313$ \end{tabular} \hspace{1pt} &
\hspace{1pt} \begin{tabular}{@{}c@{}} \cellcolor{green!25}$55.0$ \\ \cellcolor{green!25}$0.0010$ \end{tabular} \hspace{1pt} \\
\cmidrule{2-5}
& High Stake \hspace{1pt} & \hspace{1pt} \begin{tabular}{@{}rc@{}} $W$: & \cellcolor{green!25}$15.0$ \\ $p$: & \cellcolor{green!25}$0.0313$ \end{tabular} \hspace{1pt} &
\hspace{1pt} \begin{tabular}{@{}c@{}} \cellcolor{green!25}$15.0$ \\ \cellcolor{green!25}$0.0313$ \end{tabular} \hspace{1pt} &
\hspace{1pt} \begin{tabular}{@{}c@{}} \cellcolor{green!25}$55.0$ \\ \cellcolor{green!25}$0.0001$ \end{tabular} \hspace{1pt} \\
\cmidrule{2-5}
& \textbf{Combined} \hspace{1pt} & \hspace{1pt} \begin{tabular}{@{}rc@{}} $W$: & \cellcolor{green!25}$55.0$ \\ $p$: & \cellcolor{green!25}$0.0010$ \end{tabular} \hspace{1pt} & \hspace{1pt} \begin{tabular}{@{}c@{}} \cellcolor{green!25}$55.0$ \\ \cellcolor{green!25}$0.0001$ \end{tabular} \hspace{1pt} & \hspace{1pt} \begin{tabular}{@{}c@{}} \cellcolor{green!25}$210.0$ \\ \cellcolor{green!25}$<0.0001$ \end{tabular} \hspace{1pt} \\
\bottomrule
\end{tabular}
\end{table}

\begin{table}[!h]
\centering
\caption{Analogue of Table~\ref{tab:wilcoxon-treatment-individuals} of the one-sided Wilcoxon signed rank tests of individual-level efficiency gains, cf. Equation~\eqref{eq:efficiency-gain}, but excluding dropout observations.
Treatments have varying sample sizes $n$ depending on the number of dropouts, as indicated.}
\label{tab:wilcoxon-treatment-individuals-no-dropouts}
\begin{tabular}{ll||c|c|c}
\toprule
& & \multicolumn{3}{c}{\textbf{Richness of scheme}} \\
& & \multicolumn{1}{c|}{Binary} & \multicolumn{1}{c|}{Full Range} & \multicolumn{1}{c}{\textbf{Combined}} \\
\hline \hline
\multirow{4}{*}[-5pt]{\rotatebox[origin=c]{90}{\textbf{Urgency process}}} \hspace{1pt} & & & \\[-9pt]
& Low Stake \hspace{1pt} & \begin{tabular}{@{}rc@{}} $n$: & \cellcolor{green!25}$93$ \\ $W$: & \cellcolor{green!25}$3\,456.0$ \\ $p$: & \cellcolor{green!25}$<0.0001$ \end{tabular} &
\hspace{1pt} \begin{tabular}{@{}c@{}} \cellcolor{green!25}$95$ \\ \cellcolor{green!25}$3\,647.0$ \\ \cellcolor{green!25}$<0.0001$ \end{tabular} \hspace{1pt} &
\hspace{1pt} \begin{tabular}{@{}c@{}} \cellcolor{green!25}$188$ \\ \cellcolor{green!25}$14\,168.0$ \\ \cellcolor{green!25}$<0.0001$ \end{tabular} \hspace{1pt} \\
\cmidrule{2-5}
& High Stake \hspace{1pt} & \hspace{1pt} \begin{tabular}{@{}rc@{}} $n$: & \cellcolor{green!25}$93$ \\ $W$: & \cellcolor{green!25}$3\,663.0$ \\ $p$: & \cellcolor{green!25}$<0.0001$ \end{tabular} \hspace{1pt} &
\hspace{1pt} \begin{tabular}{@{}c@{}} \cellcolor{green!25}$96$ \\ \cellcolor{green!25}$3\,483.0$ \\ \cellcolor{green!25}$<0.0001$ \end{tabular} \hspace{1pt} &
\hspace{1pt} \begin{tabular}{@{}c@{}} \cellcolor{green!25}$189$ \\ \cellcolor{green!25}$14\,234.0$ \\ \cellcolor{green!25}$<0.0001$ \end{tabular} \hspace{1pt} \\
\cmidrule{2-5}
& \textbf{Combined} \hspace{1pt} & \hspace{1pt} \begin{tabular}{@{}rc@{}} $n$: & \cellcolor{green!25}$186$ \\ $W$: & \cellcolor{green!25}$14\,247.0$ \\ $p$: & \cellcolor{green!25}$<0.0001$ \end{tabular} \hspace{1pt} & \hspace{1pt} \begin{tabular}{@{}c@{}} \cellcolor{green!25}$191$ \\ \cellcolor{green!25}$14\,111.0$ \\ \cellcolor{green!25}$<0.0001$ \end{tabular} \hspace{1pt} & \hspace{1pt} \begin{tabular}{@{}c@{}} \cellcolor{green!25}$377$ \\ \cellcolor{green!25}$56\,598.0$ \\ \cellcolor{green!25}$<0.0001$ \end{tabular} \hspace{1pt} \\
\bottomrule
\end{tabular}
\end{table}

\subsubsection{Wilcoxon signed rank vs. Welch's $t$-tests}

Tables~\ref{tab:ttest-treatment-groups}--\ref{tab:ttest-treatment-individuals} mirror the results of Tables~\ref{tab:wilcoxon-treatment-groups}--\ref{tab:wilcoxon-treatment-individuals} but performing one-sided Welch's $t$-tests instead of Wilcoxon signed rank.
Results are the same.

\begin{table}[!h]
\centering
\caption{Analogue of Table~\ref{tab:wilcoxon-treatment-groups} of the one-sided Wilcoxon signed rank tests of group-level efficiency gains, cf. Equation~\eqref{eq:efficiency-gain-group}, but performing one-sided Welch's $t$-tests instead.}
\label{tab:ttest-treatment-groups}
\begin{tabular}{ll||c|c|c}
\toprule
& & \multicolumn{3}{c}{\textbf{Richness of scheme}} \\
& & \multicolumn{1}{c|}{Binary} & \multicolumn{1}{c|}{Full Range} & \multicolumn{1}{c}{\textbf{Combined}} \\
\hline \hline
\multirow{4}{*}[-5pt]{\rotatebox[origin=c]{90}{\textbf{Urgency process}}} \hspace{1pt} & & & \\[-9pt]
& Low Stake \hspace{1pt} & \begin{tabular}{@{}rc@{}} $t$: & \cellcolor{green!25}$3.9436$ \\ $p$: & \cellcolor{green!25}$0.0085$ \end{tabular} &
\hspace{1pt} \begin{tabular}{@{}c@{}} \cellcolor{green!25}$3.4986$ \\ \cellcolor{green!25}$0.0125$ \end{tabular} \hspace{1pt} &
\hspace{1pt} \begin{tabular}{@{}c@{}} \cellcolor{green!25}$5.5292$ \\ \cellcolor{green!25}$0.0002$ \end{tabular} \hspace{1pt} \\
\cmidrule{2-5}
& High Stake \hspace{1pt} & \hspace{1pt} \begin{tabular}{@{}rc@{}} $t$: & \cellcolor{green!25}$3.8893$ \\ $p$: & \cellcolor{green!25}$0.0089$ \end{tabular} \hspace{1pt} &
\hspace{1pt} \begin{tabular}{@{}c@{}} \cellcolor{green!25}$3.0848$ \\ \cellcolor{green!25}$0.0184$ \end{tabular} \hspace{1pt} &
\hspace{1pt} \begin{tabular}{@{}c@{}} \cellcolor{green!25}$5.1018$ \\ \cellcolor{green!25}$0.0003$ \end{tabular} \hspace{1pt} \\
\cmidrule{2-5}
& \textbf{Combined} \hspace{1pt} & \hspace{1pt} \begin{tabular}{@{}rc@{}} $t$: & \cellcolor{green!25}$4.6096$ \\ $p$: & \cellcolor{green!25}$0.0006$ \end{tabular} \hspace{1pt} & \hspace{1pt} \begin{tabular}{@{}c@{}} \cellcolor{green!25}$4.4451$ \\ \cellcolor{green!25}$0.0008$ \end{tabular} \hspace{1pt} & \hspace{1pt} \begin{tabular}{@{}c@{}} \cellcolor{green!25}$6.5313$ \\ \cellcolor{green!25}$<0.0001$ \end{tabular} \hspace{1pt} \\
\bottomrule
\end{tabular}
\end{table}

\begin{table}[!h]
\centering
\caption{Analogue of Table~\ref{tab:wilcoxon-treatment-individuals} of the one-sided Wilcoxon signed rank tests of individual-level efficiency gains, cf. Equation~\eqref{eq:efficiency-gain}, but performing one-sided Welch's $t$-tests instead.}
\label{tab:ttest-treatment-individuals}
\begin{tabular}{ll||c|c|c}
\toprule
& & \multicolumn{3}{c}{\textbf{Richness of scheme}} \\
& & \multicolumn{1}{c|}{Binary} & \multicolumn{1}{c|}{Full Range} & \multicolumn{1}{c}{\textbf{Combined}} \\
\hline \hline
\multirow{4}{*}[-5pt]{\rotatebox[origin=c]{90}{\textbf{Urgency process}}} \hspace{1pt} & & & \\[-9pt]
& Low Stake \hspace{1pt} & \begin{tabular}{@{}rc@{}} $t$: & \cellcolor{green!25}$2.1513$ \\ $p$: & \cellcolor{green!25}$0.0169$ \end{tabular} &
\hspace{1pt} \begin{tabular}{@{}c@{}} \cellcolor{green!25}$2.0868$ \\ \cellcolor{green!25}$0.0197$ \end{tabular} \hspace{1pt} &
\hspace{1pt} \begin{tabular}{@{}c@{}} \cellcolor{green!25}$2.9979$ \\ \cellcolor{green!25}$0.0015$ \end{tabular} \hspace{1pt} \\
\cmidrule{2-5}
& High Stake \hspace{1pt} & \hspace{1pt} \begin{tabular}{@{}rc@{}} $t$: & \cellcolor{green!25}$3.9044$ \\ $p$: & \cellcolor{green!25}$0.0001$ \end{tabular} \hspace{1pt} &
\hspace{1pt} \begin{tabular}{@{}c@{}} \cellcolor{green!25}$2.9042$ \\ \cellcolor{green!25}$0.0023$ \end{tabular} \hspace{1pt} &
\hspace{1pt} \begin{tabular}{@{}c@{}} \cellcolor{green!25}$4.8109$ \\ \cellcolor{green!25}$<0.0001$ \end{tabular} \hspace{1pt} \\
\cmidrule{2-5}
& \textbf{Combined} \hspace{1pt} & \hspace{1pt} \begin{tabular}{@{}rc@{}} $t$: & \cellcolor{green!25}$4.3340$ \\ $p$: & \cellcolor{green!25}$<0.0001$ \end{tabular} \hspace{1pt} & \hspace{1pt} \begin{tabular}{@{}c@{}} \cellcolor{green!25}$3.5457$ \\ \cellcolor{green!25}$0.0002$ \end{tabular} \hspace{1pt} & \hspace{1pt} \begin{tabular}{@{}c@{}} \cellcolor{green!25}$5.5588$ \\ \cellcolor{green!25}$<0.0001$ \end{tabular} \hspace{1pt} \\
\bottomrule
\end{tabular}
\end{table}

\subsubsection{Means vs. Medians}
\label{sec:robustness-mean-median}

\begin{figure}[!tb]
    \centering
    \begin{subfigure}[b]{0.4\textwidth}
        \centering
        \includegraphics[width=\textwidth]{figures/mean-efficiency-gains-groups.pdf}
        \caption{Means, with dropouts, cf. Figure~\ref{fig:mean-efficiency-gains-groups}}
        \label{fig:mean-efficiency-gains-groups-appendix}
    \end{subfigure}
    \hfil
    \begin{subfigure}[b]{0.4\textwidth}
        \centering
        \includegraphics[width=\textwidth]{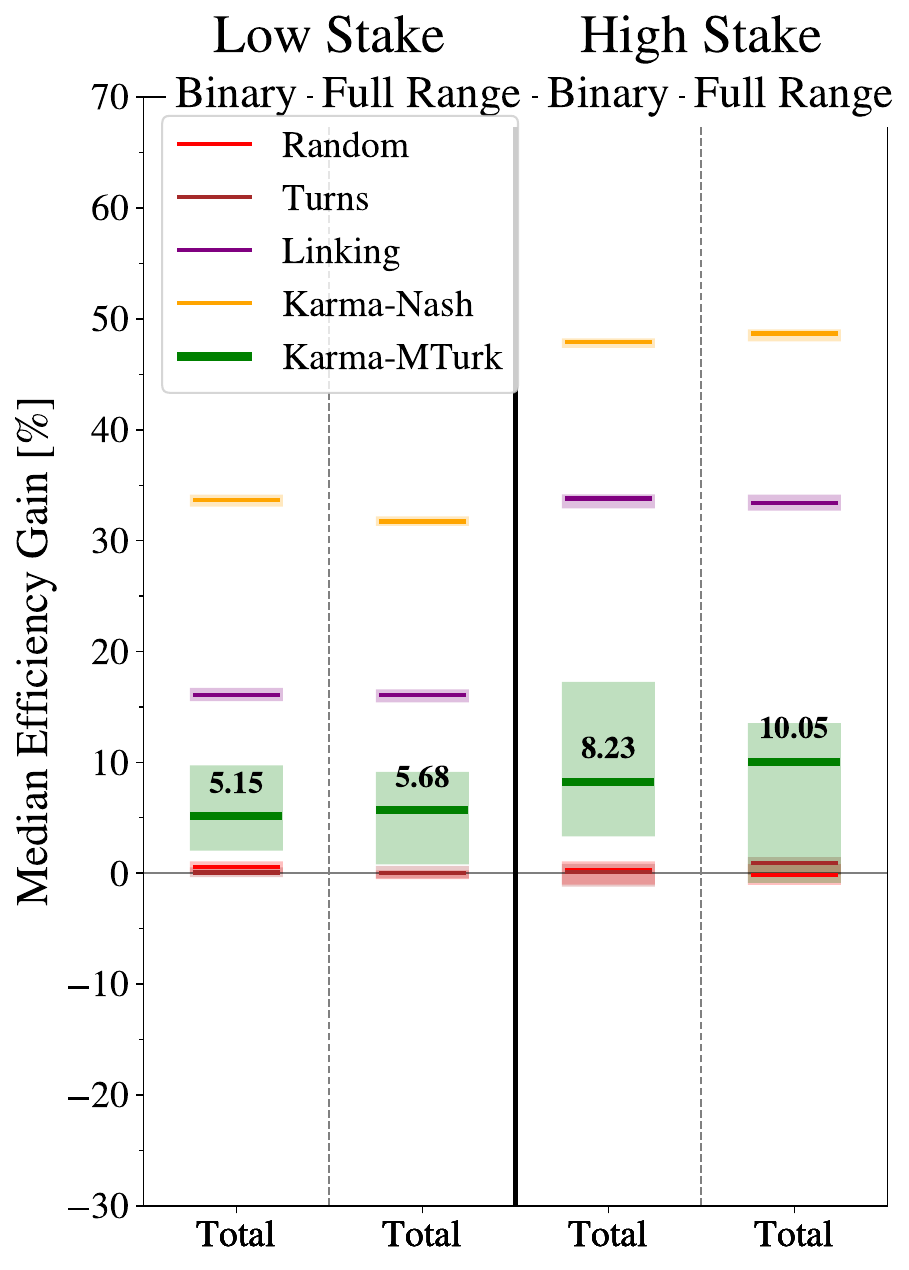}
        \caption{Medians, with dropouts}
        \label{fig:median-efficiency-gains-groups}
    \end{subfigure}

    \medskip

    \begin{subfigure}[b]{0.4\textwidth}
        \centering
        \includegraphics[width=\textwidth]{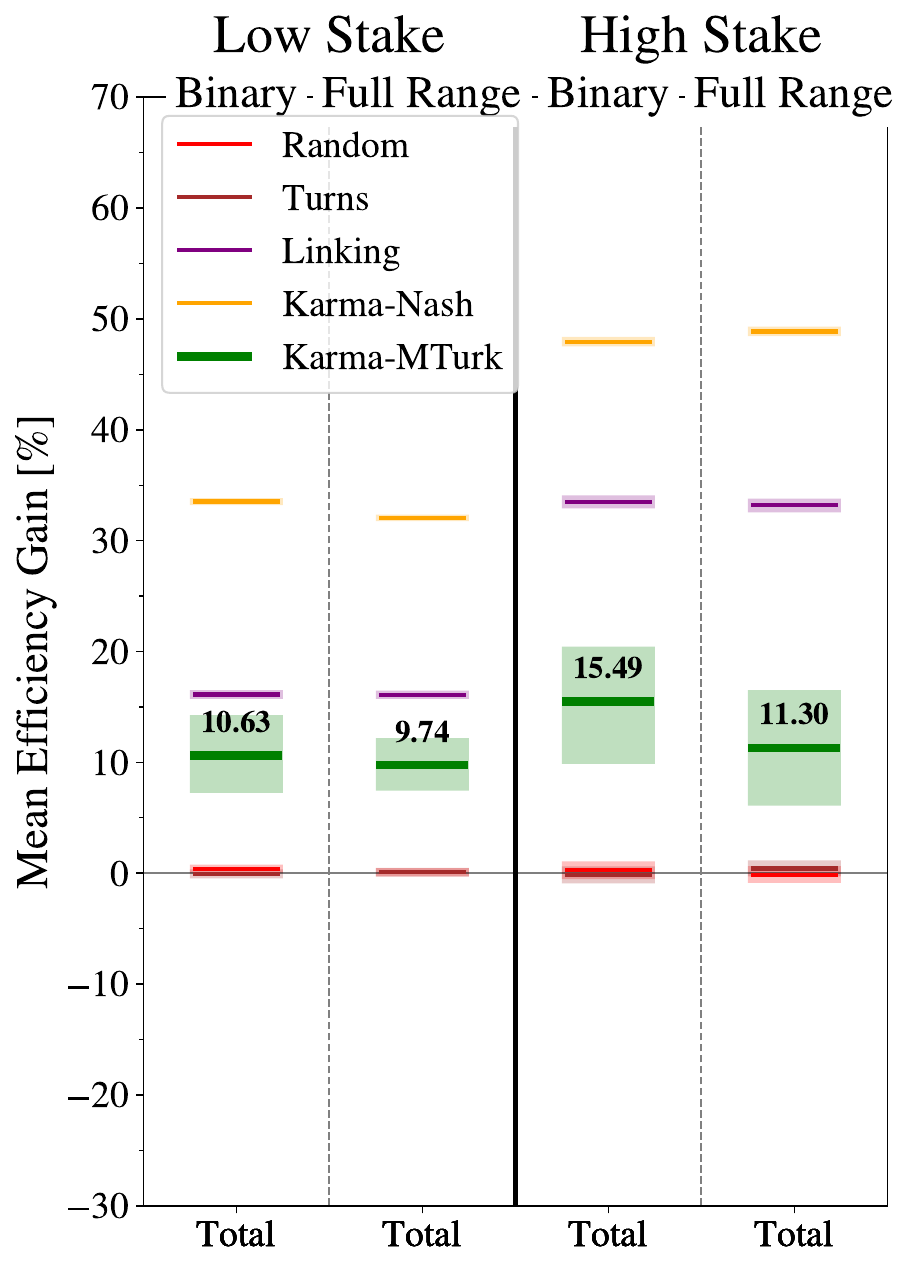}
        \caption{Means, without dropouts}
        \label{fig:mean-efficiency-gains-groups-non-dropouts}
    \end{subfigure}
    \hfil
    \begin{subfigure}[b]{0.4\textwidth}
        \centering
        \includegraphics[width=\textwidth]{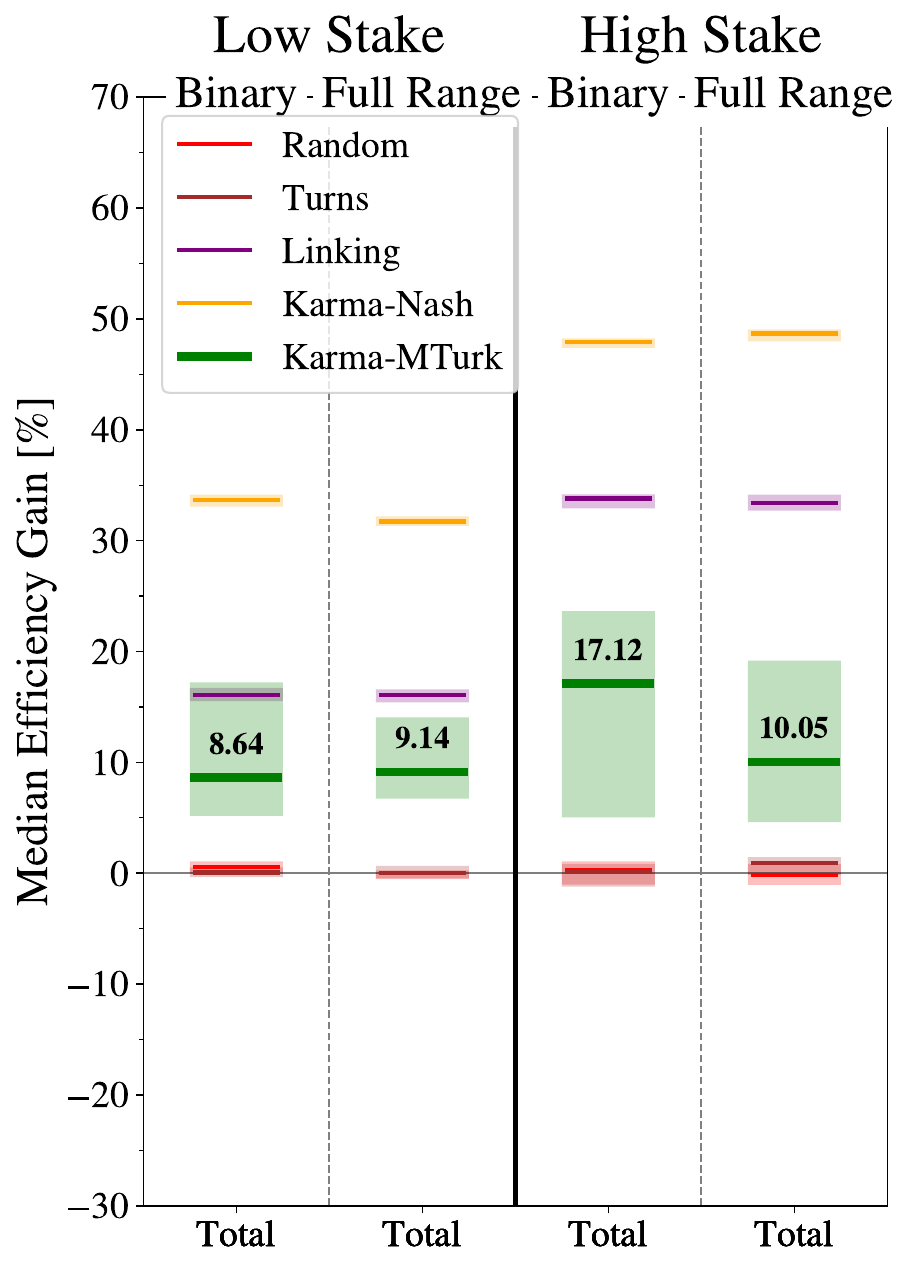}
        \caption{Medians, without dropouts}
        \label{fig:median-efficiency-gains-groups-non-dropouts}
    \end{subfigure}
    
    \caption{Robustness of Figure~\ref{fig:mean-efficiency-gains-groups} to the choice of means vs. medians, and whether dropouts are included or not in the group-level efficiency gain calculation.
    }
    \label{fig:mean-median-efficiency-gains-groups-with-without-dropouts}
\end{figure}

Figure~\ref{fig:mean-median-efficiency-gains-groups-with-without-dropouts} compares the mean group-level efficiency gains (left panel), cf. Figure~\ref{fig:mean-efficiency-gains-groups}, to the medians (right panel), both with dropouts observations included (top panel) and excluded (bottom panel).
Means and medians are found to be similar, whereas excluding dropouts leads to a Pareto improvement in both means and medians, consistently with the findings of~\ref{sec:robustness}.

\begin{figure}[!tb]
    \centering
    \begin{subfigure}[b]{0.49\textwidth}
        \centering
        \includegraphics[width=\textwidth]{figures/mean-efficiency-gains.pdf}
        \caption{Means, with dropouts, cf. Figure~\ref{fig:mean-efficiency-gains-individuals}}
        \label{fig:mean-efficiency-gains-appendix}
    \end{subfigure}
    \hfil
    \begin{subfigure}[b]{0.49\textwidth}
        \centering
        \includegraphics[width=\textwidth]{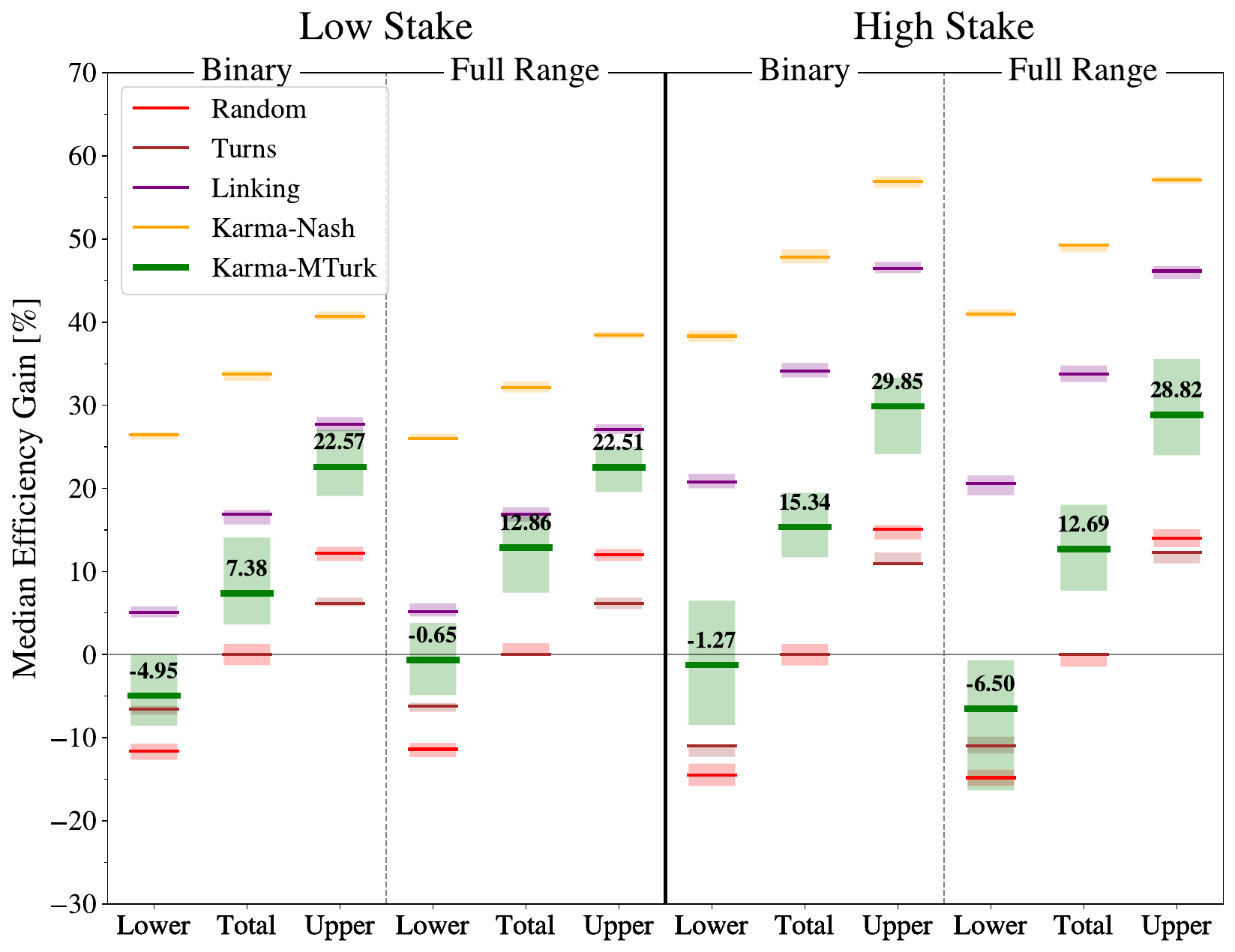}
        \caption{Medians, with dropouts}
        \label{fig:median-efficiency-gains}
    \end{subfigure}

    \medskip
    
    \begin{subfigure}[b]{0.49\textwidth}
        \centering
        \includegraphics[width=\textwidth]{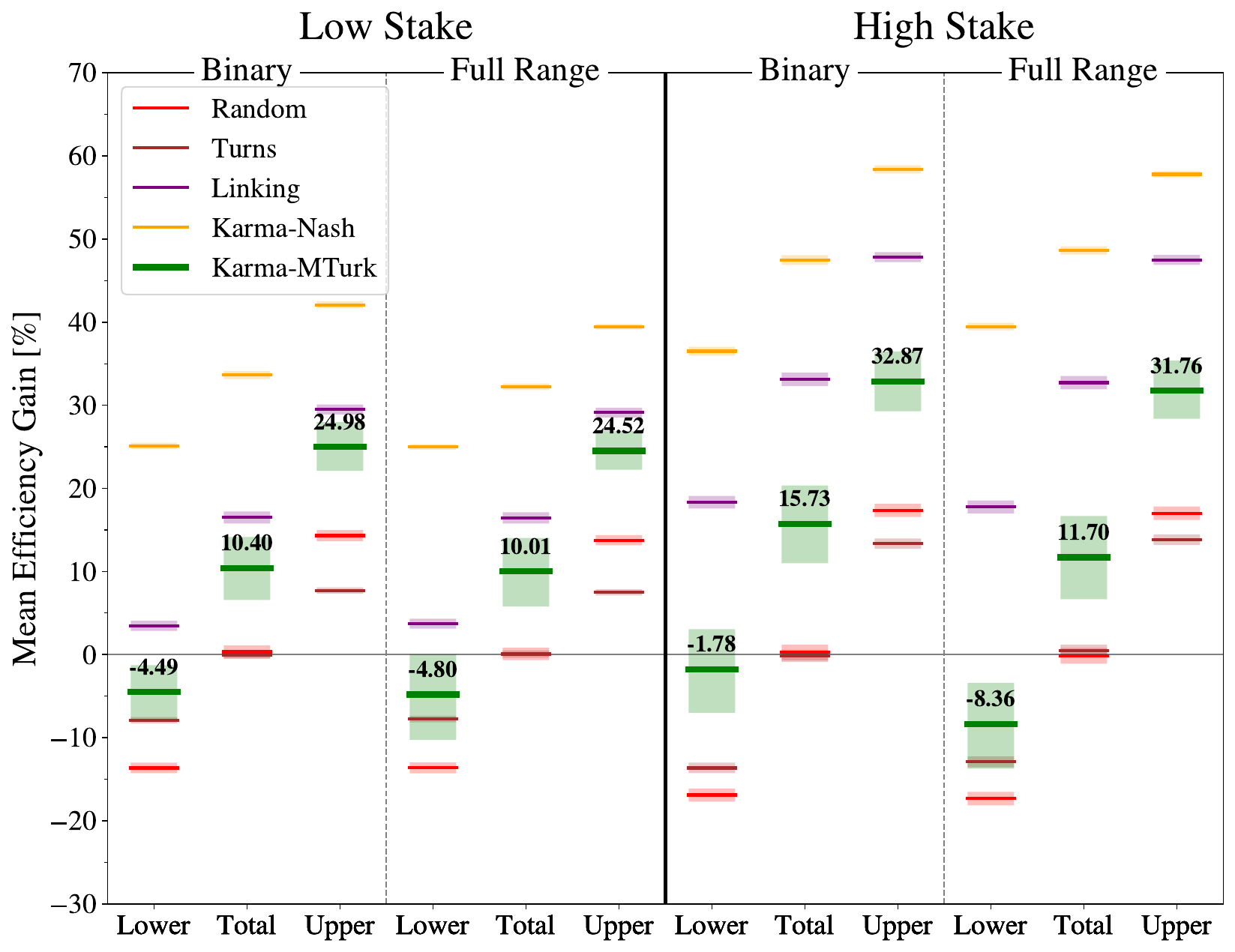}
        \caption{Means, without dropouts}
        \label{fig:mean-efficiency-gains-non-dropouts}
    \end{subfigure}
    \hfil
    \begin{subfigure}[b]{0.49\textwidth}
        \centering
        \includegraphics[width=\textwidth]{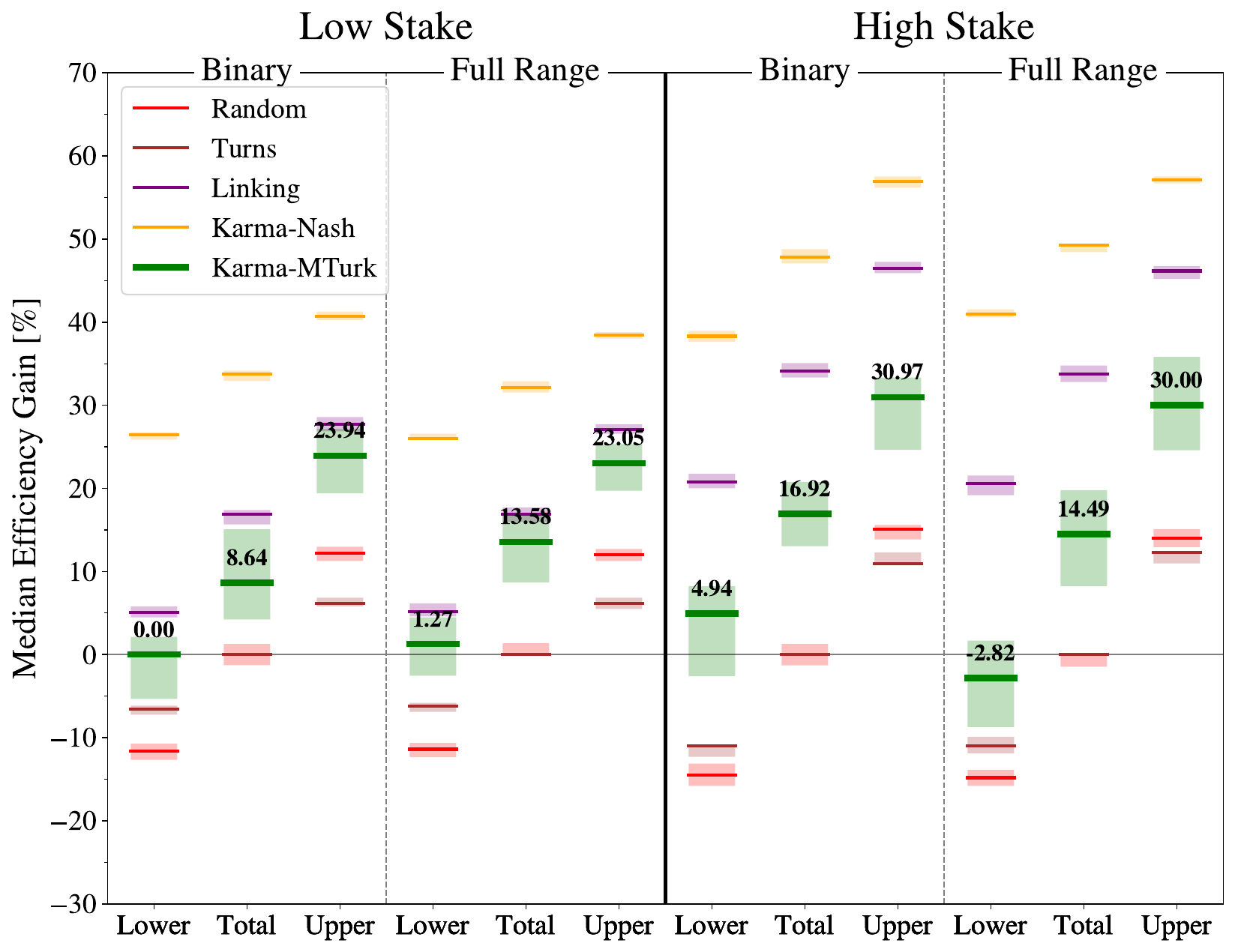}
        \caption{Medians, without dropouts}
        \label{fig:median-efficiency-gains-non-dropouts}
    \end{subfigure}
    
    \caption{Robustness of Figure~\ref{fig:mean-efficiency-gains-individuals} to the choice of means vs. medians, and whether dropout observations are included or not.
    }
    \label{fig:mean-median-efficiency-gains-with-without-dropouts}
\end{figure}

Figure~\ref{fig:mean-median-efficiency-gains-with-without-dropouts} performs the same robustness check as Figure~\ref{fig:mean-median-efficiency-gains-groups-with-without-dropouts} but at the level of individual efficiency gains, cf. Figure~\ref{fig:mean-efficiency-gains-individuals}.
When dropouts are included, medians are quantitatively slightly higher than means; while no pronounced differences are observed between means and medians when dropouts are excluded.
As before, both means and medians are higher when dropouts are excluded, and especially so in the lower half of the population, which is consistent with the fact that dropouts perform particularly poorly with persistent zero bids.

\newgeometry{left=1.5cm, right=1.5cm} 

\begin{figure}[!tb]
    \centering
    \begin{subfigure}[b]{0.49\textwidth}
        \centering
        \includegraphics[width=\textwidth]{figures/low-high-stake-mean-efficiency-gain-per-percentile.pdf}
        \caption{Means, with dropouts, cf. Figure~\ref{fig:mean-efficiency-gain-per-percentile}}
        \label{fig:mean-efficiency-gain-per-percentile-appendix}
    \end{subfigure}
    \hfil
    \begin{subfigure}[b]{0.49\textwidth}
        \centering
        \includegraphics[width=\textwidth]{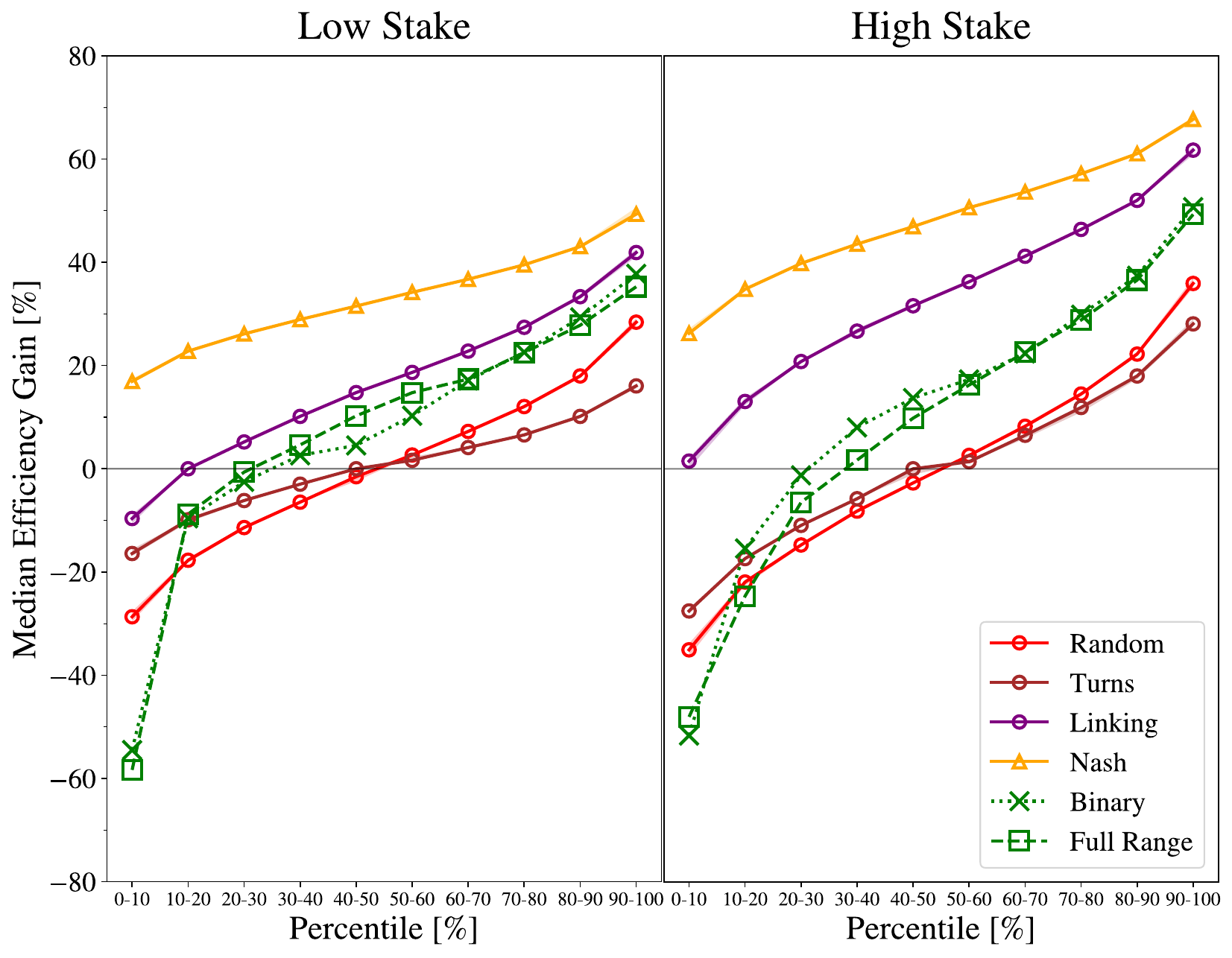}
        \caption{Medians, with dropouts}
        \label{fig:median-efficiency-gain-per-percentile}
    \end{subfigure}
    
    \begin{subfigure}[b]{0.49\textwidth}
        \centering
        \includegraphics[width=\textwidth]{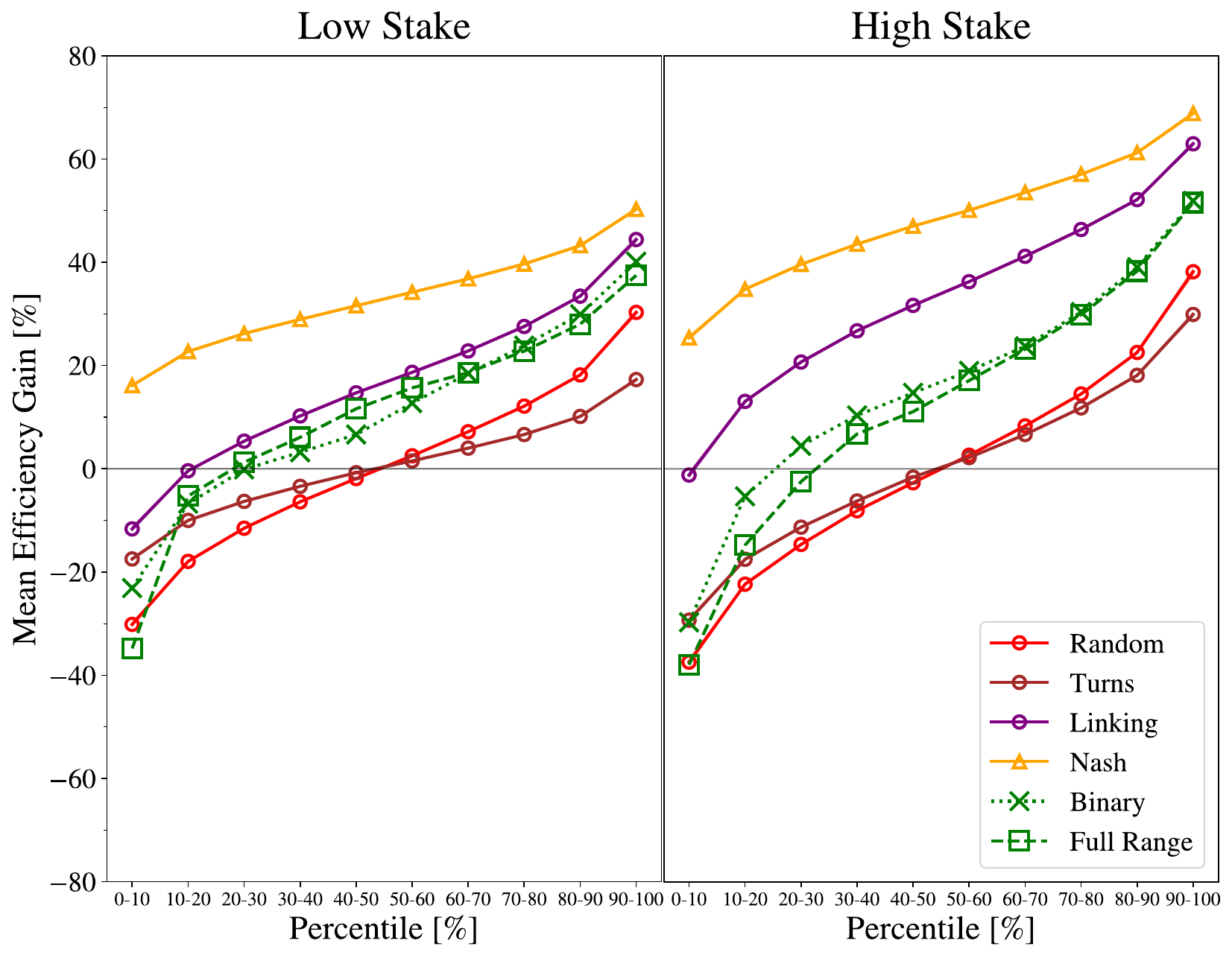}
        \caption{Means, without dropouts}
        \label{fig:mean-efficiency-gain-per-percentile-non-dropouts}
    \end{subfigure}
    \hfil
    \begin{subfigure}[b]{0.49\textwidth}
        \centering
        \includegraphics[width=\textwidth]{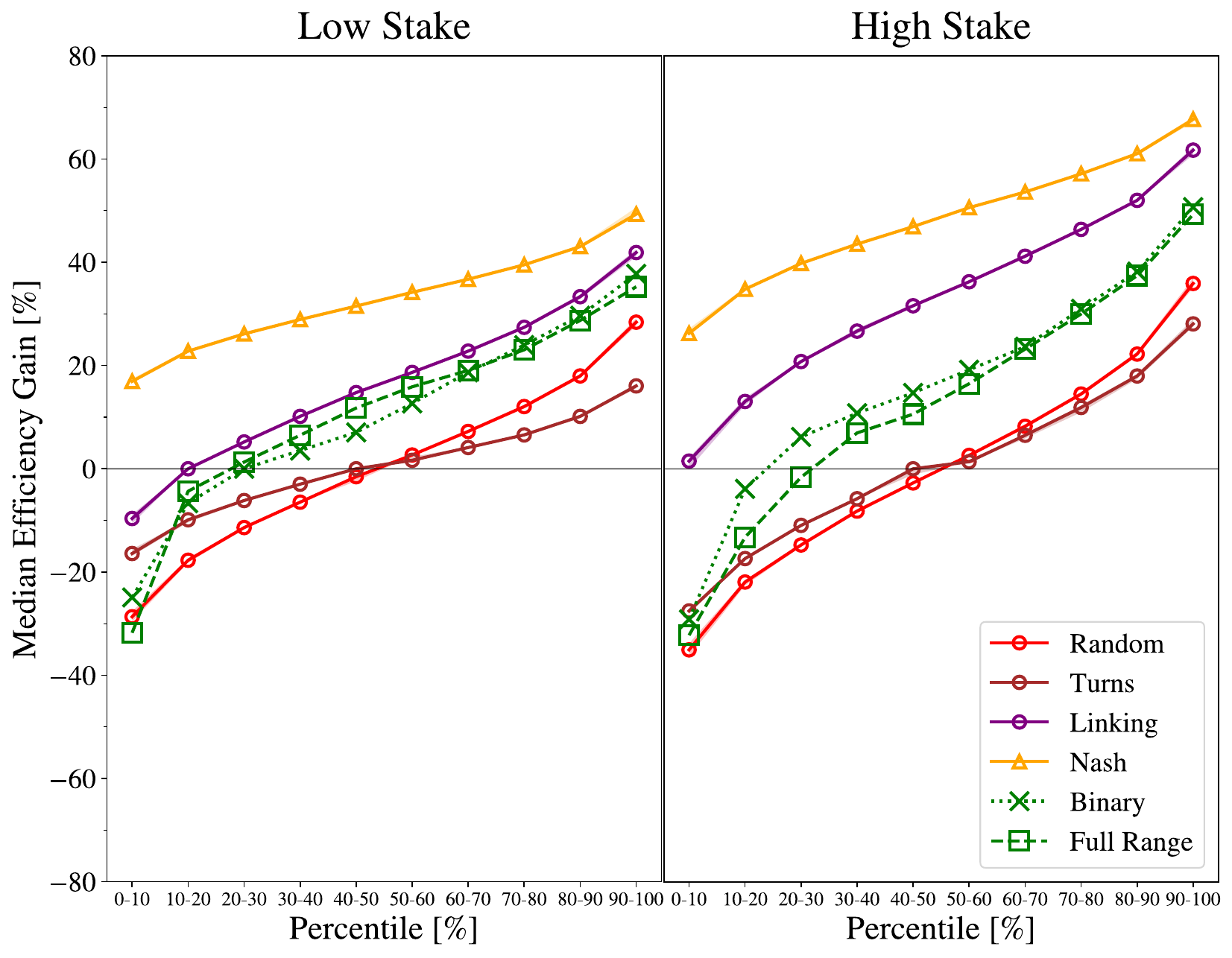}
        \caption{Medians, without dropouts}
        \label{fig:median-efficiency-gain-per-percentile-non-dropouts}
    \end{subfigure}
    
    \caption{Robustness of Figure~\ref{fig:mean-efficiency-gain-per-percentile} to the choice of means vs. medians, and whether dropout observations are included or not.
    }
    \label{fig:mean-median-efficiency-gain-per-percentile-with-without-dropouts}
\end{figure}

Figure~\ref{fig:mean-median-efficiency-gain-per-percentile-with-without-dropouts} performs the same robustness check as Figures~\ref{fig:mean-median-efficiency-gains-groups-with-without-dropouts}--\ref{fig:mean-median-efficiency-gains-with-without-dropouts}, but this time for the fairness analysis of Figure~\ref{fig:mean-efficiency-gain-per-percentile}.
In this finer-grained analysis there are no visible differences between means and medians, while excluding dropouts once again leads to higher efficiency gains especially in the bottom two deciles.
Namely, we observe an almost Pareto improvement over the random benchmark across all deciles when dropouts are excluded.

\restoregeometry

\subsection{Statistical Tests}
\label{sec:sup-tests}

\subsubsection{Power Calculation}
\label{sec:power}
We perform non-parametric \gls{MWW} or Wilcoxon signed rank tests for which there are less well-established definitions of effect sizes for the purpose of power calculations.
We follow~\cite{noether1987sample} who suggests to define effect size in one-sided tests of if distribution $X$ is stochastically greater than $Y$ (in the case of \gls{MWW}), or has central tendency greater than $Y$ (in the case of Wilcoxon signed rank), as $\prob[X > Y]$.
For one-sided \gls{MWW} tests, \cite{noether1987sample} gives the following relation
\begin{equation}
\label{eq:power}
    \prob_\textup{min}[X > Y] = \left(z_\alpha + z_\beta\right) \sqrt{\frac{n_X + n_Y}{12 \, n_X \, n_Y}} + 0.5,
\end{equation}
where $\alpha$ is the test significance level, $1-\beta$ the statistical power, $z_\alpha$ (respectively, $z_\beta$) the $\alpha$-level (respectively, $\beta$-level) significance point in the standard normal distribution, $n_X$ (respectively, $n_Y$) the sample size of $X$ (respectively, $Y$), and $\prob_\textup{min}[X > Y]$ the minimum effect size that can be detected with this test configuration.
For one-sided Wilcoxon signed-rank tests, the relation is more complex as it depends on the assumed data-generating distribution, but it coincides with Equation~\eqref{eq:power} (with $n_X=n_Y=n$) if one assumes normality~\citep{noether1987sample}.

Therefore, at standard $5\%$ significance and $80\%$ power (corresponding to $z_\alpha=1.645$, $z_\beta=0.8416$), our one-sided statistical tests can detect $\prob[X > Y] \geq 0.954$ in the group-level analysis (with $n_X=n_Y=5$, and all observations of $X$ must be greater than $Y$ to pass the test); and $\prob[X > Y] \geq 0.6015$ in the individual-level analysis (with $n_X=n_Y=100$).
For two-sided tests, Equation~\eqref{eq:power} holds using $z_{\alpha / 2}$ instead of $z_\alpha$, and our group-level analysis cannot detect differences at $5\%$ significance but can detect $\prob[X \neq Y] \geq 0.954$ at $10\%$ significance; while the individual-level analysis can detect $\prob[X \neq Y] \geq 0.6144$ at $5\%$ significance.

\subsubsection{Statistical Test Results}
Table~\ref{tab:wilcoxon-first-second-half} reports two-sided Wilcoxon signed rank test results when comparing the individual-level efficiency gains in the first versus the second half of the experiment, per treatment.
Individual-level rather than group-level gains are used in this test in order to increase the statistical power of the test.
With the exception of the high stake-binary treatment, no statistically significant differences are observed.
This is also the case if dropout observations are excluded, as reported in Table~\ref{tab:wilcoxon-first-second-half-no-dropouts}.

\begin{table}[!h]
\centering 
\caption{Two-sided Wilcoxon signed rank tests with test statistics $W$ and associated $p$-values of individual-level efficiency gains in the first versus second half of the experiment, i.e., Equation~\eqref{eq:efficiency-gain} with scores summed over rounds $t \in \{1,\dots,25\}$ versus $t \in \{26,\dots,50\}$, for all treatments.
Tests passing the $5\%$ significance level are highlighted green.
All treatments have $n=100$ observations.}
\label{tab:wilcoxon-first-second-half}
\begin{tabular}{ll||l|c|c}
\toprule
& & \multicolumn{3}{c}{\textbf{Richness of scheme}} \\
& & \multicolumn{1}{c|}{\hspace{1pt} Binary\hspace{1pt}} & \multicolumn{1}{c|}{\hspace{1pt} Full Range \hspace{1pt}} & \multicolumn{1}{c}{\hspace{1pt} \textbf{Combined} \hspace{1pt}} \\
\hline \hline
\multirow{4}{*}[-5pt]{\rotatebox[origin=c]{90}{\textbf{Urgency process}}} \hspace{1pt} & & & \\[-9pt]
& Low Stake \hspace{1pt} & \hspace{1pt} \begin{tabular}{@{}rc@{}} $W$: & $2\,506.0$ \\ $p$: & $0.9479$ \end{tabular} \hspace{1pt} &
\hspace{1pt} \begin{tabular}{@{}c@{}} $2\,197.0$ \\ $0.3319$ \end{tabular} \hspace{1pt} &
\hspace{1pt} \begin{tabular}{@{}c@{}} $9\,389.0$ \\ $0.4904$ \end{tabular} \hspace{1pt} \\
\cmidrule{2-5}
& High Stake \hspace{1pt} & \hspace{1pt} \begin{tabular}{@{}rc@{}} $W$: & \cellcolor{green!25}$1\,937.0$ \\ $p$: & \cellcolor{green!25}$0.0432$ \end{tabular} \hspace{1pt} &
\hspace{1pt} \begin{tabular}{@{}c@{}} $2\,359.0$ \\ $0.6856$ \end{tabular} \hspace{1pt} &
\hspace{1pt} \begin{tabular}{@{}c@{}} $8\,982.5$ \\ $0.2343$ \end{tabular} \hspace{1pt} \\
\cmidrule{2-5}
& \textbf{Combined} \hspace{1pt} & \hspace{1pt} \begin{tabular}{@{}rc@{}} $W$: & $8\,911.0$ \\ $p$: & $0.1646$ \end{tabular} \hspace{1pt} & \hspace{1pt} \begin{tabular}{@{}c@{}} $9\,076.5$ \\ $0.3377$ \end{tabular} \hspace{1pt} & \hspace{1pt} \begin{tabular}{@{}c@{}} $38\,850.0$ \\ $0.7111$ \end{tabular} \hspace{1pt} \\
\bottomrule
\end{tabular}
\end{table}

\begin{table}[!h]
\centering 
\caption{Analogue of Table~\ref{tab:wilcoxon-first-second-half} of the two-sided Wilcoxon signed rank tests of individual-level efficiency gains in the first versus second half of the experiment, but excluding dropout observations.
Treatments have varying sample sizes $n$ depending on the number of dropouts, as indicated.}
\label{tab:wilcoxon-first-second-half-no-dropouts}
\begin{tabular}{ll||l|c|c}
\toprule
& & \multicolumn{3}{c}{\textbf{Richness of scheme}} \\
& & \multicolumn{1}{c|}{\hspace{1pt} Binary\hspace{1pt}} & \multicolumn{1}{c|}{\hspace{1pt} Full Range \hspace{1pt}} & \multicolumn{1}{c}{\hspace{1pt} \textbf{Combined} \hspace{1pt}} \\
\hline \hline
\multirow{4}{*}[-5pt]{\rotatebox[origin=c]{90}{\textbf{Urgency process}}} \hspace{1pt} & & & \\[-9pt]
& Low Stake \hspace{1pt} & \hspace{1pt} \begin{tabular}{@{}rc@{}} $n$: & $93$ \\ $W$: & $2\,149.0$ \\ $p$: & $0.8888$ \end{tabular} \hspace{1pt} &
\hspace{1pt} \begin{tabular}{@{}c@{}} $95$ \\ $2\,024.0$ \\ $0.4317$ \end{tabular} \hspace{1pt} &
\hspace{1pt} \begin{tabular}{@{}c@{}} $188$ \\ $8\,505.0$ \\ $0.7016$ \end{tabular} \hspace{1pt} \\
\cmidrule{2-5}
& High Stake \hspace{1pt} & \hspace{1pt} \begin{tabular}{@{}rc@{}} $n$: & \cellcolor{green!25}$93$ \\ $W$: & \cellcolor{green!25}$1\,538.0$ \\ $p$: & \cellcolor{green!25}$0.0131$ \end{tabular} \hspace{1pt} &
\hspace{1pt} \begin{tabular}{@{}c@{}} $96$ \\ $2\,275.0$ \\ $0.8464$ \end{tabular} \hspace{1pt} &
\hspace{1pt} \begin{tabular}{@{}c@{}} $189$ \\ $7\,748.5$ \\ $0.1027$ \end{tabular} \hspace{1pt} \\
\cmidrule{2-5}
& \textbf{Combined} \hspace{1pt} & \hspace{1pt} \begin{tabular}{@{}rc@{}} $n$: & $186$ \\ $W$: & $7\,316.0$ \\ $p$: & $0.0606$ \end{tabular} \hspace{1pt} & \hspace{1pt} \begin{tabular}{@{}c@{}} $191$ \\ $8\,564.5$ \\ $0.5033$ \end{tabular} \hspace{1pt} & \hspace{1pt} \begin{tabular}{@{}c@{}} $377$ \\ $33\,516.0$ \\ $0.3621$ \end{tabular} \hspace{1pt} \\
\bottomrule
\end{tabular}
\end{table}

Tables~\ref{tab:mann-whitney-inter-treatment-groups}--\ref{tab:mann-whitney-inter-treatment-individuals-no-dropouts} report two-sided \gls{MWW} test results when comparing the efficiency gains of each treatment pair as well as between combined treatments, respectively for the level of groups (Table~\ref{tab:mann-whitney-inter-treatment-groups}), individuals (Table~\ref{tab:mann-whitney-inter-treatment-individuals}), and individuals excluding dropouts (Table~\ref{tab:mann-whitney-inter-treatment-individuals-no-dropouts}).
The group-level tests fail to detect statistically significant differences between treatments even at $10\%$ significance level\footnote{With the exception of combined low stake versus high stake which has a $p$-value of $0.0757$ in favor of high stake.}, cf. Table~\ref{tab:mann-whitney-inter-treatment-groups}.
The individual-level tests show significantly higher gains in high stake-binary over low-stake binary, cf. Tables~\ref{tab:mann-whitney-inter-treatment-individuals}--\ref{tab:mann-whitney-inter-treatment-individuals-no-dropouts}.

\newgeometry{left=1.5cm, right=1.5cm} 

\begin{table}[!h]
\centering
\caption{Two-sided \acrfull{MWW} tests reporting test statistics $U$ (and associated $p$-values in the brackets) of the inter-treatment group-level efficiency gains, cf. Equation~\eqref{eq:efficiency-gain-group}.
All treatments have $n=5$ observations.}
\label{tab:mann-whitney-inter-treatment-groups}
\scriptsize
\begin{tabular}{ll||c|c|c||c|c|c||c|c}
\toprule
\multicolumn{10}{c}{\textbf{Treatment Comparisons, Groups}} \\[2pt] \toprule
& & \multicolumn{3}{c||}{Low Stake} & \multicolumn{3}{c||}{High Stake} & \multicolumn{2}{c}{\textbf{Combined}}\\
& & \hspace{1pt} Binary \hspace{1pt} & \hspace{1pt} Full Range \hspace{1pt} & \hspace{1pt} \textbf{Combined} \hspace{1pt} & \hspace{1pt} Binary \hspace{1pt} &
\hspace{1pt} Full Range \hspace{1pt} & \hspace{1pt} \textbf{Combined} \hspace{1pt} & \hspace{1pt} Binary \hspace{1pt} & \hspace{1pt} Full Range \hspace{1pt} \\
\hline \hline
\multirow{4}{*}[-15.5pt]{\rotatebox[origin=c]{90}{Low Stake}} \hspace{1pt} & & & & & & & & & \\[-7pt]
& Binary \hspace{1pt} &
\hspace{1pt} -- \hspace{1pt} &
\hspace{1pt} \begin{tabular}{@{}c@{}}
$12.0$ \\ ($>0.9999$) \end{tabular} \hspace{1pt} & \hspace{1pt} -- \hspace{1pt} &
\hspace{1pt} \begin{tabular}{@{}c@{}}
$6.0$ \\ ($0.2222$) \end{tabular} \hspace{1pt} &
\hspace{1pt} \begin{tabular}{@{}c@{}}
$6.0$ \\ ($0.2222$) \end{tabular} \hspace{1pt} & \hspace{1pt} -- \hspace{1pt} & \hspace{1pt} -- \hspace{1pt} & \hspace{1pt} -- \hspace{1pt} \\
\cmidrule{2-10}
& Full Range \hspace{1pt} &
\hspace{1pt} \begin{tabular}{@{}c@{}}
$13.0$ \\ ($>0.9999$) \\ \end{tabular} \hspace{1pt} &
\hspace{1pt} -- \hspace{1pt} & \hspace{1pt} -- \hspace{1pt} &
\hspace{1pt} \begin{tabular}{@{}c@{}}
$6.0$ \\ ($0.2222$) \\ \end{tabular} \hspace{1pt} &
\hspace{1pt} \begin{tabular}{@{}c@{}}
$8.0$ \\ ($0.4206$) \\ \end{tabular} 
\hspace{1pt} & \hspace{1pt} -- \hspace{1pt} & \hspace{1pt} -- \hspace{1pt} & \hspace{1pt} -- \hspace{1pt} \\
\cmidrule{2-10}
& \textbf{Combined} \hspace{1pt} & \hspace{1pt} -- \hspace{1pt} & \hspace{1pt} -- \hspace{1pt} & \hspace{1pt} -- \hspace{1pt} & \hspace{1pt} -- \hspace{1pt} & \hspace{1pt} -- \hspace{1pt} & \hspace{1pt} \begin{tabular}{@{}c@{}}
$26.0$ \\ ($0.0757$) \\ \end{tabular} \hspace{1pt} & \hspace{1pt} -- \hspace{1pt} & \hspace{1pt} -- \hspace{1pt} \\[5pt]
\hline \hline
\multirow{4}{*}[-15.5pt]{\rotatebox[origin=c]{90}{High Stake}} \hspace{1pt} & & & & & & & & & \\[-7pt]
& Binary \hspace{1pt} &
\hspace{1pt} \begin{tabular}{@{}c@{}}
$19.0$ \\ ($0.2222$) \end{tabular} \hspace{1pt} &
\hspace{1pt} \begin{tabular}{@{}c@{}}
$19.0$ \\ ($0.2222$) \\ \end{tabular} \hspace{1pt} &
\hspace{1pt} -- \hspace{1pt} & \hspace{1pt} -- \hspace{1pt} & 
\hspace{1pt} \begin{tabular}{@{}c@{}}
$15.0$ \\ ($0.3452$) \end{tabular} \hspace{1pt} & \hspace{1pt} -- \hspace{1pt} & \hspace{1pt} -- \hspace{1pt} & \hspace{1pt} -- \hspace{1pt} \\
\cmidrule{2-10}
& Full Range \hspace{1pt} &
\hspace{1pt} \begin{tabular}{@{}c@{}}
$19.0$ \\ ($0.2222$) \\ \end{tabular} \hspace{1pt} &
\hspace{1pt} \begin{tabular}{@{}c@{}}
$17.0$ \\ ($0.4206$) \\ \end{tabular} \hspace{1pt} & \hspace{1pt} -- \hspace{1pt} &
\hspace{1pt} \begin{tabular}{@{}c@{}}
$10.0$ \\ ($0.7262$) \\ \end{tabular} \hspace{1pt} &
\hspace{1pt} -- \hspace{1pt} & \hspace{1pt} -- \hspace{1pt} & \hspace{1pt} -- \hspace{1pt} & \hspace{1pt} -- \hspace{1pt} \\
\cmidrule{2-10}
& \textbf{Combined} \hspace{1pt} & \hspace{1pt} -- \hspace{1pt} & \hspace{1pt} -- \hspace{1pt} & \hspace{1pt} \begin{tabular}{@{}c@{}}
$74.0$ \\ ($0.0757$) \\ \end{tabular} \hspace{1pt} & \hspace{1pt} -- \hspace{1pt} & \hspace{1pt} -- \hspace{1pt} & \hspace{1pt} -- \hspace{1pt} & \hspace{1pt} -- \hspace{1pt} & \hspace{1pt} -- \hspace{1pt} \\[5pt]
\hline \hline
\multirow{3}{*}[-1.5pt]{\rotatebox[origin=c]{90}{\textbf{Combined}}} \hspace{1pt} & & & & & & & & & \\[-7pt]
& Binary \hspace{1pt} & \hspace{1pt} -- \hspace{1pt} & \hspace{1pt} -- \hspace{1pt} & \hspace{1pt} -- \hspace{1pt} & \hspace{1pt} -- \hspace{1pt} & \hspace{1pt} -- \hspace{1pt} & \hspace{1pt} -- \hspace{1pt} & \hspace{1pt} -- \hspace{1pt} & \hspace{1pt} \begin{tabular}{@{}c@{}}
$52.0$ \\ ($0.9097$) \\ \end{tabular} \hspace{1pt} \\
\cmidrule{2-10}
& Full Range \hspace{1pt} & \hspace{1pt} -- \hspace{1pt} & \hspace{1pt} -- \hspace{1pt} & \hspace{1pt} -- \hspace{1pt} & \hspace{1pt} -- \hspace{1pt} & \hspace{1pt} -- \hspace{1pt} & \hspace{1pt} -- \hspace{1pt} & \hspace{1pt} \begin{tabular}{@{}c@{}}
$48.0$ \\ ($0.9097$) \\ \end{tabular} \hspace{1pt} & \hspace{1pt} -- \hspace{1pt} \\
\bottomrule
\end{tabular}
\end{table}

\begin{table}[!h]
\centering
\caption{Two-sided \acrfull{MWW} tests reporting test statistics $U$ (and associated $p$-values in brackets) of the inter-treatment individual-level efficiency gains, cf. Equation~\eqref{eq:efficiency-gain}.
Tests passing the $5\%$ significance level in favor of the row treatment are highlighted green.
All treatments have $n=100$ observations.}
\label{tab:mann-whitney-inter-treatment-individuals}
\scriptsize
\begin{tabular}{ll||c|c|c||c|c|c||c|c}
\toprule
\multicolumn{10}{c}{\textbf{Treatment Comparisons, Individuals}} \\[2pt] \toprule
& & \multicolumn{3}{c||}{Low Stake} & \multicolumn{3}{c||}{High Stake} & \multicolumn{2}{c}{\textbf{Combined}}\\
& & \hspace{1pt} Binary \hspace{1pt} & \hspace{1pt} Full Range \hspace{1pt} & \hspace{1pt} \textbf{Combined} \hspace{1pt} & \hspace{1pt} Binary \hspace{1pt} &
\hspace{1pt} Full Range \hspace{1pt} & \hspace{1pt} \textbf{Combined} \hspace{1pt} & \hspace{1pt} Binary \hspace{1pt} & \hspace{1pt} Full Range \hspace{1pt} \\
\hline \hline
\multirow{4}{*}[-15.5pt]{\rotatebox[origin=c]{90}{Low Stake}} \hspace{1pt} & & & & & & & & & \\[-7pt]
& Binary \hspace{1pt} &
\hspace{1pt} -- \hspace{1pt} &
\hspace{1pt} \begin{tabular}{@{}c@{}}
$4\,705.5$ \\ ($0.4725$) \end{tabular} \hspace{1pt} & \hspace{1pt} -- \hspace{1pt} &
\hspace{1pt} \begin{tabular}{@{}c@{}}
$4\,194.0$ \\ ($0.0490$) \end{tabular} \hspace{1pt} &
\hspace{1pt} \begin{tabular}{@{}c@{}}
$4\,534.0$ \\ ($0.2554$) \end{tabular} \hspace{1pt} & \hspace{1pt} -- \hspace{1pt} & \hspace{1pt} -- \hspace{1pt} & \hspace{1pt} -- \hspace{1pt} \\
\cmidrule{2-10}
& Full Range \hspace{1pt} &
\hspace{1pt} \begin{tabular}{@{}c@{}}
$5\,294.5$ \\ ($0.4725$) \\ \end{tabular} \hspace{1pt} &
\hspace{1pt} -- \hspace{1pt} & \hspace{1pt} -- \hspace{1pt} &
\hspace{1pt} \begin{tabular}{@{}c@{}}
$4\,418.5$ \\ ($0.1557$) \\ \end{tabular} \hspace{1pt} &
\hspace{1pt} \begin{tabular}{@{}c@{}}
$4\,752.0$ \\ ($0.5453$) \\ \end{tabular} 
\hspace{1pt} & \hspace{1pt} -- \hspace{1pt} & \hspace{1pt} -- \hspace{1pt} & \hspace{1pt} -- \hspace{1pt} \\
\cmidrule{2-10}
& \textbf{Combined} \hspace{1pt} & \hspace{1pt} -- \hspace{1pt} & \hspace{1pt} -- \hspace{1pt} & \hspace{1pt} -- \hspace{1pt} & \hspace{1pt} -- \hspace{1pt} & \hspace{1pt} -- \hspace{1pt} & \hspace{1pt} \begin{tabular}{@{}c@{}}
$17\,898.5$ \\ ($0.0692$) \\ \end{tabular} \hspace{1pt} & \hspace{1pt} -- \hspace{1pt} & \hspace{1pt} -- \hspace{1pt} \\[5pt]
\hline \hline
\multirow{4}{*}[-15.5pt]{\rotatebox[origin=c]{90}{High Stake}} \hspace{1pt} & & & & & & & & & \\[-7pt]
& Binary \hspace{1pt} &
\hspace{1pt} \begin{tabular}{@{}c@{}}
\cellcolor{green!25}$5\,806.0$ \\ \cellcolor{green!25}($0.0490$) \end{tabular} \hspace{1pt} &
\hspace{1pt} \begin{tabular}{@{}c@{}}
$5\,581.5$ \\ ($0.1557$) \\ \end{tabular} \hspace{1pt} &
\hspace{1pt} -- \hspace{1pt} & \hspace{1pt} -- \hspace{1pt} & 
\hspace{1pt} \begin{tabular}{@{}c@{}}
$5\,286.0$ \\ ($0.4854$) \end{tabular} \hspace{1pt} & \hspace{1pt} -- \hspace{1pt} & \hspace{1pt} -- \hspace{1pt} & \hspace{1pt} -- \hspace{1pt} \\
\cmidrule{2-10}
& Full Range \hspace{1pt} &
\hspace{1pt} \begin{tabular}{@{}c@{}}
$5\,466.0$ \\ ($0.2554$) \\ \end{tabular} \hspace{1pt} &
\hspace{1pt} \begin{tabular}{@{}c@{}}
$5\,248.0$ \\ ($0.5453$) \\ \end{tabular} \hspace{1pt} & \hspace{1pt} -- \hspace{1pt} &
\hspace{1pt} \begin{tabular}{@{}c@{}}
$4\,714.0$ \\ ($0.4854$) \\ \end{tabular} \hspace{1pt} &
\hspace{1pt} -- \hspace{1pt} & \hspace{1pt} -- \hspace{1pt} & \hspace{1pt} -- \hspace{1pt} & \hspace{1pt} -- \hspace{1pt} \\
\cmidrule{2-10}
& \textbf{Combined} \hspace{1pt} & \hspace{1pt} -- \hspace{1pt} & \hspace{1pt} -- \hspace{1pt} & \hspace{1pt} \begin{tabular}{@{}c@{}}
$22\,101.5$ \\ ($0.0692$) \\ \end{tabular} \hspace{1pt} & \hspace{1pt} -- \hspace{1pt} & \hspace{1pt} -- \hspace{1pt} & \hspace{1pt} -- \hspace{1pt} & \hspace{1pt} -- \hspace{1pt} & \hspace{1pt} -- \hspace{1pt} \\[5pt]
\hline \hline
\multirow{3}{*}[-1.5pt]{\rotatebox[origin=c]{90}{\textbf{Combined}}} \hspace{1pt} & & & & & & & & & \\[-7pt]
& Binary \hspace{1pt} & \hspace{1pt} -- \hspace{1pt} & \hspace{1pt} -- \hspace{1pt} & \hspace{1pt} -- \hspace{1pt} & \hspace{1pt} -- \hspace{1pt} & \hspace{1pt} -- \hspace{1pt} & \hspace{1pt} -- \hspace{1pt} & \hspace{1pt} -- \hspace{1pt} & \hspace{1pt} \begin{tabular}{@{}c@{}}
$20\,107.0$ \\ ($0.9266$) \\ \end{tabular} \hspace{1pt} \\
\cmidrule{2-10}
& Full Range \hspace{1pt} & \hspace{1pt} \hspace{1pt} & \hspace{1pt} -- \hspace{1pt} & \hspace{1pt} -- \hspace{1pt} & \hspace{1pt} -- \hspace{1pt} & \hspace{1pt} -- \hspace{1pt} & \hspace{1pt} -- \hspace{1pt} & \hspace{1pt} \begin{tabular}{@{}c@{}}
$19\,893.0$ \\ ($0.9266$) \\ \end{tabular} \hspace{1pt} & \hspace{1pt} -- \hspace{1pt} \\
\bottomrule
\end{tabular}
\end{table}

\begin{table}[!h]
\centering
\caption{Analogue of Table~\ref{tab:mann-whitney-inter-treatment-individuals} of the two-sided \acrfull{MWW} tests of individual-level efficiency gains between treatments, but excluding dropout observations.
Treatments have varying sample sizes $n$ depending on the number of dropouts, as indicated.}
\label{tab:mann-whitney-inter-treatment-individuals-no-dropouts}
\scriptsize
\begin{tabular}{ll||c|c|c||c|c|c||c|c}
\toprule
\multicolumn{10}{c}{\textbf{Treatment Comparisons, Individuals, Dropouts Excluded}} \\[2pt] \toprule
& & \multicolumn{3}{c||}{Low Stake} & \multicolumn{3}{c||}{High Stake} & \multicolumn{2}{c}{\textbf{Combined}} \\
& & \hspace{1pt} Binary \hspace{1pt} & \hspace{1pt} Full Range \hspace{1pt} & \hspace{1pt} \textbf{Combined} \hspace{1pt} & \hspace{1pt} Binary \hspace{1pt} &
\hspace{1pt} Full Range \hspace{1pt} & \hspace{1pt} \textbf{Combined} \hspace{1pt} & \hspace{1pt} Binary \hspace{1pt} & \hspace{1pt} Full Range \hspace{1pt} \\
& & \hspace{1pt} ($n=93$) \hspace{1pt} & \hspace{1pt} ($n=95$) \hspace{1pt} & \hspace{1pt} ($n=188$) \hspace{1pt} & \hspace{1pt} ($n=93$) \hspace{1pt} &
\hspace{1pt} ($n=96$) \hspace{1pt} & \hspace{1pt} ($n=189$) \hspace{1pt} & \hspace{1pt} ($n=186$) \hspace{1pt} & \hspace{1pt} ($n=191$) \hspace{1pt} \\
\hline \hline
\multirow{4}{*}[-15.5pt]{\rotatebox[origin=c]{90}{Low Stake}} \hspace{1pt} & & & & & & & & & \\[-7pt]
& \begin{tabular}{@{}l@{}} Binary \\ ($n=93$) \end{tabular} \hspace{1pt} &
\hspace{1pt} -- \hspace{1pt} &
\hspace{1pt} \begin{tabular}{@{}c@{}}
$4\,252.5$ \\ ($0.6592$) \end{tabular} \hspace{1pt} & \hspace{1pt} -- \hspace{1pt} &
\hspace{1pt} \begin{tabular}{@{}c@{}}
$3\,577.0$ \\ ($0.0419$) \end{tabular} \hspace{1pt} &
\hspace{1pt} \begin{tabular}{@{}c@{}}
$4\,139.0$ \\ ($0.3881$) \end{tabular} \hspace{1pt} & \hspace{1pt} -- \hspace{1pt} & \hspace{1pt} -- \hspace{1pt} & \hspace{1pt} -- \hspace{1pt} \\
\cmidrule{2-10}
& \begin{tabular}{@{}l@{}} Full Range \\ ($n=95$) \end{tabular} \hspace{1pt} &
\hspace{1pt} \begin{tabular}{@{}c@{}}
$4\,582.5$ \\ ($0.6592$) \\ \end{tabular} \hspace{1pt} &
\hspace{1pt} -- \hspace{1pt} & \hspace{1pt} -- \hspace{1pt} &
\hspace{1pt} \begin{tabular}{@{}c@{}}
$3\,774.0$ \\ ($0.0848$) \\ \end{tabular} \hspace{1pt} &
\hspace{1pt} \begin{tabular}{@{}c@{}}
$4\,328.0$ \\ ($0.5445$) \\ \end{tabular} 
\hspace{1pt} & \hspace{1pt} -- \hspace{1pt} & \hspace{1pt} -- \hspace{1pt} & \hspace{1pt} -- \hspace{1pt} \\
\cmidrule{2-10}
& \begin{tabular}{@{}l@{}} \textbf{Combined} \\ ($n=188$) \end{tabular} \hspace{1pt} & \hspace{1pt} -- \hspace{1pt} & \hspace{1pt} -- \hspace{1pt} & \hspace{1pt} -- \hspace{1pt} & \hspace{1pt} -- \hspace{1pt} & \hspace{1pt} -- \hspace{1pt} & \hspace{1pt} \begin{tabular}{@{}c@{}}
$15\,818.0$ \\ ($0.0656$) \\ \end{tabular} \hspace{1pt} & \hspace{1pt} -- \hspace{1pt} & \hspace{1pt} -- \hspace{1pt} \\[5pt]
\hline \hline
\multirow{4}{*}[-15.5pt]{\rotatebox[origin=c]{90}{High Stake}} \hspace{1pt} & & & & & & & & & \\[-7pt]
& \begin{tabular}{@{}l@{}} Binary \\ ($n=93$) \end{tabular} \hspace{1pt} &
\hspace{1pt} \begin{tabular}{@{}c@{}}
\cellcolor{green!25}$5\,072.0$ \\ \cellcolor{green!25}($0.0419$) \end{tabular} \hspace{1pt} &
\hspace{1pt} \begin{tabular}{@{}c@{}}
$5\,061.0$ \\ ($0.0848$) \\ \end{tabular} \hspace{1pt} &
\hspace{1pt} -- \hspace{1pt} & \hspace{1pt} -- \hspace{1pt} & 
\hspace{1pt} \begin{tabular}{@{}c@{}}
$4\,842.5$ \\ ($0.3147$) \end{tabular} \hspace{1pt} & \hspace{1pt} -- \hspace{1pt} & \hspace{1pt} -- \hspace{1pt} & \hspace{1pt} -- \hspace{1pt} \\
\cmidrule{2-10}
& \begin{tabular}{@{}l@{}} Full Range \\ ($n=96$) \end{tabular} \hspace{1pt} &
\hspace{1pt} \begin{tabular}{@{}c@{}}
$4\,789.0$ \\ ($0.3881$) \\ \end{tabular} \hspace{1pt} &
\hspace{1pt} \begin{tabular}{@{}c@{}}
$4\,792.0$ \\ ($0.5445$) \\ \end{tabular} \hspace{1pt} & \hspace{1pt} -- \hspace{1pt} &
\hspace{1pt} \begin{tabular}{@{}c@{}}
$4\,085.5$ \\ ($0.3147$) \\ \end{tabular} \hspace{1pt} &
\hspace{1pt} -- \hspace{1pt} & \hspace{1pt} -- \hspace{1pt} & \hspace{1pt} -- \hspace{1pt} & \hspace{1pt} -- \hspace{1pt} \\
\cmidrule{2-10}
& \begin{tabular}{@{}l@{}} \textbf{Combined} \\ ($n=189$) \end{tabular} \hspace{1pt} & \hspace{1pt} -- \hspace{1pt} & \hspace{1pt} -- \hspace{1pt} & \hspace{1pt} \begin{tabular}{@{}c@{}}
$19\,714.0$ \\ ($0.0656$) \\ \end{tabular} \hspace{1pt} & \hspace{1pt} -- \hspace{1pt} & \hspace{1pt} -- \hspace{1pt} & \hspace{1pt} -- \hspace{1pt} & \hspace{1pt} -- \hspace{1pt} & \hspace{1pt} -- \hspace{1pt} \\[5pt]
\hline \hline
\multirow{3}{*}[-1.5pt]{\rotatebox[origin=c]{90}{\textbf{Combined}}} \hspace{1pt} & & & & & & & & & \\[-7pt]
& \begin{tabular}{@{}l@{}} Binary \\ ($n=186$) \end{tabular} \hspace{1pt} & \hspace{1pt} -- \hspace{1pt} & \hspace{1pt} -- \hspace{1pt} & \hspace{1pt} -- \hspace{1pt} & \hspace{1pt} -- \hspace{1pt} & \hspace{1pt} -- \hspace{1pt} & \hspace{1pt} -- \hspace{1pt} & \hspace{1pt} -- \hspace{1pt} & \hspace{1pt} \begin{tabular}{@{}c@{}}
$18\,295.0$ \\ ($0.6154$) \\ \end{tabular} \hspace{1pt} \\
\cmidrule{2-10}
& \begin{tabular}{@{}l@{}} Full Range \\ ($n=191$) \end{tabular} \hspace{1pt} & \hspace{1pt} -- \hspace{1pt} & \hspace{1pt} -- \hspace{1pt} & \hspace{1pt} -- \hspace{1pt} & \hspace{1pt} -- \hspace{1pt} & \hspace{1pt} -- \hspace{1pt} & \hspace{1pt} -- \hspace{1pt} & \hspace{1pt} \begin{tabular}{@{}c@{}}
$17\,231.0$ \\ ($0.6154$) \\ \end{tabular} \hspace{1pt} & \hspace{1pt} -- \hspace{1pt} \\
\bottomrule
\end{tabular}
\end{table}

\restoregeometry


    
    

Table~\ref{tab:mann-whitney-deciles} reports two-sided Wilcoxon signed rank test results when comparing the individual-level efficiency gains of karma versus the mean gain from simulating the random benchmark in each of the lowest three deciles, for each treatment and combined.
All higher deciles show similarly significant differences in favor of karma as the third-lowest decile and are omitted for visual clarity.
The test results support the findings of Figure~\ref{fig:mean-efficiency-gain-per-percentile}.
Table~\ref{tab:mann-whitney-deciles-no-dropouts} repeats these tests but with dropout observations excluded from the karma treatments, and in this case results are shown for the lowest two deciles only as all higher deciles show similarly significant differences in favor of karma as the second-lowest decile with dropouts excluded.

\begin{table}[!h]
\centering 
\caption{Two-sided Wilcoxon signed rank tests reporting test statistics $W$ and associated $p$-values of the individual-level efficiency gains, cf. Equation~\eqref{eq:efficiency-gain}, in each of the three lowest deciles under karma versus the mean in the respective decile of the simulated random benchmark for all treatments.
Tests passing the $5\%$ significance level with central tendency in favor of karma (respectively, random) are highlighted green (respectively, red).
Most tests have $n = 10$ observations, as indicated, except where there were ties at percentile values (which were included in the lower decile). The observations are compared to the mean of $\approx200$ observations corresponding to simulating random allocation $100$ times each with a group of 20 simulated individuals.}
\label{tab:mann-whitney-deciles}
\small
\begin{tabular}{lc||c|l|c|c}
\toprule
& & & \multicolumn{3}{c}{\textbf{Richness of scheme}} \\
& & Percentile & \multicolumn{1}{c|}{\hspace{1pt} Binary\hspace{1pt}} & \multicolumn{1}{c|}{\hspace{1pt} Full Range \hspace{1pt}} & \multicolumn{1}{c}{\hspace{1pt} \textbf{Combined} \hspace{1pt}} \\
\hline \hline
\multirow{10}{*}[-144pt]{\rotatebox[origin=c]{90}{\textbf{Urgency process}}} \hspace{1pt} & & & \\[-9pt]
& \multirow{3}{*}[-33pt]{Low Stake} & $0-10\%$ & \hspace{1pt} \begin{tabular}{@{}rc@{}} $n$: & \cellcolor{red!25}$10$ \\ $W$: & \cellcolor{red!25}$0.0$ \\ $p$: & \cellcolor{red!25}$0.0020$
\end{tabular} \hspace{1pt} &
\hspace{1pt} \begin{tabular}{@{}c@{}} \cellcolor{red!25}$10$ \\ \cellcolor{red!25}$1.0$ \\ \cellcolor{red!25}$0.0039$ \end{tabular} \hspace{1pt} &
\hspace{1pt} \begin{tabular}{@{}c@{}} \cellcolor{red!25}$20$ \\ \cellcolor{red!25}$1.0$ \\ \cellcolor{red!25}$<0.0001$ \end{tabular} \hspace{1pt} \\
\cmidrule{3-6}
&  & $10-20\%$ & \hspace{1pt} \begin{tabular}{@{}rc@{}} $n$: & $10$ \\ $W$: & $18.0$ \\ $p$: & $0.3750$ \end{tabular} \hspace{1pt} &
\hspace{1pt} \begin{tabular}{@{}c@{}} \cellcolor{green!25}$10$ \\ \cellcolor{green!25}$5.0$ \\ \cellcolor{green!25}$0.0195$ \end{tabular} \hspace{1pt} &
\hspace{1pt} \begin{tabular}{@{}c@{}} \cellcolor{green!25}$20$ \\ \cellcolor{green!25}$41.0$ \\ \cellcolor{green!25}$0.0153$ \end{tabular} \hspace{1pt} \\
\cmidrule{3-6}
&  & $20-30\%$ & \hspace{1pt} \begin{tabular}{@{}rc@{}} $n$: & \cellcolor{green!25}$13$ \\ $W$: & \cellcolor{green!25}$0.0$ \\ $p$: & \cellcolor{green!25}$0.0002$ \end{tabular} \hspace{1pt} &
\hspace{1pt} \begin{tabular}{@{}c@{}} \cellcolor{green!25}$10$ \\ \cellcolor{green!25}$0.0$ \\ \cellcolor{green!25}$0.0020$ \end{tabular} \hspace{1pt} &
\hspace{1pt} \begin{tabular}{@{}c@{}} \cellcolor{green!25}$23$ \\ \cellcolor{green!25}$0.0$ \\ \cellcolor{green!25}$<0.0001$ \end{tabular} \hspace{1pt} \\
\cmidrule{2-6}
& \multirow{3}{*}[-33pt]{High Stake} & $0-10\%$ & \hspace{1pt} \begin{tabular}{@{}rc@{}} $n$: & \cellcolor{red!25}$10$ \\  $W$: & \cellcolor{red!25}$6.0$ \\ $p$: & \cellcolor{red!25}$0.0273$ \end{tabular} \hspace{1pt} &
\hspace{1pt} \begin{tabular}{@{}c@{}} \cellcolor{red!25}$10$ \\ \cellcolor{red!25}$4.0$ \\ \cellcolor{red!25}$0.0137$ \end{tabular} \hspace{1pt} &
\hspace{1pt} \begin{tabular}{@{}c@{}} \cellcolor{red!25}$20$ \\ \cellcolor{red!25}$16.0$ \\ \cellcolor{red!25}$0.0003$ \end{tabular} \hspace{1pt} \\
\cmidrule{3-6}
&  & $10-20\%$ & \hspace{1pt} \begin{tabular}{@{}rc@{}} $n$: & $10$ \\  $W$: & $10.0$ \\ $p$: & $0.0840$ \end{tabular} \hspace{1pt} &
\hspace{1pt} \begin{tabular}{@{}c@{}} $10$ \\ $26.0$ \\ $0.9219$ \end{tabular} \hspace{1pt} &
\hspace{1pt} \begin{tabular}{@{}c@{}} $20$ \\ $69.0$ \\ $0.1893$ \end{tabular} \hspace{1pt} \\
\cmidrule{3-6}
&  & $20-30\%$ & \hspace{1pt} \begin{tabular}{@{}rc@{}} $n$: & \cellcolor{green!25}$10$ \\ $W$: & \cellcolor{green!25}$0.0$ \\ $p$: & \cellcolor{green!25}$0.0020$ \end{tabular} \hspace{1pt} &
\hspace{1pt} \begin{tabular}{@{}c@{}} \cellcolor{green!25}$10$ \\ \cellcolor{green!25}$0.0$ \\ \cellcolor{green!25}$0.0020$ \end{tabular} \hspace{1pt} &
\hspace{1pt} \begin{tabular}{@{}c@{}} \cellcolor{green!25}$20$ \\ \cellcolor{green!25}$0.0$ \\ \cellcolor{green!25}$<0.0001$ \end{tabular} \hspace{1pt} \\
\cmidrule{2-6}
& \multirow{3}{*}[-33pt]{\textbf{Combined}} & $0-10\%$ & \hspace{1pt} \begin{tabular}{@{}rc@{}} $n$: & \cellcolor{red!25}$20$ \\ $W$: & \cellcolor{red!25}$4.0$ \\ $p$: & \cellcolor{red!25}$<0.0001$ \end{tabular} \hspace{1pt} &
\hspace{1pt} \begin{tabular}{@{}c@{}} \cellcolor{red!25}$20$ \\ \cellcolor{red!25}$8.0$ \\ \cellcolor{red!25}$<0.0001$ \end{tabular} \hspace{1pt} &
\hspace{1pt} \begin{tabular}{@{}c@{}} \cellcolor{red!25}$40$ \\ \cellcolor{red!25}$26.0$ \\ \cellcolor{red!25}$<0.0001$ \end{tabular} \hspace{1pt} \\
\cmidrule{3-6}
&  & $10-20\%$ & \hspace{1pt} \begin{tabular}{@{}rc@{}} $n$: & \cellcolor{green!25}$20$ \\ $W$: & \cellcolor{green!25}$52.0$ \\ $p$: & \cellcolor{green!25}$0.0484$ \end{tabular} \hspace{1pt} &
\hspace{1pt} \begin{tabular}{@{}c@{}} $20$ \\ $63.0$ \\ $0.1231$ \end{tabular} \hspace{1pt} &
\hspace{1pt} \begin{tabular}{@{}c@{}} \cellcolor{green!25}$40$ \\ \cellcolor{green!25}$217.0$ \\ \cellcolor{green!25}$0.0086$ \end{tabular} \hspace{1pt} \\
\cmidrule{3-6}
&  & $20-30\%$ & \hspace{1pt} \begin{tabular}{@{}rc@{}} $n$: & \cellcolor{green!25}$23$ \\ $W$: & \cellcolor{green!25}$0.0$ \\ $p$: & \cellcolor{green!25}$<0.0001$ \end{tabular} \hspace{1pt} &
\hspace{1pt} \begin{tabular}{@{}c@{}} \cellcolor{green!25}$20$ \\ \cellcolor{green!25}$0.0$ \\ \cellcolor{green!25}$<0.0001$ \end{tabular} \hspace{1pt} &
\hspace{1pt} \begin{tabular}{@{}c@{}} \cellcolor{green!25}$43$ \\ \cellcolor{green!25}$0.0$ \\ \cellcolor{green!25}$<0.0001$ \end{tabular} \hspace{1pt} \\
\bottomrule
\end{tabular}
\end{table}

\begin{table}[!h]
\centering 
\caption{Analogue of Table~\ref{tab:mann-whitney-deciles} of the two-sided Wilcoxon signed rank tests of individual-level efficiency gains in each of the two lowest deciles under karma versus the mean of random simulations, but excluding dropout observations from the karma treatments.
Karma treatments have varying sample sizes $n$ depending on the number of dropouts and whether there were ties at percentile values, as indicated.}
\label{tab:mann-whitney-deciles-no-dropouts}
\small
\begin{tabular}{lc||c|l|c|c}
\toprule
& & & \multicolumn{3}{c}{\textbf{Richness of scheme}} \\
& & Percentile & \multicolumn{1}{c|}{\hspace{1pt} Binary\hspace{1pt}} & \multicolumn{1}{c|}{\hspace{1pt} Full Range \hspace{1pt}} & \multicolumn{1}{c}{\hspace{1pt} \textbf{Combined} \hspace{1pt}} \\
\hline \hline
\multirow{10}{*}[-114pt]{\rotatebox[origin=c]{90}{\textbf{Urgency process}}} \hspace{1pt} & & & \\[-9pt]
& \multirow{3}{*}[-17pt]{Low Stake} & $0-10\%$ & \hspace{1pt} \begin{tabular}{@{}rc@{}} $n$: & $10$ \\ $W$: & $15.0$ \\ $p$: & $0.2324$
\end{tabular} \hspace{1pt} &
\hspace{1pt} \begin{tabular}{@{}c@{}} $10$ \\ $20.0$ \\ $0.4922$ \end{tabular} \hspace{1pt} &
\hspace{1pt} \begin{tabular}{@{}c@{}} $20$ \\ $93.0$ \\ $0.6742$ \end{tabular} \hspace{1pt} \\
\cmidrule{3-6}
&  & $10-20\%$ & \hspace{1pt} \begin{tabular}{@{}rc@{}} $n$: & \cellcolor{green!25}$9$ \\ $W$: & \cellcolor{green!25}$0.0$ \\ $p$: & \cellcolor{green!25}$0.0039$ \end{tabular} \hspace{1pt} &
\hspace{1pt} \begin{tabular}{@{}c@{}} \cellcolor{green!25}$9$ \\ \cellcolor{green!25}$0.0$ \\ \cellcolor{green!25}$0.0039$ \end{tabular} \hspace{1pt} &
\hspace{1pt} \begin{tabular}{@{}c@{}} \cellcolor{green!25}$18$ \\ \cellcolor{green!25}$0.0$ \\ \cellcolor{green!25}$<0.0001$ \end{tabular} \hspace{1pt} \\
\cmidrule{2-6}
& \multirow{3}{*}[-17pt]{High Stake} & $0-10\%$ & \hspace{1pt} \begin{tabular}{@{}rc@{}} $n$: & $10$ \\ $W$: & $11.0$ \\ $p$: & $0.1055$ \end{tabular} \hspace{1pt} &
\hspace{1pt} \begin{tabular}{@{}c@{}} $10$ \\ $24.0$ \\ $0.7695$ \end{tabular} \hspace{1pt} &
\hspace{1pt} \begin{tabular}{@{}c@{}} $20$ \\ $78.0$ \\ $0.3300$ \end{tabular} \hspace{1pt} \\
\cmidrule{3-6}
&  & $10-20\%$ & \hspace{1pt} \begin{tabular}{@{}rc@{}} $n$: & \cellcolor{green!25}$9$ \\ $W$: & \cellcolor{green!25}$0.0$ \\ $p$: & \cellcolor{green!25}$0.0039$ \end{tabular} \hspace{1pt} &
\hspace{1pt} \begin{tabular}{@{}c@{}} \cellcolor{green!25}$10$ \\ \cellcolor{green!25}$4.0$ \\ \cellcolor{green!25}$0.0137$ \end{tabular} \hspace{1pt} &
\hspace{1pt} \begin{tabular}{@{}c@{}} \cellcolor{green!25}$19$ \\ \cellcolor{green!25}$5.0$ \\ \cellcolor{green!25}$<0.0001$ \end{tabular} \hspace{1pt} \\
\cmidrule{2-6}
& \multirow{3}{*}[-17pt]{\textbf{Combined}} & $0-10\%$ & \hspace{1pt} \begin{tabular}{@{}rc@{}} $n$: & \cellcolor{green!25}$20$ \\ $W$: & \cellcolor{green!25}$50.0$ \\ $p$: & \cellcolor{green!25}$0.0400$ \end{tabular} \hspace{1pt} &
\hspace{1pt} \begin{tabular}{@{}c@{}} $20$ \\ $89.0$ \\ $0.5706$ \end{tabular} \hspace{1pt} &
\hspace{1pt} \begin{tabular}{@{}c@{}} $40$ \\ $317.0$ \\ $0.2163$ \end{tabular} \hspace{1pt} \\
\cmidrule{3-6}
&  & $10-20\%$ & \hspace{1pt} \begin{tabular}{@{}rc@{}} $n$: & \cellcolor{green!25}$18$ \\ $W$: & \cellcolor{green!25}$0.0$ \\ $p$: & \cellcolor{green!25}$<0.0001$ \end{tabular} \hspace{1pt} &
\hspace{1pt} \begin{tabular}{@{}c@{}} \cellcolor{green!25}$19$ \\ \cellcolor{green!25}$8.0$ \\ \cellcolor{green!25}$0.0001$ \end{tabular} \hspace{1pt} &
\hspace{1pt} \begin{tabular}{@{}c@{}} \cellcolor{green!25}$37$ \\ \cellcolor{green!25}$9.0$ \\ \cellcolor{green!25}$<0.0001$ \end{tabular} \hspace{1pt} \\
\bottomrule
\end{tabular}
\end{table}

Table~\ref{tab:wilcoxon-diff-to-Nash} reports two-sided Wilcoxon signed rank test results testing if the individual-level efficiency gains of karma in each bin of mean absolute differences to Nash, portrayed in Figure~\ref{fig:mean-efficiency-gain-per-mean-equilibrium-abs-diff}, are statistically significantly different from zero, for each treatment.
The test results support the findings of Figure~\ref{fig:mean-efficiency-gain-per-mean-equilibrium-abs-diff}.

\begin{table}[!h]
\centering 
\caption{Two-sided Wilcoxon signed rank tests reporting test statistics $W$ and associated $p$-values of the individual-level efficiency gains, cf. Equation~\eqref{eq:efficiency-gain}, in each bin of mean absolute differences to Nash portrayed in Figure~\ref{fig:mean-efficiency-gain-per-mean-equilibrium-abs-diff}.
Tests passing the $5\%$ significance level with positive (respectively, negative) central tendency are highlighted green (respectively, red).
Mean absolute difference to Nash bins have varying sample sizes $n$, as indicated.
Tests that had insufficient observations to detect significant differences, i.e., $n < 6$, are highlighted orange.}
\label{tab:wilcoxon-diff-to-Nash}
\scriptsize
\begin{tabular}{lc||c|l|c|c}
\toprule
& & Mean abs. diff. & \multicolumn{3}{c}{\textbf{Richness of scheme}} \\
& & to Nash [karma] & \multicolumn{1}{c|}{\hspace{1pt} Binary\hspace{1pt}} & \multicolumn{1}{c}{\hspace{1pt} Full Range \hspace{1pt}} & \multicolumn{1}{c}{\hspace{1pt} \textbf{Combined} \hspace{1pt}} \\
\hline \hline
\multirow{14}{*}[-172pt]{\rotatebox[origin=c]{90}{\textbf{Urgency process}}} \hspace{1pt} & & & \\[-5pt]
& \multirow{5}{*}[-50pt]{Low Stake} & $0-1$ & \hspace{1pt} \begin{tabular}{@{}rc@{}} $n$: & \cellcolor{green!25}14 \\ $W$: & \cellcolor{green!25}$0.0$ \\ $p$: & \cellcolor{green!25}$0.0001$ \end{tabular} \hspace{1pt} &
\hspace{1pt} \begin{tabular}{@{}c@{}} \cellcolor{orange!25}$2$ \\ \cellcolor{orange!25}$0.0$ \\ \cellcolor{orange!25}$0.5000$ \end{tabular} \hspace{1pt} & \hspace{1pt} \begin{tabular}{@{}c@{}} \cellcolor{green!25}$16$ \\ \cellcolor{green!25}$0.0$ \\ \cellcolor{green!25}$<0.0001$ \end{tabular} \hspace{1pt} \\
\cmidrule{3-6}
&  & $1-2$ & \hspace{1pt} \begin{tabular}{@{}rc@{}} $n$: & \cellcolor{green!25}55 \\ $W$: & \cellcolor{green!25}$74.5$ \\ $p$: & \cellcolor{green!25}$<0.0001$ \end{tabular} \hspace{1pt} &
\hspace{1pt} \begin{tabular}{@{}c@{}} \cellcolor{green!25}$27$ \\ \cellcolor{green!25}$1.0$ \\ \cellcolor{green!25}$<0.0001$ \end{tabular} \hspace{1pt} & \hspace{1pt} \begin{tabular}{@{}c@{}} \cellcolor{green!25}$82$ \\ \cellcolor{green!25}$94.5$ \\ \cellcolor{green!25}$<0.0001$ \end{tabular} \hspace{1pt} \\
\cmidrule{3-6}
&  & $2-3$ & \hspace{1pt} \begin{tabular}{@{}rc@{}} $n$: & 21 \\ $W$: & $58.0$ \\ $p$: & $0.0793$ \end{tabular} \hspace{1pt} &
\hspace{1pt} \begin{tabular}{@{}c@{}} \cellcolor{green!25}$50$ \\ \cellcolor{green!25}$123.5$ \\ \cellcolor{green!25}$<0.0001$ \end{tabular} \hspace{1pt} & \hspace{1pt} \begin{tabular}{@{}c@{}} \cellcolor{green!25}$71$ \\ \cellcolor{green!25}$661.0$ \\ \cellcolor{green!25}$0.0018$ \end{tabular} \hspace{1pt} \\
\cmidrule{3-6}
&  & $3-4$ & \hspace{1pt} \begin{tabular}{@{}rc@{}} $n$: & \cellcolor{orange!25}4 \\ $W$: & \cellcolor{orange!25}$1.0$ \\ $p$: & \cellcolor{orange!25}$0.2500$ \end{tabular} \hspace{1pt} &
\hspace{1pt} \begin{tabular}{@{}c@{}} $10$ \\ $25.0$ \\ $0.8457$ \end{tabular} \hspace{1pt} & \hspace{1pt} \begin{tabular}{@{}c@{}} $14$ \\ $40.0$ \\ $0.4631$ \end{tabular} \hspace{1pt} \\
\cmidrule{3-6}
&  & $4-5$ & \hspace{1pt} \begin{tabular}{@{}rc@{}} $n$: & \cellcolor{orange!25}4 \\ $W$: & \cellcolor{orange!25}$0.0$ \\ $p$: & \cellcolor{orange!25}$0.1250$ \end{tabular} \hspace{1pt} &
\hspace{1pt} \begin{tabular}{@{}c@{}} \cellcolor{orange!25}$3$ \\ \cellcolor{orange!25}$0.0$ \\ \cellcolor{orange!25}$0.2500$ \end{tabular} \hspace{1pt} & \hspace{1pt} \begin{tabular}{@{}c@{}} \cellcolor{red!25}$7$ \\ \cellcolor{red!25}$0.0$ \\ \cellcolor{red!25}$0.0156$ \end{tabular} \hspace{1pt} \\
\cmidrule{2-6}
& \multirow{4}{*}[-36pt]{High Stake} & $0-1$ & \hspace{1pt} \begin{tabular}{@{}rc@{}} $n$: & \cellcolor{green!25}$11$ \\ $W$: & \cellcolor{green!25}$0.0$ \\ $p$: & \cellcolor{green!25}$0.0010$ \end{tabular} \hspace{1pt} &
\hspace{1pt} \begin{tabular}{@{}c@{}} \cellcolor{orange!25}$2$ \\ \cellcolor{orange!25}$0.0$ \\ \cellcolor{orange!25}$0.5000$ \end{tabular} \hspace{1pt} & \hspace{1pt} \begin{tabular}{@{}c@{}} \cellcolor{green!25}$13$ \\ \cellcolor{green!25}$0.0$ \\ \cellcolor{green!25}$0.0002$ \end{tabular} \hspace{1pt} \\
\cmidrule{3-6}
&  & $1-2$ & \hspace{1pt} \begin{tabular}{@{}rc@{}} $n$: & $21$ \\ $W$: & $62.0$ \\ $p$: & $0.0647$ \end{tabular} \hspace{1pt} &
\hspace{1pt} \begin{tabular}{@{}c@{}} \cellcolor{green!25}$14$ \\ \cellcolor{green!25}$11.0$ \\ \cellcolor{green!25}$0.0067$ \end{tabular} \hspace{1pt} & \hspace{1pt} \begin{tabular}{@{}c@{}} \cellcolor{green!25}$35$ \\ \cellcolor{green!25}$124.0$ \\ \cellcolor{green!25}$0.0013$ \end{tabular} \hspace{1pt} \\
\cmidrule{3-6}
&  & $2-3$ & \hspace{1pt} \begin{tabular}{@{}rc@{}} $n$: & \cellcolor{green!25}$59$ \\ $W$: & \cellcolor{green!25}$333.5$ \\ $p$: & \cellcolor{green!25}$<0.0001$ \end{tabular} \hspace{1pt} &
\hspace{1pt} \begin{tabular}{@{}c@{}} \cellcolor{green!25}$61$ \\ \cellcolor{green!25}$258.0$ \\ \cellcolor{green!25}$<0.0001$ \end{tabular} \hspace{1pt} & \hspace{1pt} \begin{tabular}{@{}c@{}} \cellcolor{green!25}$120$ \\ \cellcolor{green!25}$1\,169.5$ \\ \cellcolor{green!25}$<0.0001$ \end{tabular} \hspace{1pt} \\
\cmidrule{3-6}
&  & $3-4$ & \hspace{1pt} \begin{tabular}{@{}rc@{}} $n$: & \cellcolor{red!25}$9$ \\ $W$: & \cellcolor{red!25}$5.0$ \\ $p$: & \cellcolor{red!25}$0.0391$ \end{tabular} \hspace{1pt} &
\hspace{1pt} \begin{tabular}{@{}c@{}} \cellcolor{red!25}$21$ \\ \cellcolor{red!25}$13.0$ \\ \cellcolor{red!25}$0.0010$ \end{tabular} \hspace{1pt} & \hspace{1pt} \begin{tabular}{@{}c@{}} \cellcolor{red!25}$30$ \\ \cellcolor{red!25}$35.0$ \\ \cellcolor{red!25}$0.0001$ \end{tabular} \hspace{1pt} \\
\cmidrule{2-6}
& \multirow{4}{*}[-36pt]{\textbf{Combined}} & $0-1$ & \hspace{1pt} \begin{tabular}{@{}rc@{}} $n$: & \cellcolor{green!25}$25$ \\ $W$: & \cellcolor{green!25}$0.0$ \\ $p$: & \cellcolor{green!25}$<0.0001$ \end{tabular} \hspace{1pt} &
\hspace{1pt} \begin{tabular}{@{}c@{}} \cellcolor{orange!25}$4$ \\ \cellcolor{orange!25}$0.0$ \\ \cellcolor{orange!25}$0.1250$ \end{tabular} \hspace{1pt} & \hspace{1pt} \begin{tabular}{@{}c@{}} \cellcolor{green!25}$29$ \\ \cellcolor{green!25}$0.0$ \\ \cellcolor{green!25}$<0.0001$ \end{tabular} \hspace{1pt} \\
\cmidrule{3-6}
&  & $1-2$ & \hspace{1pt} \begin{tabular}{@{}rc@{}} $n$: & \cellcolor{green!25}$76$ \\ $W$: & \cellcolor{green!25}$325.0$ \\ $p$: & \cellcolor{green!25}$<0.0001$ \end{tabular} \hspace{1pt} &
\hspace{1pt} \begin{tabular}{@{}c@{}} \cellcolor{green!25}$41$ \\ \cellcolor{green!25}$39.0$ \\ \cellcolor{green!25}$<0.0001$ \end{tabular} \hspace{1pt} & \hspace{1pt} \begin{tabular}{@{}c@{}} \cellcolor{green!25}$117$ \\ \cellcolor{green!25}$567.0$ \\ \cellcolor{green!25}$<0.0001$ \end{tabular} \hspace{1pt} \\
\cmidrule{3-6}
&  & $2-3$ & \hspace{1pt} \begin{tabular}{@{}rc@{}} $n$: & \cellcolor{green!25}$80$ \\ $W$: & \cellcolor{green!25}$928.0$ \\ $p$: & \cellcolor{green!25}$0.0014$ \end{tabular} \hspace{1pt} &
\hspace{1pt} \begin{tabular}{@{}c@{}} \cellcolor{green!25}$111$ \\ \cellcolor{green!25}$743.5$ \\ \cellcolor{green!25}$<0.0001$ \end{tabular} \hspace{1pt} & \hspace{1pt} \begin{tabular}{@{}c@{}} \cellcolor{green!25}$191$ \\ \cellcolor{green!25}$3\,480.5$ \\ \cellcolor{green!25}$<0.0001$ \end{tabular} \hspace{1pt} \\
\cmidrule{3-6}
&  & $3-4$ & \hspace{1pt} \begin{tabular}{@{}rc@{}} $n$: & \cellcolor{red!25}$13$ \\ $W$: & \cellcolor{red!25}$8.0$ \\ $p_{\textup k > 0}$: & \cellcolor{red!25}$0.0061$ \end{tabular} \hspace{1pt} &
\hspace{1pt} \begin{tabular}{@{}c@{}} \cellcolor{red!25}$31$ \\ \cellcolor{red!25}$84.0$ \\ \cellcolor{red!25}$0.0039$ \end{tabular} \hspace{1pt} & \hspace{1pt} \begin{tabular}{@{}c@{}} \cellcolor{red!25}$44$ \\ \cellcolor{red!25}$141.0$ \\ \cellcolor{red!25}$0.0001$ \end{tabular} \hspace{1pt} \\
\bottomrule
\end{tabular}
\end{table}

Table~\ref{tab:mann-whitney-stationarity} reports two-sided Wilcoxon signed rank test results when comparing the variations in karma and bid distributions of successive rounds, as measured by the Wasserstein-1 distance, between the experimental karma observations and the mean variation from the simulated stationary Nash benchmark.
The test results support the findings of Figure~\ref{fig:karma-bid-distribution-diff}.

\begin{table}[!h]
\centering 
\caption{Two-sided Wilcoxon signed rank tests reporting test statistics $W$ and associated $p$-values of the Wasserstein-1 variations in karma and bid distributions between successive rounds, for the experimental \gls{MTurk} subjects versus the respective mean variation of the simulated Nash benchmark in all treatments.
Tests passing the $5\%$ significance level with central tendency \gls{MTurk} $<$ Nash (respectively, \gls{MTurk} $>$ Nash) are highlighted green (respectively, red).
All treatments have $n=245$ observations (corresponding to 49 rounds in 5 groups), which are compared to the mean of $4\,900$ observations corresponding to simulating stationary Nash behavior $100$ times each with a group of 20 simulated individuals.
}
\label{tab:mann-whitney-stationarity}
\begin{tabular}{lc||c|l|c}
\toprule
& & Karma / Bid & \multicolumn{2}{c}{\textbf{Richness of scheme}} \\
& & Distribution & \multicolumn{1}{c|}{\hspace{1pt} Binary\hspace{1pt}} & \multicolumn{1}{c}{\hspace{1pt} Full Range \hspace{1pt}} \\
\hline \hline
\multirow{4}{*}[-24pt]{\rotatebox[origin=c]{90}{\textbf{Urgency process}}} \hspace{1pt} & & \\[-9pt]
& \multirow{2}{*}[-10pt]{Low Stake} & Karma & \hspace{1pt} \begin{tabular}{@{}rc@{}} $W$: & \cellcolor{green!25}$5\,726.0$ \\ $p$: & \cellcolor{green!25}$<0.0001$ \end{tabular} \hspace{1pt} &
\hspace{1pt} \begin{tabular}{@{}c@{}} $13\,194.0$ \\ $0.0916$ \end{tabular} \hspace{1pt} \\
\cmidrule{3-5}
&  & Bid & \hspace{1pt} \begin{tabular}{@{}rc@{}} $W$: & \cellcolor{green!25}$6\,126.0$ \\ $p$: & \cellcolor{green!25}$<0.0001$ \end{tabular} \hspace{1pt} &
\hspace{1pt} \begin{tabular}{@{}c@{}} \cellcolor{red!25}$2\,088.0$ \\ \cellcolor{red!25}$<0.0001$ \end{tabular} \hspace{1pt} \\
\cmidrule{2-5}
& \multirow{2}{*}[-10pt]{High Stake} & Karma & \hspace{1pt} \begin{tabular}{@{}rc@{}} $W$: & \cellcolor{green!25}$9\,465.0$ \\ $p$: & \cellcolor{green!25}$<0.0001$ \end{tabular} \hspace{1pt} &
\hspace{1pt} \begin{tabular}{@{}c@{}} \cellcolor{red!25}$9\,327.0$ \\ \cellcolor{red!25}$<0.0001$ \end{tabular} \hspace{1pt} \\
\cmidrule{3-5}
&  & Bid & \hspace{1pt} \begin{tabular}{@{}rc@{}} $W$: & \cellcolor{green!25}$12\,025.0$ \\ $p$: & \cellcolor{green!25}$0.0061$ \end{tabular} \hspace{1pt} &
\hspace{1pt} \begin{tabular}{@{}c@{}} \cellcolor{red!25}$196.0$ \\ \cellcolor{red!25}$<0.0001$ \end{tabular} \hspace{1pt} \\
\bottomrule
\end{tabular}
\end{table}

\end{document}